\documentclass[
 amsmath,amssymb,
reprint,%
aps,
pra,
]{revtex4-2}

\usepackage{graphicx}

\begin{document}

\title{Phenomenology and Dynamics of Competitive Ecosystems Beyond the Niche-Neutral Regimes }

\author{Nava Leibovich}
\author{Jeremy Rothschild}%
\affiliation{%
 Department of Physics, University of Toronto
}%

\author{Sidhartha Goyal}
 \email{goyal@physics.utoronto.ca}
\author{Anton Zilman}
 \email{zilmana@physics.utoronto.ca}
\affiliation{%
 Department of Physics, University of Toronto
}%
\affiliation{
 Institute for Biomedical Engineering, University of Toronto
}%

\keywords{ecological drift $|$ neutral-niche  theory $|$ phase diagram $|$ } 

\begin{abstract}
Structure, composition and stability of ecological populations are shaped by the inter- and intra-species interactions within these communities.
It remains to be fully understood how the interplay of these interactions with other factors, such as immigration, control the structure, diversity and the long term stability of ecological systems in the presence of noise and fluctuations. We address this problem using a minimal model of interacting multi-species ecological communities that incorporates competition, immigration and demographic noise. We find that the complete phase diagram exhibits rich behavior with multiple regimes that go beyond the classical `niche' and `neutral' regimes, extending and modifying the  `neutral-like' or `niche-like' dichotomy. In particular, we observe novel regimes that cannot be characterized as either `niche' or `neutral' where a multimodal species abundance distribution is observed. We characterize the transitions between the different regimes and show how they arise from the underlying kinetics of the species turnover, extinction and invasion. Our model serves as a minimal null model  of noisy competitive ecological systems, against which more complex models that include factors such as mutations and environmental noise can be compared. 

\end{abstract}

\maketitle

\section{Introduction}
\label{sec:introduction}

Composition and behavior of ecological communities are shaped by direct and indirect interactions between the occupant species, such as the competition for the physical space and the intrinsic and the extrinsic resources. 
Examples of such competitive ecosystems are microbial communities in various biomes such as the soil~\cite{ratzke2020strength}, the ocean~\cite{tilman1977resource,strom2008microbial} and the human body~\cite{foster2017evolution} - in particular the human gut which hosts a diverse microbiome whose dynamics are important for human health~\cite{coyte2015ecology,gorter2020understanding}. 
In the context of cellular populations within organisms, the evolution of neoplasms and tumor cells \cite{merlo2006cancer,kareva2015cancer,smart2021roles}, interactions within the immune system ~\cite{tauber2008immune,schmid2003evolutionary}, as well as the appearance of dominant clones during cell reprogramming~\cite{shakiba2019cell}, exhibit phenomenology akin to ecological competition. 
Beyond biology \cite{tilman1982resource,morin2009community,tuljapurkar2013population}, competitive interactions shape behaviors in a vast array of systems such as competition economics \cite{budzinski2007monoculture} and social networks \cite{koura2017competitive}.

A classical example of the effects of inter-species competition - which inspired important ecological competition paradigms  - is the differentiation in beak forms of finches in the Gal\'apagos islands \cite{lewin1983finches,lack1983darwin}. 
On these islands, dissimilar finch species possess beaks of varying shapes and sizes allowing them to consume different food sources and thus occupy distinct niches; this type of ecosystem structure is commonly referred to as an ecological niche model \cite{grant1979darwin,pocheville2015ecological}. 
Various niche models have been used to describe the community structures observed in diverse ecosystems such as plant grassland communities \cite{zuppinger2014selection,silvertown2004plant}, marine plankton~\cite{cullen1998behavior} and conservation ecology~\cite{melo2020ecological,aguirre2015similar}.
Commonly, niche specialization results in weaker competition for resources between individuals occupying separate niches (inter-species competition) compared to the competition between individuals of the same kind residing in the same niche (intra-species competition)\cite{badali2020effects,capitan2015similar,chesson2000mechanisms}. 

Another paradigmatic class of ecological models {that are used to describe noisy ecosystems comprises} neutral models. 
In contrast to niche models, {in neutral models individuals of all species are considered functionally equivalent, and interactions between them} are identical regardless of their species \cite{bell2001neutral,hubbell2001unified,chave2004neutral,marquet2017proportional}.
{One classical example of a neutral model is the Hubbell model, that showed that a neutral process underlying the population dynamics of an ecosystem recovers experimental observations of species abundances in tree communities~\cite{hubbell2001unified}.}
Subsequently, neutral models have commonly served as the paradigmatic null hypotheses for the exploration of ecological processes in {which} the differences between inter-specific and intra-specific interaction are functionally negligible \cite{bell2001neutral,gotelli2006null,blythe2012neutral,mckane2004analytic}.
Neutral theories may be viewed as a limit of niche theories where inter-specific and intra-specific interactions are equal: in other words, all species reside in completely overlapping niches ~\cite{grover1997resource,begon2006ecology,pocheville2015ecological}.

In multi-species communities, the intra- and inter-species interactions as well as interactions with the environment, can lead to complex community composition and population dynamics; some species survive in the long term, while others are driven to extinction.
However, in large communities with high numbers of competing species, it is often impractical or impossible to characterize the entire system composition by the assemblage of abundances for each species. 
Hence, coarse-grained paradigmatic descriptions are often used to provide general insights into the common behavior of these ecological communities.

Two variables commonly used to characterize complex ecological communities are 1) the richness, reflecting the number of co-occurring species \cite{adams2009species,kery2020applied}, and 2) the species abundance distributions (SAD) - the number of species present at a given abundance. { The latter is  closely related to the species proportional abundance distribution (SPAD) as well as} {to the }species rank abundance (SRA) - the species ranked in terms of their abundance \cite{marquet2017proportional,nias1968clone, rulands2018universality, de2020naive, mcgill2007species, matthews2015species}. 
These aggregate variables are observable experimentally and serve as the reporters on the underlying community structure, dynamics and the interaction network \cite{rahbek2001multiscale,hong2006predicting,adler2011productivity,valencia2020synchrony}. 
Richness, for example, is commonly considered to be an indicator of the competition strength and stability of the ecosystem \cite{pimm1984complexity, ives2000stability, jousset2011intraspecific,mallon2015microbial,capitan2017stochastic}. 


The shape of the SAD is also used as a proxy for the structure of the underlying interactions' network.  For high immigration or weak inter-species competition, the SAD commonly has a peak at high species abundance, away from extinction.
This community structure is closely related to the niche models whereby different species co-exist: most species inhabit their own niches with their species abundance fluctuating around the peak of the SAD. Conversely, other ecosystems, such as many microbial communities and T-cell repertoires, commonly comprise few high-abundance species  alongside highly diverse populations of low-abundance species \cite{lynch2015ecology, de2020naive}.
This unimodal, monotonically decreasing SAD - sometimes called a `hollow-curved distribution' - we refer to as the `rare biosphere' SAD. Interestingly, this behaviour is empirically observed in many different ecosystems and is often considered universal (see \cite{leidinger2017biodiversity} and references therein). Neutral models have been championed to describe the emergence of this universality, although other theoretical explanations for the `rare biosphere' SAD in competitive ecosystems have been suggested \cite{mcgill2007species,magurran2013measuring}.

Theoretical studies {commonly employ a small number of paradigmatic models} to quantify the competitive dynamics, the richness and the abundance distributions in ecological populations. 
One common model of ecological competition is the deterministic, competitive Lotka-Voltera (LV) model, which has been especially useful in characterizing the niche regime by describing stable species coexistence as stable fixed points of the model.  
Depending on the ratios of inter- and intra-species competition strengths, deterministic LV models provide examples of both the `niche-like' regimes of multiple species coexistence, and the competitive exclusion where species with weaker intra-species interactions drive others to extinction \cite{hardin1960competitive,macarthur1967limiting,MacArthur1969species, gause2019struggle}.
In complex scenarios, such as when the strengths of inter-specific interactions are randomly distributed among different species pairs, multi-species deterministic LV models can exhibit not only deterministic fixed point coexistence but also chaotic behavior reflected in the SAD shapes and richness \cite{scheffer2006self,vergnon2012emergent,kessler2015generalized,bunin2016interaction,roy2020complex}. 
Beyond disorder in the interaction network, dynamical noise from various sources - both extrinsic and intrinsic - has important effects on the system composition and dynamics, especially in the neutral regime.
In order to capture experimentally observed stochastic fluctuations of population abundances, environmental noise is often introduced into the mathematical models \cite{fisher2014transition,lynch2015ecology,verberk2011explaining,fowler2013colonization,barabas2016effect}.
In particular, by tuning the strength of environmental noise the shape of the SAD can change from unimodal to bimodal \cite{fisher2014transition}, indicating a transition between `niche-like' and `rare biosphere' regimes.
{Incorporating both asymmetric interactions and environmental noise, `patch models' of communities have also been utilized to study coexistence and abundances~\cite{evans2013stochastic,tejo2021coexistence} in island metacommunities.}

Regardless of the presence of the external environmental noise or randomness in the interaction network, the demographic noise - the inherent randomness of birth and death events - is ever-present and has fundamental impact on the community structure and stochastic population dynamics \cite{marquet2017proportional,hubbell2001unified,alonso2006merits,haegeman2011mathematical}.  
%
In particular, demographic noise in neutral systems has been shown to result in an SAD shape characterized by {a monotonically decreasing distribution often referred to as a `rare biosphere' distribution}. 
{Consequently, it has been suggested that the `rare biosphere' SAD observed in many experimental systems is the outcome of neutral dynamics of ecological communities\cite{marquet2017proportional,hubbell2001unified,baxter2007exact,mckane2004analytic}}. 
On the other hand, neutral birth-death-immigration processes with demographic noise have also been shown to exhibit bimodal SADs at very low immigration rates \cite{xu2018immigration} breaking from the paradigm wherein neutrality synonymously refers to an SAD of the `rare biosphere' type.
Although demographic noise models have been shown to reproduce the observed features of a number of ecological systems \cite{haegeman2011mathematical,capitan2015similar,capitan2017stochastic,capitan2020competitive}, a complete picture of the different regimes of community structures, is still missing. 
In particular, it remains to be fully understood how the interplay of the competition strength, the immigration rate, demographic noise and the resulting dynamics of species turnover shape transitions between these different community structure regimes.

\begin{figure}[t!]
   \begin{flushleft}
        A
   \end{flushleft}
    \includegraphics[width=\columnwidth]{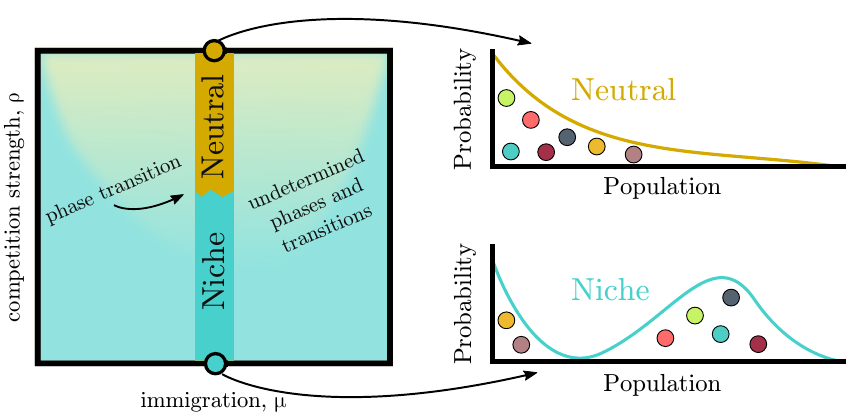}
    \begin{flushleft}
        B
   \end{flushleft}
    \includegraphics[width=\columnwidth]{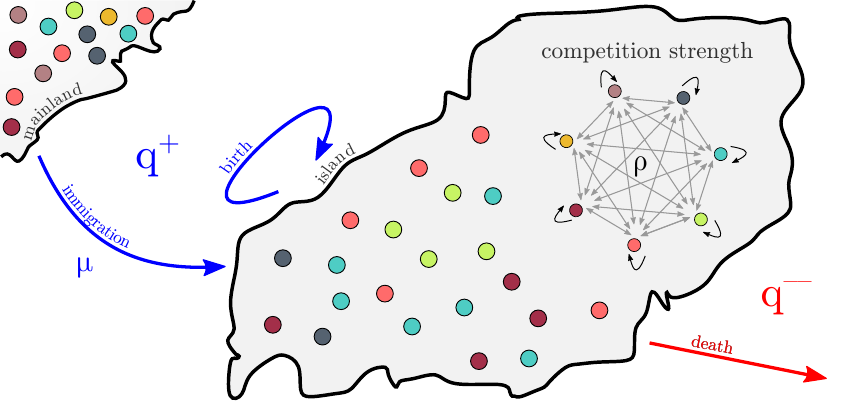}
    \caption{Island model.  Panel \textbf{A}: Conventionally, weak competition is associated with `niche-like' bimodal SAD, while strong competition is linked to `rare biosphere' monotonically decreasing SAD. However, this paradigm is not complete, since the dependence on other parameters, such as immigration rate $\mu$ or diversity $S$, is not fully investigated. Thus, the entire phase space, e.g. $(\mu, \rho)$ or $(S, \rho)$,  remains unexplored. Panel \textbf{B}: The model illustration. An island with $J$ individuals from $S^*$ species. Each individual may proliferate and die with some rate corresponding to inter- and intraspecific interactions within the island. Here we consider deterministic, symmetric, fully-connected interspecific interactions network, governed by single parameter; the competition strength $\rho$. Additionally, individuals may migrate from a cloud/mainland, contains $S$ species, into the island with a constant rate $\mu$.       }
    \label{fig:fig1}
\end{figure}

In this paper, we systematically investigate the full parameter space of the community composition and structure using a competitive LV model with the demographic noise and an interaction network of minimal complexity structure; more complex scenarios may be examined by building on this paradigmatic null model. We show that, beyond the perception of dichotomous neutral-niche regimes, many different regimes of richness and SAD shape emerge from the interplay between the competition strength and immigration in the presence of stochasticity as illustrated in Fig.~\ref{fig:fig1}.
These regimes exhibit contrasting dynamics that underpin the differences in the community structures in different regimes, and the transitions between them. 

The paper is structured as follows. In Section \ref{sec:model} we introduce the minimal model. In Section \ref{sec:results} we present our main results, including the regimes boundaries, their richness and the abundance distributions, as well as their associated underlying dynamics. Lastly, in Section \ref{sec:discussion}, we discuss our results in the context of experimental observations.



\section{Mathematical models and methods}
\label{sec:model}
The minimal model studied in this paper incorporates three essential features of the ecological processes: competitive interactions, immigration and intrinsic demographic noise~\cite{black2012stochastic,haegeman2011mathematical}.
In the model, illustrated in Fig.~\ref{fig:fig1}B, the community composition is characterized by the  species abundances, $\vec{n}=(n_1,\dots n_i\dots n_S)$ where the discrete random variable $n_i$ represents the  number of individuals of the $i$-th species, and $S$ is the total number of species.
The dynamics of the system are described by a birth-death process with interactions, whereby the abundance (number of individuals) of any species can increase by one with the birth rate $q^+$ or decrease by one with the death rate $q^-$ defined as 
\begin{align}
q_i^+(\vec{n})&=r^+ n_i +\mu,  \\
q_i^-(\vec{n})&=r^- n_i + \frac{r}{K} n_i \left(n_i +\sum_{j\neq i} \rho _{j,i} n_j\right) \nonumber
\end{align}
for each species $i\in \{1,2,\dots,S\}$.
{These rates recover LV models which have been extensively used to study deterministic multispecies coexistence, and in particular to explore various stabilizing and equalizing mechanisms~\cite{chesson2018updates,barabas2016effect}.}

The birth rate incorporates two factors: the per-capita birth rate $r^+$ corresponding to procreation, and the constant and positive immigration rate $\mu$ from an external basin which ensures that the system possesses no global absorbing extinction state \cite{capitan2015similar}. 
The death rates include the `bare' per-capita death rate of the organisms $r^-$ and the competitive interactions effects that increase the mortality at high population numbers, incorporated through a quadratic term in the death rates; Parameter $\rho_{j,i}$ quantifies the competition strength  between species $i$ and $j$; the {competition strength (analogous to the niche overlap \cite{badali2020effects,capitan2015similar}) is  defined as the ratio between the inter-specific and intra-specific competition strengths. }
The carrying capacity for each species is represented by $K$. 
The per-capita turnover rate is $r=r^+-r^-$. 

These aggregate coarse-grained parameters are determined by a variety of system factors such as the efficiency of resource consumption, interactions with the environment and external forces. Although it is possible to derive these rates from explicit resource competition models in several special cases, the expressions are highly model-dependent and are not explicitly modeled here
~\cite{macArthur1970species,chesson1990macarthur,o2018whence}.
For biological reasons,  $K$, $r^+$, $r^- > 0$ are all positive, which results in strictly positive transition rates for all $ n_i\geq 0$. In this paper, we focus on the homogeneous case where the parameters ($\mu$, $K$, $\rho$, $r^+$, and $r^-$) are identical for all species and the competitive interactions  $\forall i,j :\rho_{j,i}=\rho$ for all species pairs. 
This symmetric and homogeneous interaction network 
has been used in \cite{badali2020effects,capitan2017stochastic,capitan2020competitive,haegeman2011mathematical} in contrast to the models   wherein the competition strengths are inhomogeneous and drawn from a distribution \cite{fisher2014transition,allesina2012stability}. 
This minimal complexity model allows us to investigate the full phase space of the system to examine the underlying principle without {extensive and }impractical multi-parameter sweeps.

The stochastic evolution of the system is described by the master equation
\begin{multline}
\label{master-eq}
\partial_t  {\rm \mathcal{P}}(\vec{n};t)= \sum_{i}\left\{ -\left[q^+_i(\vec{n})+q^-_i(\vec{n})\right]{\rm \mathcal{P}}(\vec{n};t) \vphantom{\left[ \sum q^+_i\right]} \right.\\
\left. \vphantom{\left[ \sum q^+_i\right]} +q^+_i (\vec{n}-\vec{e}_i){\rm \mathcal{P}}(\vec{n}-\vec{e}_i;t)+q^-_i (\vec{n}+\vec{e}_i){\rm \mathcal{P}}(\vec{n}+\vec{e}_i;t)\right\},
\end{multline}
where $\vec{e}_i$ is the standard basis vector and ${\rm \mathcal{P}}(\vec{n},t)$ is the joint probability density function for the system to exhibit the species composition $\vec{n}$ at time $t$ \cite{gardiner1985handbook}. 
In the long time limit, the system reaches a stationary state where $\partial_t \mathcal{P}=0$, see SI Section 2 \cite{grimmett2001probability,schnakenberg1976network,meyn1993stability,gupta2014scalable}.

The species abundance distribution (SAD) describing the mean fractions of species with $n$ individuals, can be related to the marginal single species probability distribution $P(n)$:
\begin{align}
    {\rm SAD}&(n) = \frac{1}{S}\left\langle \sum_{i=1}^{S}\delta (n_i -n) \right\rangle\\
   \nonumber &=\frac{1}{S}\sum_{i=1}^S\left[ \sum_{n_1=0}^{\infty}\cdots \sum_{n_{i-1}=0}^{\infty}\sum_{n_{i+1}=0}^{\infty}\cdots \sum_{n_S=0}^{\infty}{\rm \mathcal{P}}(\vec{n})|_{n_i=n}\right] \\  &=\nonumber  P_i(n) \equiv  P(n),
\end{align}
where $\delta$ is the Kronecker delta function, and using the fact that in this homogeneous system the marginal distributions $P_i(n)=P(n)$ of population abundance are identical for all species. 


{Dynamics of ecological populations can also be described using continuous approximations (see the Supplementary Information; SI Section 1) \cite{fisher2014transition,lynch2015ecology,verberk2011explaining,fowler2013colonization,barabas2016effect, marquet2017proportional}}.
In particular, in the Fokker-Planck approximation, the continuous deterministic limit of the master equation (Eq.~\ref{master-eq}) recovers the well-known competitive Lotka-Volterra (LV) equations
\begin{align}
    \frac{\partial x_i}{\partial t}&= q_i^+(\vec{x}) - q_i^-(\vec{x})\nonumber\\
    &=r x_i \left( 1 - \frac{x_i}{K} - \sum_{j\neq i} \rho \frac{x_j}{K} \right) + \mu 
    \label{eq:LV}
\end{align}
for the variable $x_i$, which corresponds to the continuous deterministic limit 
of the discrete variable $n_i$  \cite{gardiner1985handbook}; see SI Section 1 for further details.

The deterministic steady state is given by
\begin{equation}
    \tilde{x}(S) = \frac{K}{2[1+\rho(S-1)]}\left\{1 +  \sqrt{1+\frac{4\mu[1+\rho(S -1)]}{r K}}\right\}.
    \label{eq:fixed-LV}
\end{equation}
Note that in the deterministic LV process all species survive with abundance $\tilde{x}$ as long as $\rho\leq 1$ and $\mu>0$ \cite{capitan2015similar}.
Conversely, in the stochastic competitive environment the numbers of individuals of each species fluctuate, occasionally reaching extinction. 
Thus, the number of co-existing species $S^*$ is a stochastic variable as well, and may be smaller than the overall number species $S$ in the immigration flux from the larger basin, with $S^*\leq S$.
{The number of co-existing species has a corresponding probability distribution whose evolution is governed by a master equation derived from~Eq.~\ref{master-eq} (see SI Section 2)~\cite{dobson2020unsolved}.}
The richness, denoted as $\langle S^* \rangle $, is defined as the average number of the (co-)existing species, and is related to the SAD via
\begin{equation}
\label{eq:richness}
\langle S^* \rangle = S(1-P(0)).
\end{equation}
{Intuitively, this is the sum of the expectation of $S$ random indicator variables; the richness }is determined by $S$ times the probability that a species is present in the system, $1-P(0)$ (see SI Section 3.A).

No exact analytical solution for the high-dimensional master equation Eq.~\ref{master-eq} is known for a general competition strength $\rho$. To understand the principles of the community organization and the impact of competition, immigration and demographic noise, we developed approximate analytical solutions to the master equation verified by Gillespie simulations (see SI Section 2 for details).

\section{Results}
\label{sec:results}

\subsection{Mean-Field Approximation}
The full master equation Eq.~\ref{master-eq} can be reduced to a one dimensional approximation for the marginal distribution $P(n)$ with effective birth-death rates (see SI Section 2.A). The SAD, $P(n)$, is obtained as a self-consistent stationary solution of this equation as
\begin{multline}
   P(n) \equiv P_i(n_i=n)\\
   =P(0)\frac{(r^+)^{n}(\mu/r^+)_{n}}{n!\prod_{n_i=1}^{n}\left(r^-+r n_i/K+r\rho \sum_{j\neq i}^S\langle n_j |n_i \rangle /K\right)}.\\
   \label{eq:mean-field}
\end{multline}
To obtain an analytical approximation to $P(n)$ we use a mean field closure for the unknown conditional averages $\langle n_j |n_i\rangle$ as
$\left\langle \sum_{j\neq i} n_j |n_i\right\rangle\approx  (S-1) \langle n \rangle$ (see SI Section 2.A for discussion and alternative approximations). 
Thus, Eq.~\ref{eq:mean-field} becomes a closed-form implicit equation for the probability distribution $P(n)$ which can be solved numerically. 
We have found a good agreement between exact stochastic simulation results and this mean-field approximation for most of the parameter space examined. 

Following Eq.~\ref{eq:richness}, the average richness in the mean-field approximation is
\begin{eqnarray}
\langle S^*\rangle = S \left( 1 - \frac{1}{{_1}F_1[a,b+1;c]}\right ) , 
\label{eq:mf-richness}
\end{eqnarray}
where $P(0)=1/{_1F_1}[a,b+1+1;c]$ is the normalization constant of $P(n)$ where ${_1F_1}[a,b;c] $ is the hypergeometric Kummer confluent function, with $a=\mu/r^+$, ${b}= 
[r^-K+r\rho (S-1)\langle n\rangle  ]/r$, and ${c}=
{r^+ K}/{r}$.
{The exact solution for the distribution of the number of co-existing species, $S^*$, can be obtained for $\rho=0$ (a binomial distribution) and $\rho=1$ (a sum of hypergeometric functions; see SI Section 3)\cite{haegeman2011mathematical}.
For intermediate competition strengths, $0 < \rho < 1$, a mean-field approximation results in a binomial distribution for the species richness distribution as in~\cite{dobson2020unsolved}; however, we discuss how this mean-field solution differs from the richness distribution from simulations in the SI Section 3.}

\subsection{The system exhibits rich behavior with distinct regimes of population structures controlled by competition strength, immigration rate and the species number} 
\label{sec:Phases}

Depending on the values of the competition strength and the immigration rate, the number of species and the system size, the population can exhibit a number of different regimes of behavior which can be categorized by their richness and the shape of their SAD, as visualized in Fig.~\ref{fig:phases_sim} and described below.

\subsubsection{Richness regimes}
In the classical deterministic LV model, the systems exhibits either an interior fixed-point with full coexistence of all species at abundances given by Eq.~\ref{eq:fixed-LV}, or mass extinction with a single surviving species, in agreement with the well-known Gause's law of deterministic competitive exclusion \cite{capitan2015similar}. By contrast, the stochastic model may exhibit partial coexistence due to the temporary extinctions of some species driven by the
abundance fluctuations arising from the demographic noise. Overall, the number of co-existing species and their abundances are determined by the balance between the immigration and the competition induced stochastic extinction events.
Three distinct richness regimes can be discerned as shown in Fig.~\ref{fig:phases_sim}, based on the variations of the richness of the system {$\langle S^* \rangle$} in different regions of the ($\rho$,$\mu$,$S$) parameter space.

At low competition strength - region (a) in Fig.~\ref{fig:phases_sim}A - all species co-exist so that the richness of the system is equal to the total number of species $\langle S^*\rangle \approx S$, {similar to} the deterministic regime. 
In this regime, each species effectively inhabits its own niche because the inter-species competition is not sufficiently strong to drive any of the species to extinction in the presence of abundance fluctuations arising from the demographic noise.
The probability for a species to be present is determined by the balance of its immigration rate and the extinction rate.
At higher immigration rates this regime extends into regions with higher competition strength $\rho$: high immigration rates stabilize full richness populations even with a relatively high competition strength.



In the second regime - region (b) in Fig.~\ref{fig:phases_sim}A - only a fraction of the species are simultaneously present on average, which we denote as the partial coexistence regime.
In this regime, the immigration influx is not high enough to prevent temporary stochastic extinctions of some species resulting from the competition. 

At very high competition strengths a complete exclusion regime - region (c) in Fig.~\ref{fig:phases_sim}A - is found. 
High competition along with the very low immigration rates act in unison to reduce the richness to below two species on average.
Although regime (c) may appear similar to regime (b) since both present partial coexistence, they are distinguished by key behaviors as explained below. 

Note that the stochasticity is central to the  effect of the competition on the observed richness.
Stochastic fluctuations increase the risk of extinction with increasing competition strength, unlike in the deterministic case where the richness is independent of the competition strength for $\rho < 1$ \cite{capitan2015similar}.

\subsubsection{SAD shape and modality regimes}
\begin{figure*}[t!]
    \begin{minipage}{.3\linewidth}
   \begin{flushleft}
        A
   \end{flushleft}
   \includegraphics[width=\textwidth]{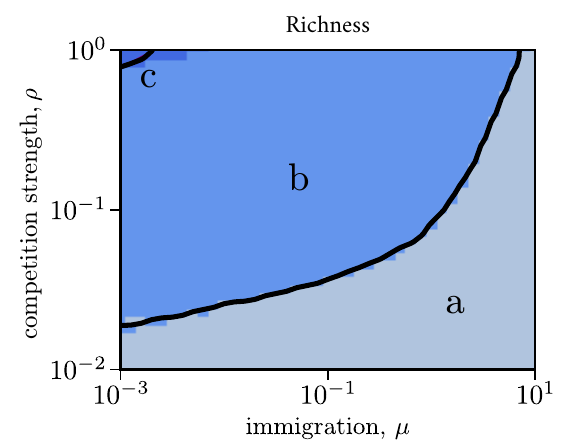}
   \end{minipage}
    \begin{minipage}{.65\linewidth}
   \begin{flushleft}
        B
   \end{flushleft}
   \includegraphics[width=\textwidth]{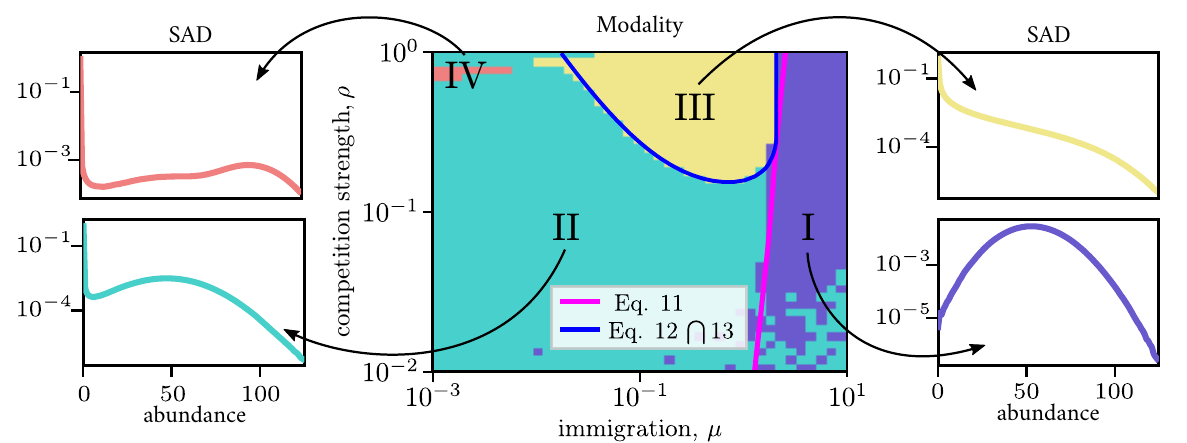}
   \end{minipage}
    \begin{minipage}[t]{.45\textwidth}
   \begin{flushleft}
    C
    \end{flushleft}
    \vspace{0pt}
    \includegraphics[width=\textwidth]{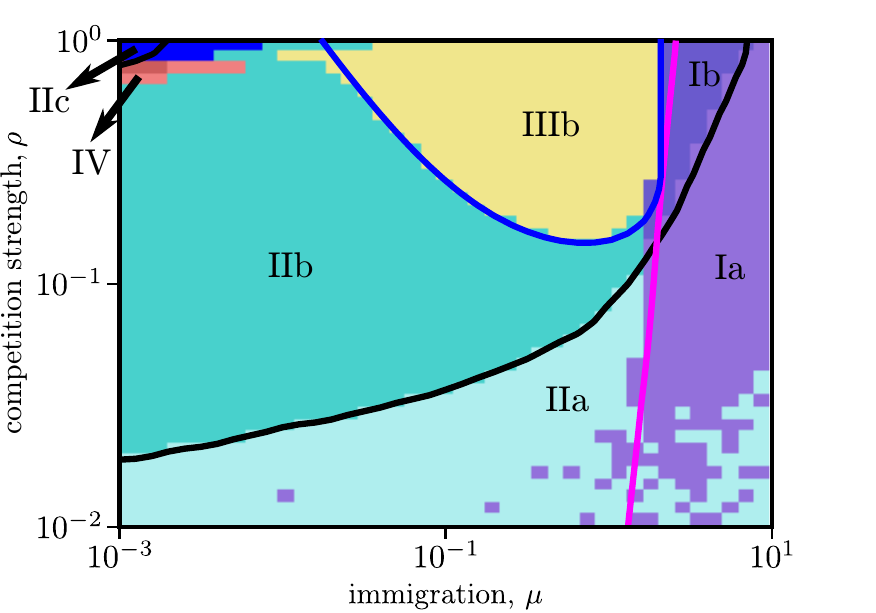}
  \end{minipage}
  \hfill
  \begin{minipage}[t]{.45\textwidth}
    \begin{flushright}
        \begin{flushleft}
        D
        \end{flushleft}
    \vspace{8pt}
    \includegraphics[width=\textwidth]{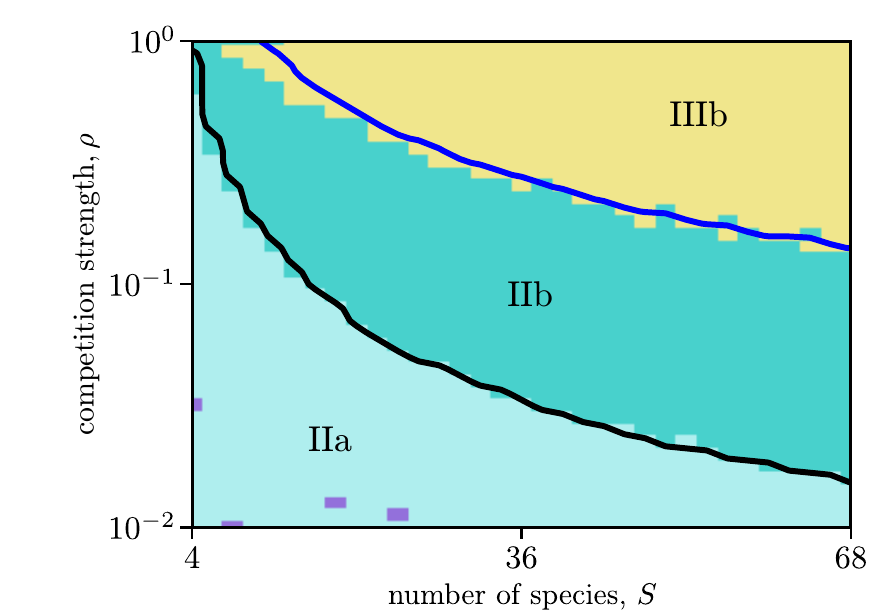}
    \end{flushright} 
  \end{minipage}
    \caption{Phenomenology of the population structures. Panel \textbf{A}: The system possesses three distinct richness phases. (a): full coexistence of all the species $\langle S^* \rangle \approx S$; (b): partial coexistence with $\langle S^* \rangle < S$; (c): a single species exists on average. Panel \textbf{B}: Different population regimes are distinguished by different SAD modalities. (I): immigration dominated regime with unimodal SAD at a typical abundance given by the positive root of $\tilde{n}$; (II): bimodal regime with species at non-zero abundance $\tilde{n}$ and a rapid species turnover peak a zero abundance; (III): `rare biosphere' regime of a unimodal SAD with peak at zero abundance resulting from the rapid turnover of the temporarily extinct species; (IV) multimodal regime.
    Panels \textbf{C} and \textbf{D}: Intersection of the modality and richness regimes in the $(\mu,\rho)$ plane (\textbf{ C}) and $(S,\rho)$ plane (\textbf{D}); see text for discussion. In panels \textbf{ A}, \textbf{B} and \textbf{C} the number of  species $S=30$.  In panel \textbf{D} the immigration rate is $\mu=10^{-1}$. For all panels; Colored regions represent data from simulation (see Methods), whereas boundaries from the mean-field approximation are represented by solid black lines. 
    The solution for the master equation Eq.~\ref{master-eq} is simulated using the Gillespie algorithm with $6 \cdot 10^8$ time steps, $r^+=2$, $r^-=1$, and $K=100$.
    }
    \label{fig:phases_sim}
\end{figure*}
\begin{figure}[t!]
   \begin{flushleft}
        A
   \end{flushleft}
    \includegraphics{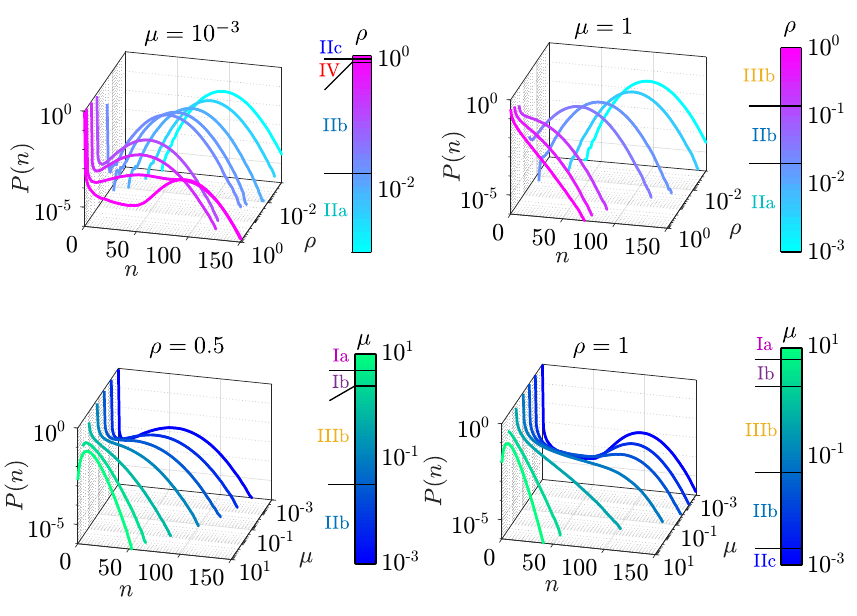}
    \begin{flushleft}
        B
   \end{flushleft}
    \includegraphics[width=\columnwidth, trim= 0 0 0 15,clip]{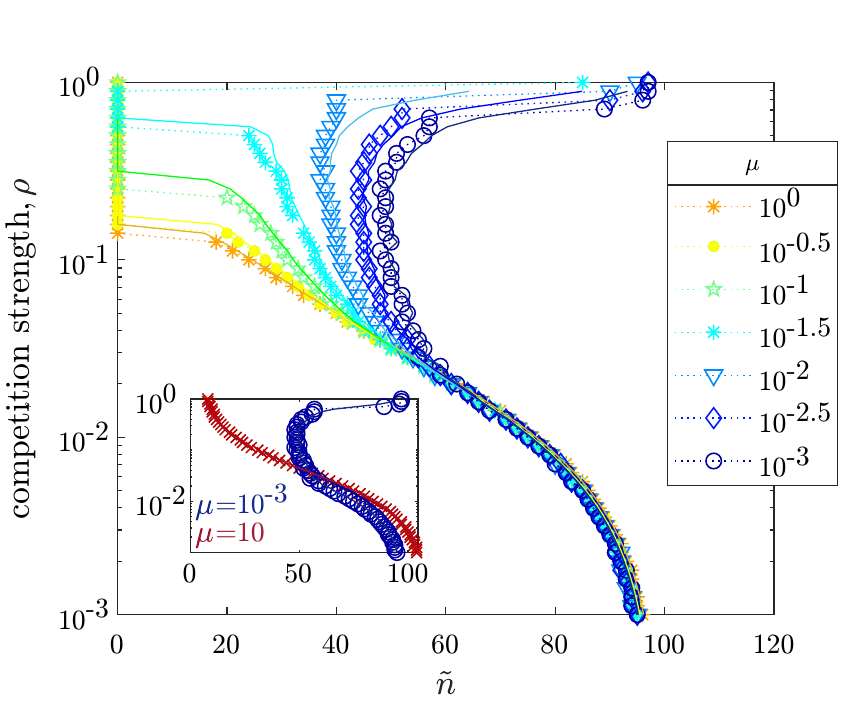}
    \caption{SAD changes between different regimes. Panel \textbf{A}: (upper left) Simulation results for species abundance distributions (SADs) for fixed $\mu=10^{-3}$ as a function of $\rho$. (upper right) same for $\mu=1$. Different values of the competition strength  $\rho$ are emphasized with different colors indicated in the color-bar. (lower left) Simulation results for SADs as a function of $\mu$ for fixed $\rho=0.5$  (lower right) same for $\rho=1$. Different immigration rates $\mu$ are emphasised with different color shown in the color-bar. 
    Panel \textbf{B}: The non-zero mode of the SAD  given by the positive solution of  $\tilde{n}$ representing the dominant species abundance as a function of $\rho$ for different values of $\mu$. Markers and dotted lines represent simulation results, while solid lines are given from analytic analysis, Eq.~\ref{eq:dom-level}.}
 \label{fig:Fig3}
\end{figure}

%
%
Besides {determining} the richness, the balance between immigration and stochastic competitive extinctions also dictates the mean abundances of the individual species and the species abundance distribution (SAD).
When the immigration influx of individuals into the system is higher than the average out-flux due to the transient extinctions, shown in Fig.~\ref{fig:phases_sim}B as region (I), most species are forced away from extinction.
In this regime, the SAD is unimodal with a peak at relatively high species abundances $\tilde{n}$  approximately located at 
\begin{equation}
    \label{eq:dom-level}
    \tilde{n} =
    \frac{K-\rho(S-1)\langle n\rangle}{2}\left\{1 \pm \sqrt{1+4\frac{(\mu-r^+) K}{r(K-\rho(S-1)\langle n\rangle)^2}}\right\},
\end{equation}
which agrees with the simulation results, as shown in Fig.~\ref{fig:Fig3}; see also  SI Section 2.B. 

At lower immigration rates - regime (II) in Fig.~\ref{fig:phases_sim}B - the immigration rate is insufficiently strong to overcome the competition-driven temporary extinctions of some species, and the SAD develops an additional peak around $n=0$ corresponding to the temporarily extinct species. 
The subset of the `quasi-stable' co-existing species {dominate the population number with abundances that fluctuate around the `niche-like' abundance peak, }$\tilde{n}$.
{Their persistence at dominant abundances is} punctuated by rare fluctuation-driven extinctions and the occasional invasion of a temporarily extinct species into the dominant population.
By contrast, the dynamics of species in the $n=0$ zero peak is characterized by the rapid turnover of the remaining species close to extinction. 
{This balance between the immigration and the stochastic competitive extinctions may be related to the trade-offs in competitive ability and dispersal (immigration) in meta-community population dynamics \cite{tejo2021coexistence, macarthur1967theory}. }

At low immigration rates, the peak at Eq.~\ref{eq:dom-level} coincides with the deterministic stable solution in Eq.~\ref{eq:fixed-LV}  (see SI Section 4.B) 
\begin{equation}
\lim_{\mu\rightarrow 0}\tilde{n}=\lim_{\mu\rightarrow 0}\tilde{x}\left(\langle S^* \rangle \right) = \frac{K}{1+\rho(\langle S^*\rangle -1)}.
\label{eq:Lim_ss}
\end{equation}
Namely, in the bimodal regime the coexisting dominant species are fluctuating around $\tilde{n}$ which, at low immigration, is the deterministic fixed point with $\langle S^* \rangle$ species. 
{In this regime, the dynamics of the fluctuations of the} abundant species around $\tilde{n}$ can be heuristically understood as a spatially dependent diffusion in an effective potential well of the Fokker-Plank Equation (See Section \ref{sec:model} and SI Section 1). 

Somewhat unexpectedly, at low immigration rate $\mu \lesssim .05$, the bimodal regime extends onto the neutral line at $\rho = 1$  where the SAD has been commonly believed to have the monotonically decreasing `rare biosphere' shape \cite{hubbell2001unified,baxter2007exact}.
{Surprisingly}, in this regime the competition is so strong that most of the time either no species are present at high abundance, or only one species survives in a kinetically `frozen' and long lived quasi-stable state with an abundance $\tilde{n}\simeq K$, as observed previously \cite{xu2018immigration}{; this is region (IIc) in Fig.~\ref{fig:fig1}C.} 

Furthermore, at the intermediate immigration rates and relatively high competition strengths we observe a unimodal behaviour with a peak at zero rather than at a finite $\tilde{n}$ - region (III) in Fig.~\ref{fig:phases_sim}B. 
In this regime, the competition is strong enough {so} that the fluctuations competitively drive species
to temporary extinction before any of them is able to establish a `quasi-stable' state at a high abundance.
All species undergo rapid turnover around zero resulting from the balance between random immigration and extinction events. 
This regime corresponds to what was previously described as the `rare biosphere': fewer number of species are found at higher abundances resulting in a monotonically decreasing SAD. 
This SAD shape is classically recognized as a hallmark of a `rare biosphere' regime. However, as shown in Fig.~\ref{fig:phases_sim} the unimodal regime (III) unexpectedly extends substantially beyond the neutral manifold $\rho=1$, into the non-neutral regions with $\rho < 1$, and the monotonic-decreasing SAD persists even for competition strengths as low as $\rho \approx 0.1$ - an order of magnitude weaker than the classical neutral regime.
This challenges the common perception that the `rare biosphere' SAD is an indicator of neutrality.
{On the neutral line $\rho=1$, for large $n$ the SAD asymptotically tends to a power law with an exponential cutoff in line with similar functional forms found in previous works (see SI Section 2.C)\cite{hubbell2001unified,mckane2004analytic,baxter2007exact, goyal2015mechanisms}}. 

Finally, we found an entirely novel multimodal regime with more than two peaks - regime (IV) in Fig.~\ref{fig:phases_sim} -  which possesses one rapid turnover peak around extinction and multiple peaks at non-zero abundances. 
Similar to region (IIc), the peak at $n=0$ comprises species which rapidly turnover around extinction. However, in addition to the peak at positive abundances $K$ formed by one surviving species ($S^*=1$) in a meta-stable frozen state, this regime possesses a second peak at $\sim K/(1+\rho)$ with two simultaneously surviving quasi-stable species ($S^*=2$. {The abundance at these peaks are solutions to Eq.~\ref{eq:Lim_ss} wherein $\langle S^* \rangle$ is replaced by the momentary $S^*$}. The slow fluctuations between the states with $S^*=1$ and $S^*=2$ result in the appearance of the SAD with two non-zero modes at quasi-stable dominance abundance, $\tilde{n} \sim K$ and $\tilde{n}\sim K/(1+\rho)$ observed in the region (IV). These two peaks are only visibly separated when the richness is low and carrying capacity is high, {since solutions of Eq.~\ref{eq:Lim_ss} for different $S^*$ are more distant in this regime}.

The transitions between the different modality regimes and the corresponding changes in the SAD shapes are illustrated in Fig.~\ref{fig:Fig3}. Generally, {at low competition strength $\rho$ } the species are practically independent of each other,{ residing in largely non-overlapping niches and with }their typical abundance $\tilde{n}$ close to the carrying capacity $K$. 
Increasing competition strength $\rho$ makes it harder to sustain the co-existing species at high abundances, and accordingly $\tilde{n}$ decreases, as illustrated in the top panels of Fig.~\ref{fig:Fig3}A and  Fig.~\ref{fig:Fig3}B. With further increase in $\rho$ the system behavior bifurcates depending on the immigration rate $\mu$. 
At high immigration rates, $\mu \gtrsim 0.05$, the competition-driven decrease in $\tilde{n}$ continues up to the critical competition strength (calculated in the next section) where the peak around $\tilde{n}$ disappears (top right panel of Fig.~\ref{fig:Fig3}A) and  Fig.~\ref{fig:Fig3}B), as the system is not able to sustain `quasi-stable' niche-like species co-existence. This corresponds to the transition from the bimodal region (II) to the `rare biosphere' region (III) in Fig.~\ref{fig:phases_sim}).
At lower immigration rates (top left panel of Fig.~\ref{fig:Fig3}A and  Fig.~\ref{fig:Fig3}B), further increases in the competition strength eventually cause mass species extinctions which allow the remaining few dominant species to maintain higher abundances (region (III) Fig.~\ref{fig:phases_sim}). As $\rho \rightarrow 1$, the system transitions to the region (IIc) of the Fig.~\ref{fig:phases_sim}: only one dominant species remains, as described in \cite{xu2018immigration}, with abundance {fluctuating around }$K$. 

\subsubsection{Global Phase Diagram and Regime Boundaries}
\label{sec:regime-bound}

In this section we describe the complete phase diagram of the system defined by the intersection of the different richness and the SAD shape/modality regimes, derive the regime boundaries and discuss the transitions between them, as shown in the ($\mu,\rho$) space in Fig.~\ref{fig:phases_sim}C, and in ($S,\rho$) space in Fig.~\ref{fig:phases_sim}D.
We show that the boundaries between different regimes observed in simulations can be understood within simple mean field theories, and discuss the underlying physical factors responsible for the transitions between different regimes.

We define the boundary between the full coexistence (a) and partial coexistence (b) regimes to be at $\langle S^* \rangle=S-1/2$: the midpoint between full richness $S^*=S$ and the loss of 1 species on average.
Similarly, the boundary between the partial coexistence (b) and exclusion (c) regimes is located at $\langle S^* \rangle=3/2$, that is to say where the richness is between one and two species such that on average only 1 species is present in regime (c).

To derive the boundaries corresponding to the transitions of the SAD modality regimes, we use discrete derivatives of the approximated SAD to determine the existence of peaks and their location (see SI Section 4.C).
The immigration dominated regime (I) is characterized by a unimodal SAD with a peak at the positive root of $\tilde{n}$ given in Eq.~\ref{eq:dom-level}.
Compared to this immigration dominated regime, the neighboring bimodal and monotonically-decreasing unimodal regimes - regions (II) and (III) respectively - differ by the emergence of a new mode at zero abundance. 

Thus, the boundary that defines transitions to either regime (II) or (III) from the immigration dominated regime (I) is described by a flattening of SAD at $n=0$: $\partial P(n)/ \partial n|_{n=0} = 0$. {In the discrete case, this }heuristically corresponds to $P(0)=P(1)$. Combining this condition for the boundary with the global-balance of the master equation Eq.~\ref{master-eq}  results in the rate balance equation,
$\langle q_i^+(\vec{n})|n_i=0 \rangle=\langle q_i^-(\vec{n})|n_i=1 \rangle$.

In the mean-field approximation, this boundary is found at
\begin{equation}
    \label{eq:boundary-I}
    \mu = r^- +\frac{r}{K}[1+\rho(S-1)\langle n \rangle ].
\end{equation}
This equation recovers the similar transition for $\rho=1$ derived independently in \cite{xu2018immigration}. 

The boundary between the bimodal regime (II) and the `rare biosphere' regime (III) is characterized by the disappearance of the peak at high abundance $\tilde{n}$ in Eq.~\ref{eq:dom-level}. 
In the bimodal regime at least one solution to $\tilde{n}$ is real and positive;{ as such} a maximal, real peak exists. 
Conversely, in the `rare biosphere' regime, both solutions of $\tilde{n}$ are negative or imaginary.  
We find that the boundary between the real and imaginary $\tilde{n}$ is
\begin{equation}
\label{eq:boundary-IIIa}
r(K-\rho (S-1) \langle n \rangle )^2=4(r^+-\mu){K}
\end{equation}
and the transition line between positive and negative solutions, $\tilde{n}=0$, is
\begin{equation}
    \frac{(K- \rho (S-1) \langle n \rangle )^4}{16}=1+ \frac{K(\mu-r^+)}{r}.
    \label{eq:boundary-IIIb}
\end{equation}
The intersection of these two conditions defines the `rare biosphere' regime and is shown as the blue line in Fig.~\ref{fig:phases_sim}B,C.

The modality and the richness of the system are also affected by the number of species $S$ as shown in  Fig.~\ref{fig:phases_sim}D.
In brief, the frequency of the immigration events rises as more species are present in the immigration flux.
Increased immigration causes the total population to rise without providing more room for each species in the system; this  increases the stochastic competition, driving more species to extinction. 
Hence, as $S$ increases, the transition from the bimodal regime (II) to the unimodal regime (III) occurs at lower values of competition strength $\rho$, and the fraction of the concurrently surviving species decreases. This effect has been qualitatively observed experimentally \cite{gore2021}, and we return to it in the Discussion.

These analytical expressions for the regime boundaries - confirmed by stochastic simulations - provide insights into the effects of different control parameters on the regime boundaries. In particular, using the low $\mu$ deterministic approximation for $\langle n \rangle \approx K / \left[ 1+\rho (S-1) \right]$, shows that the location of the boundary of the `rare biosphere' regime grows proportionally to the carrying capacity and is a decreasing function of the number of species $S$. Thus, the size of the `rare biopshere' regime increases with the number of species $S$ as shown in Fig.~\ref{fig:phases_sim}D,
whereas increasing the carrying capacity shrinks this regime (see SI Section 5).
    
\subsection{Kinetics of the species turnover, extinction and recovery underlie the transitions between different regimes}
\label{sec:Dynamics}

\begin{figure}[t!]
   \begin{flushleft}
        A
   \end{flushleft}
   \vspace{-30pt}
   \begin{center}
        \includegraphics[width=0.70\columnwidth]{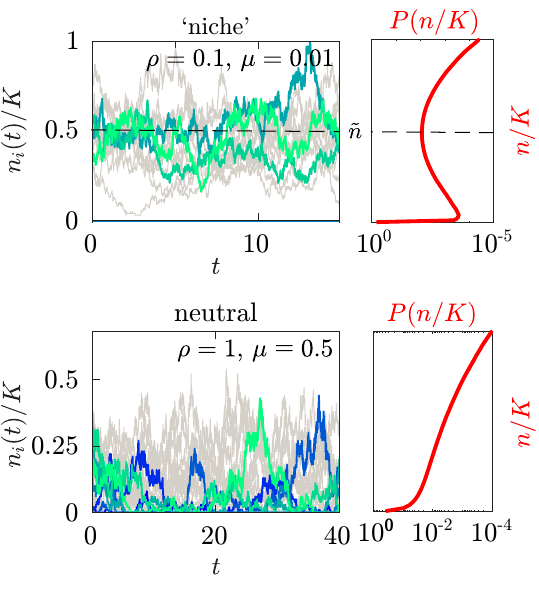}
   \end{center}
   \vspace{-10pt}
   \begin{flushleft}
        B
   \end{flushleft}
   \vspace{-30pt}
   \begin{center}
       \includegraphics[width=0.70\columnwidth]{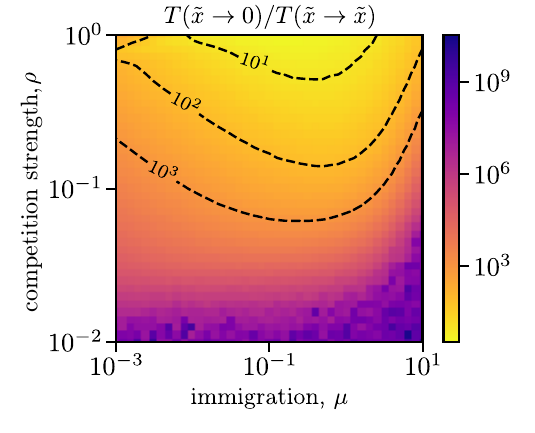}
   \end{center}
   \vspace{-10pt}
    \begin{flushleft}
        C
   \end{flushleft}
   \vspace{-30pt}
   \begin{center}
       \includegraphics[width=0.70\columnwidth]{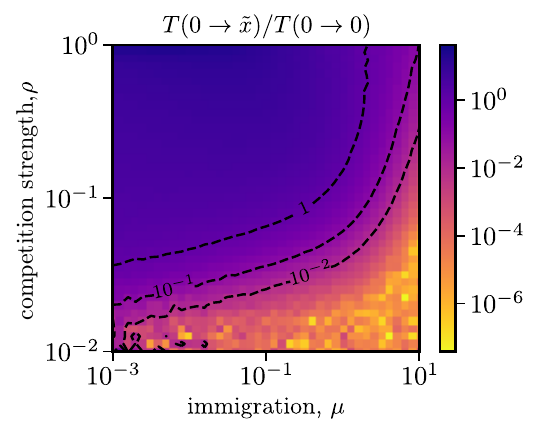}
   \end{center}
   \vspace{-20pt}
    \caption{Kinetics of species extinction, invasion and turnover. Panel \textbf{A}:  Sample trajectories of the species abundances. 
    (Upper panel): stable `niche-like' dynamics, where the dominant species fluctuate about $\tilde{n}$. The red curve represents the corresponding bimodal SAD. (Lower panel): the erratic dynamics in the `rare biosphere' regime, where species fluctuate close to extinction. The SAD is a monotonically decreasing function. Panel \textbf{B}: The MFPT ratio $ T(\tilde{x}\rightarrow 0) / T(\tilde{x}\rightarrow \tilde{x})$ {as a function of $\mu$ and $\rho$}. Select contour lines are highlighted as dashed lines. This ratio qualitatively captures the transition from `rare biosphere' to `'niche-like' regimes Fig.~\ref{fig:phases_sim}A. For weak immigration rates $\mu\approx 10^{-3}$ the ratio is non-monotonic in the competition strength, revealing regime (c). Panel \textbf{C}: The MFPT ratio $ T(0\rightarrow \tilde{x}) / T(0\rightarrow 0)$. This ratio qualitatively captures the richness behaviour.}
    \label{fig:turnover}
\end{figure}

{To better understand the driving forces for the transitions between the different regimes, we focused on the underlying kinetics of species turnover and fluctuations.}
{There is a stark contrast in the} kinetics of an individual species in the unimodal `rare biosphere' regime (III) and the `niche-like' regimes with a peak in SAD at non-zero abundance, as shown in Fig.~\ref{fig:turnover}A. In regime (III), all species undergo rapid turnover in the relatively broad range of abundances around extinction. By contrast, in regimes (I, II, and IV) the `quasi-stable' dominant species undergo fast fluctuations around the co-existence peak at $\tilde{n}$ in addition to fast turnover of the remaining species near extinction. {These fluctuations around the `quasi-stable' abundance are punctuated by the temporary extinctions and the reverse invasions of temporarily extinct species into the dominant `niche-like' peak.}

{To characterize the kinetics in different regimes, we calculate the mean first-passage times $T(a\rightarrow b)$ (MFPT) of the transitions between different abundance levels ($a$ and $b$), using the one-dimensional backward Master equation (see SI Section 6)~\cite{iyer2016first,redner2001guide}.}

We first focus on the ratio of the MFPT of the transition from dominance to exclusion to the MFPT of return to the dominant abundance level (starting from the dominant abundance level), $T(\tilde{x}(\langle S^* \rangle )\rightarrow 0) / T(\tilde{x}(\langle S^* \rangle )\rightarrow \tilde{x}(\langle S^* \rangle ))$, shown in (Fig.~\ref{fig:turnover}B).
{Here, $\tilde{x}$, given in ~Eq.~\ref{eq:fixed-LV} is the deterministic extension of the peak abundance $\tilde{n}$ in regimes without a non-zero abundance peak}.
Large values of this ratio signify that the extinction rate from $\tilde{x}(\langle S^* \rangle )$ is much slower than the rate of local fluctuations in the effective potential well around $\tilde{x}(\langle S^* \rangle )$.
Accordingly, Fig.~\ref{fig:turnover}B shows that this ratio is high in the bimodal and immigration-dominated regimes.
Conversely, this ratio is lower within the `rare biosphere' regime which does not possess a high abundance peak with `quasi-stable' co-existing species.
As shown in Fig.~\ref{fig:turnover}B, this ratio approximately delineates the `rare biosphere' regime from the `niche-like' regimes and its contour lines qualitatively recover the boundaries of region IIIb in Fig.~\ref{fig:phases_sim}C{; see SI Section 6 for further discussion}.

The second ratio, which underlies the richness transitions in the system, 
$T(0\rightarrow \tilde{x}(\langle S^* \rangle ))/T(0\rightarrow0)$ (Fig.~\ref{fig:turnover} panel C) relates MFPT from extinction at zero abundance to dominance at $\tilde{x}$ to the the mean return time to extinction.
{This ratio gives a rough measure of the number of species present in the system: $T(0\rightarrow0)$ is inversely proportional to $P(0)$ and $T(0\rightarrow \tilde{x}(\langle S^* \rangle ))$ is heuristically inversely proportional to the number of co-existing species (see SI Sections 3A and 6).}
As such, this MFPT ratio approximates the ratio of the average number of temporarily extinct species, $S - \langle S^* \rangle$ to the average number of existing species, $\langle S^* \rangle$, see Fig.~\ref{fig:turnover}C.
{As shown in Fig.~\ref{fig:turnover}C} ratio quantitatively recovers the boundaries of richness regimes in Fig.~\ref{fig:phases_sim} in most regions of the parameter space.

{These MFPTs not only serve to interpret the underlying dynamics that give rise to different regimes, they may also be more easily experimentally measured than steady state distributions.
Further discussion on the dynamical features are presented in the SI Section 6.}

\section{Discussion}
\label{sec:discussion}

Ecological systems display a wide variety of different behavior regimes that have been commonly analysed through a limited number of paradigmatic models such as the `niche' and `neutral' theories. However, it remains incompletely understood what features of ecological population structure and dynamics are universal and which are system specific, how different models relate to  each other, and what behavior is expected in the full range of the parameter space. Using a minimal model of the competitive population dynamics with demographic noise, we have investigated
the different regimes of the population structures and dynamics as a function of the immigration rate $\mu$, the competition strength $\rho$, {as well as} the number of species $S$.
Although this minimal model may not fully capture the more complex interaction structures of many ecological communities, it exhibits rich and unexpected behaviours paralleling many experimental observ{ations} (see Table \ref{tab:my_label}), and illuminates the underlying mechanisms that shape population structures in different ecosystems.

We have focused on the system richness reflecting the number of the co-existing species, and the SAD shape as the characteristics of the different population regimes, using a combination of simulations and analytical mean-field approaches. Our analysis shows that the ecosystem behaviors can be partitioned into different regimes of richness and SAD shape/modality, parameterized by the immigration rate and the {competition strength - which is analogous to the competitive overlap discussed in other studies and references therein \cite{capitan2020competitive, badali2020effects}}.

Our model recovers the limits of the well known `rare biosphere' and the `niche-like' regimes. In particular, at $\rho = 1$ and intermediate values of $\mu$, the SAD has the monotonically decreasing shape characteristic of the classical neutral regime. On the other hand, at low competition strength, the system SAD exhibits a peak at high species abundance where all species co-exist. {We recover the expected regime where different species effectively occupy distinct ecological niches.} Notably, even independent species with no inter-species competition with $\rho=0$ may present either a unimodal or bimodal SAD depending on the immigration rate, as shown in Fig.~\ref{fig:phases_sim}B, C. Unlike the immigration dominated high abundance peak at high immigration rates, at the very low immigration rates the SAD is peaked around zero due to high extinction probability solely from the intra-species competition.

{We found that, contrary to the common expectation that different species inhabit separate niches away from neutrality}, the system can maintain the monotonically decaying `rare biosphere' SAD even at low competition strength (up to $\rho \approx 0.1$) as shown in the regime (III) in Fig.~\ref{fig:phases_sim}. Similarly, unexpectedly, at the very low immigration rates, the system SAD maintains the peak at non-zero abundance characteristic of `niche-like' regimes even for the high values of the competition strength $\rho$ usually considered to be in the `rare biosphere' domain (regime (IIc)) in Fig.~\ref{fig:phases_sim} and  Section \ref{sec:results}B.

We have also uncovered an unusual - and to the best of our knowledge hitherto not described - regime characterized by the multi-modal SAD with more than one positive, `quasi-stable' abundance peak (Regime (IV) in Fig~\ref{fig:phases_sim}). This multi-modality arises from the richness fluctuations in this regime: the number of co-existing species is switching randomly between two relatively long-lasting states with  $S^*=1$ and $S^*=2$. Thus, one peak of the SAD is found around $\sim K$ and the other one in the vicinity of $\sim K/2$, as explained in Section \ref{sec:results}B. We observe that for low $K$, the multimodal regime is non-existent and appears as $K$ increases; see the corresponding phase diagrams in SI Section 5.

{Transitions of the SAD between different regimes occur through different routes. }
In particular,  {as the immigration rate increases, the SAD peak of the bimodal regime (II) at $\rho=1$, gradually decreases in height without significantly shifting its location until it disappears at the boundary of the `rare biosphere' regime (III)}.  By contrast, at lower competition strengths $\rho<1$, the transition from the bi-modality to the `rare biosphere' regime occurs via  simultaneous changes in the peak's height and location. This is discussed in Section \ref{sec:results}B.

{We show that the population structures in different regimes and the transitions between them, stem from  the underlying dynamics of species fluctuations, extinctions and invasions.} In the `rare biosphere' regimes, all species undergo relatively fast turnover around extinction. This is reflected in the low ratio of the turnover to the extinction mean first-passage times. Conversely, in the `niche-like' regimes the system develops two additional time scales: relatively fast fluctuations about the high abundance peak, and the long waiting times for the transitions from the `quasi-stable' co-existence at high abundance to extinction.  {This is reflected in the fact that the ratio of the mean extinction time to the mean time of return to dominance is higher in the `niche-like' regime, as discussed in Section \ref{sec:results}C.} 

Interestingly, ecological regimes  akin to those predicted by our demographic noise model (except for the multimodal SAD regime) have been also found using deterministic, noiseless LV models with a random matrix of inter-species competitive competition strengths  \cite{may1972will,allesina2008network,allesina2012stability,kessler2015generalized,gore2021}. 
However, the underlying mechanisms that give rise to the apparently similar regimes in the two model types are very different. 
{In the demographic noise model, the partial richness `niche-like' regime (IIb) (Fig.~\ref{fig:phases_sim}C) comprises the quasi-stable coexistence of a subset of species at a positive abundance in parallel with the temporary stochastic extinctions of other species. }
By contrast, in the deterministic LV models with random asymmetric interactions, the partial richnes `niche-like' regime comprises large number of saddle fixed points where different sets of species are competitively excluded deterministically. At higher competition strengths, the deterministic system transitions to the chaotic behavior that resembles the `rare biosphere' regime (III) (Fig.~\ref{fig:phases_sim}C), however the nature of the species turnover and the shape of the SAD are different from the results we presented in Section \ref{sec:results} \cite{bunin2017ecological,kessler2015generalized,gore2021}.

The existence of the predicted regimes and the transitions between them can be tested experimentally by measuring the SAD and the dynamics of the species abundances in ecosystems with varying immigration and competition strengths,{ numbers of species and effective carrying capacities. Measurements of the SADs and the community compositions have become more attainable due to the advances in single cell gene sequencing techniques \cite{ratzke2020strength, gore2021,shakiba2019cell}, overcoming the difficulties of SAD estimation due to data limitations.}
Long-term observations may provide measurements of the stationary species abundance distributions \cite{weigelt2010jena}. 
Although it may be difficult to experimentally determine and control the immigration rate, the competition strength, and the carrying capacity, practical proxies for these parameters exist. By way of an example, the flow rate carrying bacteria into a chamber of a microfluidic device is a well controlled quantity that approximates well the immigration rate for populations encased in the chamber~\cite{duran2021slipstreaming}.
Another commonly used and robustly estimated experimental observable is the species rank abundance (SRA), which can be used to infer the SAD to which it is closely mathematically related (see SI Section 7), although in practice the conversion might be constrained by limitations of noise and quantity of the experimental data. 

The asymptotic behaviour of the SADs may {show qualitative dissimilarities between distributions allowing one to discern} different regimes of behavior among the experimental observations. In {the mean-field approximation, the asymptotic behaviour of the model's} SAD on the neutral line $\rho=1$ is well approximated by a power law with an exponential cutoff (see SI Section 2.C){. This asymptotic is similar in functional form to the SADs commonly found by Hubbell models of a neutral birth-death process with a fixed total population size} \cite{baxter2007exact,mckane2004analytic}. 
Notably, the Yule process that is often used to model neutral processes also results in the SAD of a similar form.
However, the Yule process is substantially different from the model of this paper because it does not include inter-species interactions and reaches the steady state SAD only if the effective death rate is higher than the effective birth/immigration rate~\cite{bacaer2011yule}. 

In Table 1, we qualitatively compare the family of the regimes predicted by our model to the various behaviors inferred from experimental findings based on the SAD measurements and population abundance time series. 
The apparent abundance of the neutral ecosystems observed experimentally - such as gastrointestinal microbiomes - may pertain to our finding (Section~\ref{sec:results}B) that the `rare biosphere' regime extends substantially beyond the neutral line $\rho =1$: non-neutral communities may appear neutral as they exhibit SAD's characteristic of neutral communities \cite{jeraldo2012quantification}. 
Furthermore, multimodal SAD's predicted by our model that are related to the richness fluctuations may provide an explanation for the multimodal SADs observed in some ecological data,  complementary to the existing explanations such as spatial heterogeneity or emergent neutrality \cite{dornelas2008multiple,vergnon2012emergent}

\begin{table*}[ht!]
    \centering
    \begin{tabular}{|l|p{7.5cm}|p{5cm}|}
       System (Ref.) & Regimes  & Observations  \\ \hline 
microbial competition \cite{gore2021} & stable full coexistence (IIa), stable partial coexistence (IIb), persistent fluctuation (IIIb) & 	Community composition/ richness/  fluctuating communities
\\
global birds species \cite{callaghan2021global} &	unimodal - log skew (I) & SAD   \\
plankton \cite{ser2018ubiquitous}  &	power-law decay (III) & SAD and SRA \\
coral \cite{dornelas2008multiple}	& multimodal (IV) &	 SAD 
\\
arthropods \cite{matthews2014multimodal} &	multimodal (IV) & SAD 
\\
T-cell receptors \cite{oakes2017quantitative}	& bimodal (II) and unimodal (III)
& SAD	
\\
microbial competition  \cite{descheemaeker2020stochastic} & `rare biosphere' (III) and `niche-like' (I \& II) 
& SRA and time series
\\ 
gastrointestinal microbiomes \cite{jeraldo2012quantification} &	`rare biosphere' (III) 
& SRA and operational taxonomic units (OTUs)
\\ 
    \end{tabular}
    \caption{Qualitative classification of observed population regimes in various ecological systems.  }
    \label{tab:my_label}
\end{table*}


One quantity that is relatively easy to control {experimentally} is the total number of species $S$. The regimes predicted by the model and the transitions between them  are shown in Fig.~\ref{fig:phases_sim}D: { our model yields `rare biosphere' regimes for high $S$ and $\rho$, which are characterized by high-turnover dynamics, and `niche-like' regines with more stable behavior at lower $S$ or $\rho$. These predictions qualitatively agree with the experimentally observed phase-space  in \cite{gore2021}, which were previously explained within the deterministic LV models with a random interaction matrix~\cite{gore2021}. The fact that both the deterministic LV model with a random interaction matrix and the homogeneous LV model with demographic noise are in qualitative agreement with the experimental data raises interesting and important questions concerning the interplay of stochastic and deterministic dynamics in determining the community composition.} 

Another quantity that may enable qualitative and quantitative testing of different models is the carrying capacity $K$ which may be controlled experimentally in some systems.
As shown in SM, {the} `rare biosphere' regime shrinks in size with increasing $K$ because a higher carrying capacity can sustain higher average abundance, and larger (less likely) fluctuations are needed for the extinction events to occur.
Higher average abundance together with insufficiently
strong fluctuations result in longer MFPTs from 
dominance to extinction abundances and vice-versa. These effects will be investigated in future work 

{In the context of other ecological theories, the competition strength, as defined in this work, can be viewed as a quantification of the heuristic notion of the niche overlap, and we observe that decreasing niche overlap results in richness increases as suggested previously \cite{li2018seed,chesson2008interaction}.
Our model also serves as a quantitative example of some of the coexistence promoting mechanisms of the contemporary ecological theory; we explore stabilizing mechanisms that increase richness via decreases in niche overlap, such as fluctuation-dependent processes and fitness-density covariance~\cite{chesson2000mechanisms,hillerislambers2012rethinking}. 
In particular, the demographic noise model studied here exhibits fluctuation-dependent mechanisms that promote richness as species are able to coexist at high abundance in our model. 
}


We expect that the minimal model of this paper can be used for more complicated scenarios, including more complex distributions of the interaction network $\rho_{i,j}$,  speciation to probe the interaction of the natural selection, and inter-species interactions and population diversity and structure.

{Finally, our model of a local island community in the mainland-island ecosystem (see Fig.~\ref{fig:fig1}),} 
can be expanded to many-island models {or many-patch dynamics \cite{evans2013stochastic,tejo2021coexistence}.
These many-island and many-patch models examine the interplay between competition and dispersal rate \cite{tejo2021coexistence} and its effects on the diversity of the metacommunity, a prominent topic in conservation ecology and the study of the human microbiome. 
The patch models may address how coexistence and persistence are influenced by spatial heterogeneity and environmental noise in a demographic noise formulation.}
{Future work will explore integration of our model into other scenarios to predict species fitness, non-equilibrium coexistence and their connection to broader qualitative ideas in ecology.}
    
\section*{Methods}
The solution for the master equation Eq.~\ref{master-eq} is simulated using the Gillespie algorithm with $10^8$ time steps. We use $r^+=2$, $r^-=1$, $K=100$. 
Modalities' classification   is numerically executed after smoothing the simulated SAD. The MFPT is evaluated via the simulated SAD ($\tilde{x}(S^*)$ is rounded), where a uni-dimensional approximation of the process is considered, see details in SI Section 6.

\begin{acknowledgments}
The authors acknowledge helpful discussions and comments from all the members of the Goyal and Zilman Groups. AZ acknowledges the support from the National Science and Engineering Research Council of Canada (NSERC) through the Discovery Grant Program. SG acknowledges the support from the National Science and Engineering Research Council of Canada (NSERC) through the Discovery Grant Program and from the Medicine by DEsign Program at the University of Toronto.
\end{acknowledgments}

\bibliography{bibliography}

\begin{thebibliography}{110}%
\makeatletter
\providecommand \@ifxundefined [1]{%
 \@ifx{#1\undefined}
}%
\providecommand \@ifnum [1]{%
 \ifnum #1\expandafter \@firstoftwo
 \else \expandafter \@secondoftwo
 \fi
}%
\providecommand \@ifx [1]{%
 \ifx #1\expandafter \@firstoftwo
 \else \expandafter \@secondoftwo
 \fi
}%
\providecommand \natexlab [1]{#1}%
\providecommand \enquote  [1]{``#1''}%
\providecommand \bibnamefont  [1]{#1}%
\providecommand \bibfnamefont [1]{#1}%
\providecommand \citenamefont [1]{#1}%
\providecommand \href@noop [0]{\@secondoftwo}%
\providecommand \href [0]{\begingroup \@sanitize@url \@href}%
\providecommand \@href[1]{\@@startlink{#1}\@@href}%
\providecommand \@@href[1]{\endgroup#1\@@endlink}%
\providecommand \@sanitize@url [0]{\catcode `\\12\catcode `\$12\catcode
  `\&12\catcode `\#12\catcode `\^12\catcode `\_12\catcode `\%12\relax}%
\providecommand \@@startlink[1]{}%
\providecommand \@@endlink[0]{}%
\providecommand \url  [0]{\begingroup\@sanitize@url \@url }%
\providecommand \@url [1]{\endgroup\@href {#1}{\urlprefix }}%
\providecommand \urlprefix  [0]{URL }%
\providecommand \Eprint [0]{\href }%
\providecommand \doibase [0]{https://doi.org/}%
\providecommand \selectlanguage [0]{\@gobble}%
\providecommand \bibinfo  [0]{\@secondoftwo}%
\providecommand \bibfield  [0]{\@secondoftwo}%
\providecommand \translation [1]{[#1]}%
\providecommand \BibitemOpen [0]{}%
\providecommand \bibitemStop [0]{}%
\providecommand \bibitemNoStop [0]{.\EOS\space}%
\providecommand \EOS [0]{\spacefactor3000\relax}%
\providecommand \BibitemShut  [1]{\csname bibitem#1\endcsname}%
\let\auto@bib@innerbib\@empty
\bibitem [{\citenamefont {Ratzke}\ \emph {et~al.}(2020)\citenamefont {Ratzke},
  \citenamefont {Barrere},\ and\ \citenamefont {Gore}}]{ratzke2020strength}%
  \BibitemOpen
  \bibfield  {author} {\bibinfo {author} {\bibfnamefont {C.}~\bibnamefont
  {Ratzke}}, \bibinfo {author} {\bibfnamefont {J.}~\bibnamefont {Barrere}},\
  and\ \bibinfo {author} {\bibfnamefont {J.}~\bibnamefont {Gore}},\ }\bibfield
  {title} {\bibinfo {title} {Strength of species interactions determines
  biodiversity and stability in microbial communities},\ }\href@noop {}
  {\bibfield  {journal} {\bibinfo  {journal} {Nature ecology \& evolution}\
  }\textbf {\bibinfo {volume} {4}},\ \bibinfo {pages} {376} (\bibinfo {year}
  {2020})}\BibitemShut {NoStop}%
\bibitem [{\citenamefont {Tilman}(1977)}]{tilman1977resource}%
  \BibitemOpen
  \bibfield  {author} {\bibinfo {author} {\bibfnamefont {D.}~\bibnamefont
  {Tilman}},\ }\bibfield  {title} {\bibinfo {title} {Resource competition
  between plankton algae: an experimental and theoretical approach},\
  }\href@noop {} {\bibfield  {journal} {\bibinfo  {journal} {Ecology}\ }\textbf
  {\bibinfo {volume} {58}},\ \bibinfo {pages} {338} (\bibinfo {year}
  {1977})}\BibitemShut {NoStop}%
\bibitem [{\citenamefont {Strom}(2008)}]{strom2008microbial}%
  \BibitemOpen
  \bibfield  {author} {\bibinfo {author} {\bibfnamefont {S.~L.}\ \bibnamefont
  {Strom}},\ }\bibfield  {title} {\bibinfo {title} {Microbial ecology of ocean
  biogeochemistry: a community perspective},\ }\href@noop {} {\bibfield
  {journal} {\bibinfo  {journal} {Science}\ }\textbf {\bibinfo {volume}
  {320}},\ \bibinfo {pages} {1043} (\bibinfo {year} {2008})}\BibitemShut
  {NoStop}%
\bibitem [{\citenamefont {Foster}\ \emph {et~al.}(2017)\citenamefont {Foster},
  \citenamefont {Schluter}, \citenamefont {Coyte},\ and\ \citenamefont
  {Rakoff-Nahoum}}]{foster2017evolution}%
  \BibitemOpen
  \bibfield  {author} {\bibinfo {author} {\bibfnamefont {K.~R.}\ \bibnamefont
  {Foster}}, \bibinfo {author} {\bibfnamefont {J.}~\bibnamefont {Schluter}},
  \bibinfo {author} {\bibfnamefont {K.~Z.}\ \bibnamefont {Coyte}},\ and\
  \bibinfo {author} {\bibfnamefont {S.}~\bibnamefont {Rakoff-Nahoum}},\
  }\bibfield  {title} {\bibinfo {title} {The evolution of the host microbiome
  as an ecosystem on a leash},\ }\href@noop {} {\bibfield  {journal} {\bibinfo
  {journal} {Nature}\ }\textbf {\bibinfo {volume} {548}},\ \bibinfo {pages}
  {43} (\bibinfo {year} {2017})}\BibitemShut {NoStop}%
\bibitem [{\citenamefont {Coyte}\ \emph {et~al.}(2015)\citenamefont {Coyte},
  \citenamefont {Schluter},\ and\ \citenamefont {Foster}}]{coyte2015ecology}%
  \BibitemOpen
  \bibfield  {author} {\bibinfo {author} {\bibfnamefont {K.~Z.}\ \bibnamefont
  {Coyte}}, \bibinfo {author} {\bibfnamefont {J.}~\bibnamefont {Schluter}},\
  and\ \bibinfo {author} {\bibfnamefont {K.~R.}\ \bibnamefont {Foster}},\
  }\bibfield  {title} {\bibinfo {title} {The ecology of the microbiome:
  networks, competition, and stability},\ }\href@noop {} {\bibfield  {journal}
  {\bibinfo  {journal} {Science}\ }\textbf {\bibinfo {volume} {350}},\ \bibinfo
  {pages} {663} (\bibinfo {year} {2015})}\BibitemShut {NoStop}%
\bibitem [{\citenamefont {Gorter}\ \emph {et~al.}(2020)\citenamefont {Gorter},
  \citenamefont {Manhart},\ and\ \citenamefont
  {Ackermann}}]{gorter2020understanding}%
  \BibitemOpen
  \bibfield  {author} {\bibinfo {author} {\bibfnamefont {F.~A.}\ \bibnamefont
  {Gorter}}, \bibinfo {author} {\bibfnamefont {M.}~\bibnamefont {Manhart}},\
  and\ \bibinfo {author} {\bibfnamefont {M.}~\bibnamefont {Ackermann}},\
  }\bibfield  {title} {\bibinfo {title} {Understanding the evolution of
  interspecies interactions in microbial communities},\ }\href@noop {}
  {\bibfield  {journal} {\bibinfo  {journal} {Philosophical Transactions of the
  Royal Society B}\ }\textbf {\bibinfo {volume} {375}},\ \bibinfo {pages}
  {20190256} (\bibinfo {year} {2020})}\BibitemShut {NoStop}%
\bibitem [{\citenamefont {Merlo}\ \emph {et~al.}(2006)\citenamefont {Merlo},
  \citenamefont {Pepper}, \citenamefont {Reid},\ and\ \citenamefont
  {Maley}}]{merlo2006cancer}%
  \BibitemOpen
  \bibfield  {author} {\bibinfo {author} {\bibfnamefont {L.~M.}\ \bibnamefont
  {Merlo}}, \bibinfo {author} {\bibfnamefont {J.~W.}\ \bibnamefont {Pepper}},
  \bibinfo {author} {\bibfnamefont {B.~J.}\ \bibnamefont {Reid}},\ and\
  \bibinfo {author} {\bibfnamefont {C.~C.}\ \bibnamefont {Maley}},\ }\bibfield
  {title} {\bibinfo {title} {Cancer as an evolutionary and ecological
  process},\ }\href@noop {} {\bibfield  {journal} {\bibinfo  {journal} {Nature
  reviews cancer}\ }\textbf {\bibinfo {volume} {6}},\ \bibinfo {pages} {924}
  (\bibinfo {year} {2006})}\BibitemShut {NoStop}%
\bibitem [{\citenamefont {Kareva}(2015)}]{kareva2015cancer}%
  \BibitemOpen
  \bibfield  {author} {\bibinfo {author} {\bibfnamefont {I.}~\bibnamefont
  {Kareva}},\ }\bibfield  {title} {\bibinfo {title} {Cancer ecology: Niche
  construction, keystone species, ecological succession, and ergodic theory},\
  }\href@noop {} {\bibfield  {journal} {\bibinfo  {journal} {Biological
  Theory}\ }\textbf {\bibinfo {volume} {10}},\ \bibinfo {pages} {283} (\bibinfo
  {year} {2015})}\BibitemShut {NoStop}%
\bibitem [{\citenamefont {Smart}\ \emph {et~al.}(2021)\citenamefont {Smart},
  \citenamefont {Goyal},\ and\ \citenamefont {Zilman}}]{smart2021roles}%
  \BibitemOpen
  \bibfield  {author} {\bibinfo {author} {\bibfnamefont {M.}~\bibnamefont
  {Smart}}, \bibinfo {author} {\bibfnamefont {S.}~\bibnamefont {Goyal}},\ and\
  \bibinfo {author} {\bibfnamefont {A.}~\bibnamefont {Zilman}},\ }\bibfield
  {title} {\bibinfo {title} {Roles of phenotypic heterogeneity and
  microenvironment feedback in early tumor development},\ }\href@noop {}
  {\bibfield  {journal} {\bibinfo  {journal} {Physical Review E}\ }\textbf
  {\bibinfo {volume} {103}},\ \bibinfo {pages} {032407} (\bibinfo {year}
  {2021})}\BibitemShut {NoStop}%
\bibitem [{\citenamefont {Tauber}(2008)}]{tauber2008immune}%
  \BibitemOpen
  \bibfield  {author} {\bibinfo {author} {\bibfnamefont {A.~I.}\ \bibnamefont
  {Tauber}},\ }\bibfield  {title} {\bibinfo {title} {The immune system and its
  ecology},\ }\href@noop {} {\bibfield  {journal} {\bibinfo  {journal}
  {Philosophy of Science}\ }\textbf {\bibinfo {volume} {75}},\ \bibinfo {pages}
  {224} (\bibinfo {year} {2008})}\BibitemShut {NoStop}%
\bibitem [{\citenamefont {Schmid-Hempel}\ and\ \citenamefont
  {Ebert}(2003)}]{schmid2003evolutionary}%
  \BibitemOpen
  \bibfield  {author} {\bibinfo {author} {\bibfnamefont {P.}~\bibnamefont
  {Schmid-Hempel}}\ and\ \bibinfo {author} {\bibfnamefont {D.}~\bibnamefont
  {Ebert}},\ }\bibfield  {title} {\bibinfo {title} {On the evolutionary ecology
  of specific immune defence},\ }\href@noop {} {\bibfield  {journal} {\bibinfo
  {journal} {Trends in Ecology \& Evolution}\ }\textbf {\bibinfo {volume}
  {18}},\ \bibinfo {pages} {27} (\bibinfo {year} {2003})}\BibitemShut {NoStop}%
\bibitem [{\citenamefont {Shakiba}\ \emph {et~al.}(2019)\citenamefont
  {Shakiba}, \citenamefont {Fahmy}, \citenamefont {Jayakumaran}, \citenamefont
  {McGibbon}, \citenamefont {David}, \citenamefont {Trcka}, \citenamefont
  {Elbaz}, \citenamefont {Puri}, \citenamefont {Nagy}, \citenamefont {van~der
  Kooy} \emph {et~al.}}]{shakiba2019cell}%
  \BibitemOpen
  \bibfield  {author} {\bibinfo {author} {\bibfnamefont {N.}~\bibnamefont
  {Shakiba}}, \bibinfo {author} {\bibfnamefont {A.}~\bibnamefont {Fahmy}},
  \bibinfo {author} {\bibfnamefont {G.}~\bibnamefont {Jayakumaran}}, \bibinfo
  {author} {\bibfnamefont {S.}~\bibnamefont {McGibbon}}, \bibinfo {author}
  {\bibfnamefont {L.}~\bibnamefont {David}}, \bibinfo {author} {\bibfnamefont
  {D.}~\bibnamefont {Trcka}}, \bibinfo {author} {\bibfnamefont
  {J.}~\bibnamefont {Elbaz}}, \bibinfo {author} {\bibfnamefont {M.~C.}\
  \bibnamefont {Puri}}, \bibinfo {author} {\bibfnamefont {A.}~\bibnamefont
  {Nagy}}, \bibinfo {author} {\bibfnamefont {D.}~\bibnamefont {van~der Kooy}},
  \emph {et~al.},\ }\bibfield  {title} {\bibinfo {title} {Cell competition
  during reprogramming gives rise to dominant clones},\ }\href@noop {}
  {\bibfield  {journal} {\bibinfo  {journal} {Science}\ }\textbf {\bibinfo
  {volume} {364}} (\bibinfo {year} {2019})}\BibitemShut {NoStop}%
\bibitem [{\citenamefont {Tilman}(1982)}]{tilman1982resource}%
  \BibitemOpen
  \bibfield  {author} {\bibinfo {author} {\bibfnamefont {D.}~\bibnamefont
  {Tilman}},\ }\href@noop {} {\emph {\bibinfo {title} {Resource competition and
  community structure}}}\ (\bibinfo  {publisher} {Princeton university press},\
  \bibinfo {year} {1982})\BibitemShut {NoStop}%
\bibitem [{\citenamefont {Morin}(2009)}]{morin2009community}%
  \BibitemOpen
  \bibfield  {author} {\bibinfo {author} {\bibfnamefont {P.~J.}\ \bibnamefont
  {Morin}},\ }\href@noop {} {\emph {\bibinfo {title} {Community ecology}}}\
  (\bibinfo  {publisher} {John Wiley \& Sons},\ \bibinfo {year}
  {2009})\BibitemShut {NoStop}%
\bibitem [{\citenamefont {Tuljapurkar}(2013)}]{tuljapurkar2013population}%
  \BibitemOpen
  \bibfield  {author} {\bibinfo {author} {\bibfnamefont {S.}~\bibnamefont
  {Tuljapurkar}},\ }\href@noop {} {\emph {\bibinfo {title} {Population dynamics
  in variable environments}}},\ Vol.~\bibinfo {volume} {85}\ (\bibinfo
  {publisher} {Springer Science \& Business Media},\ \bibinfo {year}
  {2013})\BibitemShut {NoStop}%
\bibitem [{\citenamefont {Budzinski}(2007)}]{budzinski2007monoculture}%
  \BibitemOpen
  \bibfield  {author} {\bibinfo {author} {\bibfnamefont {O.}~\bibnamefont
  {Budzinski}},\ }\bibfield  {title} {\bibinfo {title} {Monoculture versus
  diversity in competition economics},\ }\href@noop {} {\bibfield  {journal}
  {\bibinfo  {journal} {Cambridge Journal of Economics}\ }\textbf {\bibinfo
  {volume} {32}},\ \bibinfo {pages} {295} (\bibinfo {year} {2007})}\BibitemShut
  {NoStop}%
\bibitem [{\citenamefont {Koura}\ \emph {et~al.}(2017)\citenamefont {Koura},
  \citenamefont {Zhang},\ and\ \citenamefont {Liu}}]{koura2017competitive}%
  \BibitemOpen
  \bibfield  {author} {\bibinfo {author} {\bibfnamefont {Y.~H.}\ \bibnamefont
  {Koura}}, \bibinfo {author} {\bibfnamefont {Y.}~\bibnamefont {Zhang}},\ and\
  \bibinfo {author} {\bibfnamefont {H.}~\bibnamefont {Liu}},\ }\bibfield
  {title} {\bibinfo {title} {Competitive interaction model for online social
  networks’ users’ data forwarding at a subnet},\ }\href@noop {} {\bibfield
   {journal} {\bibinfo  {journal} {Mathematical Problems in Engineering}\
  }\textbf {\bibinfo {volume} {2017}} (\bibinfo {year} {2017})}\BibitemShut
  {NoStop}%
\bibitem [{\citenamefont {Lewin}(1983)}]{lewin1983finches}%
  \BibitemOpen
  \bibfield  {author} {\bibinfo {author} {\bibfnamefont {R.}~\bibnamefont
  {Lewin}},\ }\bibfield  {title} {\bibinfo {title} {Finches show competition in
  ecology},\ }\href@noop {} {\bibfield  {journal} {\bibinfo  {journal}
  {Science}\ }\textbf {\bibinfo {volume} {219}},\ \bibinfo {pages} {1411}
  (\bibinfo {year} {1983})}\BibitemShut {NoStop}%
\bibitem [{\citenamefont {Lack}(1983)}]{lack1983darwin}%
  \BibitemOpen
  \bibfield  {author} {\bibinfo {author} {\bibfnamefont {D.}~\bibnamefont
  {Lack}},\ }\href@noop {} {\emph {\bibinfo {title} {Darwin's finches}}}\
  (\bibinfo  {publisher} {Cambridge University Press},\ \bibinfo {year}
  {1983})\BibitemShut {NoStop}%
\bibitem [{\citenamefont {Grant}\ and\ \citenamefont
  {Grant}(1979)}]{grant1979darwin}%
  \BibitemOpen
  \bibfield  {author} {\bibinfo {author} {\bibfnamefont {B.}~\bibnamefont
  {Grant}}\ and\ \bibinfo {author} {\bibfnamefont {P.}~\bibnamefont {Grant}},\
  }\bibfield  {title} {\bibinfo {title} {Darwin's finches: population variation
  and sympatric speciation},\ }\href@noop {} {\bibfield  {journal} {\bibinfo
  {journal} {Proceedings of the National Academy of Sciences}\ }\textbf
  {\bibinfo {volume} {76}},\ \bibinfo {pages} {2359} (\bibinfo {year}
  {1979})}\BibitemShut {NoStop}%
\bibitem [{\citenamefont {Pocheville}(2015)}]{pocheville2015ecological}%
  \BibitemOpen
  \bibfield  {author} {\bibinfo {author} {\bibfnamefont {A.}~\bibnamefont
  {Pocheville}},\ }\bibfield  {title} {\bibinfo {title} {The ecological niche:
  history and recent controversies},\ }in\ \href@noop {} {\emph {\bibinfo
  {booktitle} {Handbook of evolutionary thinking in the Sciences}}}\ (\bibinfo
  {publisher} {Springer},\ \bibinfo {year} {2015})\ pp.\ \bibinfo {pages}
  {547--586}\BibitemShut {NoStop}%
\bibitem [{\citenamefont {Zuppinger-Dingley}\ \emph {et~al.}(2014)\citenamefont
  {Zuppinger-Dingley}, \citenamefont {Schmid}, \citenamefont {Petermann},
  \citenamefont {Yadav}, \citenamefont {De~Deyn},\ and\ \citenamefont
  {Flynn}}]{zuppinger2014selection}%
  \BibitemOpen
  \bibfield  {author} {\bibinfo {author} {\bibfnamefont {D.}~\bibnamefont
  {Zuppinger-Dingley}}, \bibinfo {author} {\bibfnamefont {B.}~\bibnamefont
  {Schmid}}, \bibinfo {author} {\bibfnamefont {J.~S.}\ \bibnamefont
  {Petermann}}, \bibinfo {author} {\bibfnamefont {V.}~\bibnamefont {Yadav}},
  \bibinfo {author} {\bibfnamefont {G.~B.}\ \bibnamefont {De~Deyn}},\ and\
  \bibinfo {author} {\bibfnamefont {D.~F.}\ \bibnamefont {Flynn}},\ }\bibfield
  {title} {\bibinfo {title} {Selection for niche differentiation in plant
  communities increases biodiversity effects},\ }\href@noop {} {\bibfield
  {journal} {\bibinfo  {journal} {Nature}\ }\textbf {\bibinfo {volume} {515}},\
  \bibinfo {pages} {108} (\bibinfo {year} {2014})}\BibitemShut {NoStop}%
\bibitem [{\citenamefont {Silvertown}(2004)}]{silvertown2004plant}%
  \BibitemOpen
  \bibfield  {author} {\bibinfo {author} {\bibfnamefont {J.}~\bibnamefont
  {Silvertown}},\ }\bibfield  {title} {\bibinfo {title} {Plant coexistence and
  the niche},\ }\href@noop {} {\bibfield  {journal} {\bibinfo  {journal}
  {Trends in Ecology \& evolution}\ }\textbf {\bibinfo {volume} {19}},\
  \bibinfo {pages} {605} (\bibinfo {year} {2004})}\BibitemShut {NoStop}%
\bibitem [{\citenamefont {Cullen}\ and\ \citenamefont
  {MacIntyre}(1998)}]{cullen1998behavior}%
  \BibitemOpen
  \bibfield  {author} {\bibinfo {author} {\bibfnamefont {J.~J.}\ \bibnamefont
  {Cullen}}\ and\ \bibinfo {author} {\bibfnamefont {J.~G.}\ \bibnamefont
  {MacIntyre}},\ }\bibfield  {title} {\bibinfo {title} {Behavior, physiology
  and the niche of depth-regulating phytoplankton},\ }\href@noop {} {\bibfield
  {journal} {\bibinfo  {journal} {Nato Asi Series G Ecological Sciences}\
  }\textbf {\bibinfo {volume} {41}},\ \bibinfo {pages} {559} (\bibinfo {year}
  {1998})}\BibitemShut {NoStop}%
\bibitem [{\citenamefont {Melo-Merino}\ \emph {et~al.}(2020)\citenamefont
  {Melo-Merino}, \citenamefont {Reyes-Bonilla},\ and\ \citenamefont
  {Lira-Noriega}}]{melo2020ecological}%
  \BibitemOpen
  \bibfield  {author} {\bibinfo {author} {\bibfnamefont {S.~M.}\ \bibnamefont
  {Melo-Merino}}, \bibinfo {author} {\bibfnamefont {H.}~\bibnamefont
  {Reyes-Bonilla}},\ and\ \bibinfo {author} {\bibfnamefont {A.}~\bibnamefont
  {Lira-Noriega}},\ }\bibfield  {title} {\bibinfo {title} {Ecological niche
  models and species distribution models in marine environments: A literature
  review and spatial analysis of evidence},\ }\href@noop {} {\bibfield
  {journal} {\bibinfo  {journal} {Ecological Modelling}\ }\textbf {\bibinfo
  {volume} {415}},\ \bibinfo {pages} {108837} (\bibinfo {year}
  {2020})}\BibitemShut {NoStop}%
\bibitem [{\citenamefont {Aguirre-Guti{\'e}rrez}\ \emph
  {et~al.}(2015)\citenamefont {Aguirre-Guti{\'e}rrez}, \citenamefont
  {Serna-Chavez}, \citenamefont {Villalobos-Arambula}, \citenamefont {Perez
  de~la Rosa},\ and\ \citenamefont {Raes}}]{aguirre2015similar}%
  \BibitemOpen
  \bibfield  {author} {\bibinfo {author} {\bibfnamefont {J.}~\bibnamefont
  {Aguirre-Guti{\'e}rrez}}, \bibinfo {author} {\bibfnamefont {H.~M.}\
  \bibnamefont {Serna-Chavez}}, \bibinfo {author} {\bibfnamefont {A.~R.}\
  \bibnamefont {Villalobos-Arambula}}, \bibinfo {author} {\bibfnamefont
  {J.~A.}\ \bibnamefont {Perez de~la Rosa}},\ and\ \bibinfo {author}
  {\bibfnamefont {N.}~\bibnamefont {Raes}},\ }\bibfield  {title} {\bibinfo
  {title} {Similar but not equivalent: ecological niche comparison across
  closely--related m exican white pines},\ }\href@noop {} {\bibfield  {journal}
  {\bibinfo  {journal} {Diversity and distributions}\ }\textbf {\bibinfo
  {volume} {21}},\ \bibinfo {pages} {245} (\bibinfo {year} {2015})}\BibitemShut
  {NoStop}%
\bibitem [{\citenamefont {Badali}\ and\ \citenamefont
  {Zilman}(2020)}]{badali2020effects}%
  \BibitemOpen
  \bibfield  {author} {\bibinfo {author} {\bibfnamefont {M.}~\bibnamefont
  {Badali}}\ and\ \bibinfo {author} {\bibfnamefont {A.}~\bibnamefont
  {Zilman}},\ }\bibfield  {title} {\bibinfo {title} {Effects of niche overlap
  on coexistence, fixation and invasion in a population of two interacting
  species},\ }\href@noop {} {\bibfield  {journal} {\bibinfo  {journal} {Royal
  Society open Science}\ }\textbf {\bibinfo {volume} {7}},\ \bibinfo {pages}
  {192181} (\bibinfo {year} {2020})}\BibitemShut {NoStop}%
\bibitem [{\citenamefont {Capit{\'a}n}\ \emph {et~al.}(2015)\citenamefont
  {Capit{\'a}n}, \citenamefont {Cuenda},\ and\ \citenamefont
  {Alonso}}]{capitan2015similar}%
  \BibitemOpen
  \bibfield  {author} {\bibinfo {author} {\bibfnamefont {J.~A.}\ \bibnamefont
  {Capit{\'a}n}}, \bibinfo {author} {\bibfnamefont {S.}~\bibnamefont
  {Cuenda}},\ and\ \bibinfo {author} {\bibfnamefont {D.}~\bibnamefont
  {Alonso}},\ }\bibfield  {title} {\bibinfo {title} {How similar can
  co-occurring species be in the presence of competition and ecological
  drift?},\ }\href@noop {} {\bibfield  {journal} {\bibinfo  {journal} {Journal
  of the Royal Society Interface}\ }\textbf {\bibinfo {volume} {12}},\ \bibinfo
  {pages} {20150604} (\bibinfo {year} {2015})}\BibitemShut {NoStop}%
\bibitem [{\citenamefont {Chesson}(2000)}]{chesson2000mechanisms}%
  \BibitemOpen
  \bibfield  {author} {\bibinfo {author} {\bibfnamefont {P.}~\bibnamefont
  {Chesson}},\ }\bibfield  {title} {\bibinfo {title} {Mechanisms of maintenance
  of species diversity},\ }\href@noop {} {\bibfield  {journal} {\bibinfo
  {journal} {Annual review of Ecology and Systematics}\ }\textbf {\bibinfo
  {volume} {31}},\ \bibinfo {pages} {343} (\bibinfo {year} {2000})}\BibitemShut
  {NoStop}%
\bibitem [{\citenamefont {Bell}(2001)}]{bell2001neutral}%
  \BibitemOpen
  \bibfield  {author} {\bibinfo {author} {\bibfnamefont {G.}~\bibnamefont
  {Bell}},\ }\bibfield  {title} {\bibinfo {title} {Neutral macroecology},\
  }\href@noop {} {\bibfield  {journal} {\bibinfo  {journal} {Science}\ }\textbf
  {\bibinfo {volume} {293}},\ \bibinfo {pages} {2413} (\bibinfo {year}
  {2001})}\BibitemShut {NoStop}%
\bibitem [{\citenamefont {Hubbell}(2001)}]{hubbell2001unified}%
  \BibitemOpen
  \bibfield  {author} {\bibinfo {author} {\bibfnamefont {S.~P.}\ \bibnamefont
  {Hubbell}},\ }\href@noop {} {\emph {\bibinfo {title} {The unified neutral
  theory of biodiversity and biogeography (MPB-32)}}},\ Vol.~\bibinfo {volume}
  {32}\ (\bibinfo  {publisher} {Princeton University Press},\ \bibinfo {year}
  {2001})\BibitemShut {NoStop}%
\bibitem [{\citenamefont {Chave}(2004)}]{chave2004neutral}%
  \BibitemOpen
  \bibfield  {author} {\bibinfo {author} {\bibfnamefont {J.}~\bibnamefont
  {Chave}},\ }\bibfield  {title} {\bibinfo {title} {Neutral theory and
  community ecology},\ }\href@noop {} {\bibfield  {journal} {\bibinfo
  {journal} {Ecology letters}\ }\textbf {\bibinfo {volume} {7}},\ \bibinfo
  {pages} {241} (\bibinfo {year} {2004})}\BibitemShut {NoStop}%
\bibitem [{\citenamefont {Marquet}\ \emph {et~al.}(2017)\citenamefont
  {Marquet}, \citenamefont {Espinoza}, \citenamefont {Abades}, \citenamefont
  {Ganz},\ and\ \citenamefont {Rebolledo}}]{marquet2017proportional}%
  \BibitemOpen
  \bibfield  {author} {\bibinfo {author} {\bibfnamefont {P.~A.}\ \bibnamefont
  {Marquet}}, \bibinfo {author} {\bibfnamefont {G.}~\bibnamefont {Espinoza}},
  \bibinfo {author} {\bibfnamefont {S.~R.}\ \bibnamefont {Abades}}, \bibinfo
  {author} {\bibfnamefont {A.}~\bibnamefont {Ganz}},\ and\ \bibinfo {author}
  {\bibfnamefont {R.}~\bibnamefont {Rebolledo}},\ }\bibfield  {title} {\bibinfo
  {title} {On the proportional abundance of species: Integrating population
  genetics and community ecology},\ }\href@noop {} {\bibfield  {journal}
  {\bibinfo  {journal} {Scientific reports}\ }\textbf {\bibinfo {volume} {7}},\
  \bibinfo {pages} {1} (\bibinfo {year} {2017})}\BibitemShut {NoStop}%
\bibitem [{\citenamefont {Gotelli}\ and\ \citenamefont
  {McGill}(2006)}]{gotelli2006null}%
  \BibitemOpen
  \bibfield  {author} {\bibinfo {author} {\bibfnamefont {N.~J.}\ \bibnamefont
  {Gotelli}}\ and\ \bibinfo {author} {\bibfnamefont {B.~J.}\ \bibnamefont
  {McGill}},\ }\bibfield  {title} {\bibinfo {title} {Null versus neutral
  models: what's the difference?},\ }\href@noop {} {\bibfield  {journal}
  {\bibinfo  {journal} {Ecography}\ }\textbf {\bibinfo {volume} {29}},\
  \bibinfo {pages} {793} (\bibinfo {year} {2006})}\BibitemShut {NoStop}%
\bibitem [{\citenamefont {Blythe}(2012)}]{blythe2012neutral}%
  \BibitemOpen
  \bibfield  {author} {\bibinfo {author} {\bibfnamefont {R.~A.}\ \bibnamefont
  {Blythe}},\ }\bibfield  {title} {\bibinfo {title} {Neutral evolution: a null
  model for language dynamics},\ }\href@noop {} {\bibfield  {journal} {\bibinfo
   {journal} {Advances in complex systems}\ }\textbf {\bibinfo {volume} {15}},\
  \bibinfo {pages} {1150015} (\bibinfo {year} {2012})}\BibitemShut {NoStop}%
\bibitem [{\citenamefont {McKane}\ \emph {et~al.}(2004)\citenamefont {McKane},
  \citenamefont {Alonso},\ and\ \citenamefont {Sol{\'e}}}]{mckane2004analytic}%
  \BibitemOpen
  \bibfield  {author} {\bibinfo {author} {\bibfnamefont {A.~J.}\ \bibnamefont
  {McKane}}, \bibinfo {author} {\bibfnamefont {D.}~\bibnamefont {Alonso}},\
  and\ \bibinfo {author} {\bibfnamefont {R.~V.}\ \bibnamefont {Sol{\'e}}},\
  }\bibfield  {title} {\bibinfo {title} {Analytic solution of hubbell's model
  of local community dynamics},\ }\href@noop {} {\bibfield  {journal} {\bibinfo
   {journal} {Theoretical Population Biology}\ }\textbf {\bibinfo {volume}
  {65}},\ \bibinfo {pages} {67} (\bibinfo {year} {2004})}\BibitemShut {NoStop}%
\bibitem [{\citenamefont {Grover}\ \emph {et~al.}(1997)\citenamefont {Grover},
  \citenamefont {Hudziak},\ and\ \citenamefont {Grover}}]{grover1997resource}%
  \BibitemOpen
  \bibfield  {author} {\bibinfo {author} {\bibfnamefont {J.~P.}\ \bibnamefont
  {Grover}}, \bibinfo {author} {\bibfnamefont {J.}~\bibnamefont {Hudziak}},\
  and\ \bibinfo {author} {\bibfnamefont {J.~D.}\ \bibnamefont {Grover}},\
  }\href@noop {} {\emph {\bibinfo {title} {Resource competition}}},\
  Vol.~\bibinfo {volume} {19}\ (\bibinfo  {publisher} {Springer Science \&
  Business Media},\ \bibinfo {year} {1997})\BibitemShut {NoStop}%
\bibitem [{\citenamefont {Begon}\ \emph {et~al.}(2006)\citenamefont {Begon},
  \citenamefont {Townsend},\ and\ \citenamefont {Harper}}]{begon2006ecology}%
  \BibitemOpen
  \bibfield  {author} {\bibinfo {author} {\bibfnamefont {M.}~\bibnamefont
  {Begon}}, \bibinfo {author} {\bibfnamefont {C.~R.}\ \bibnamefont
  {Townsend}},\ and\ \bibinfo {author} {\bibfnamefont {J.~L.}\ \bibnamefont
  {Harper}},\ }\href@noop {} {\emph {\bibinfo {title} {Ecology: from
  individuals to ecosystems}}},\ \bibinfo {number} {Sirsi) i9781405111171}\
  (\bibinfo {year} {2006})\BibitemShut {NoStop}%
\bibitem [{\citenamefont {Adams}(2009)}]{adams2009species}%
  \BibitemOpen
  \bibfield  {author} {\bibinfo {author} {\bibfnamefont {J.}~\bibnamefont
  {Adams}},\ }\href@noop {} {\emph {\bibinfo {title} {Species richness:
  patterns in the diversity of life}}}\ (\bibinfo  {publisher} {Springer},\
  \bibinfo {year} {2009})\BibitemShut {NoStop}%
\bibitem [{\citenamefont {K{\'e}ry}\ and\ \citenamefont
  {Royle}(2020)}]{kery2020applied}%
  \BibitemOpen
  \bibfield  {author} {\bibinfo {author} {\bibfnamefont {M.}~\bibnamefont
  {K{\'e}ry}}\ and\ \bibinfo {author} {\bibfnamefont {J.~A.}\ \bibnamefont
  {Royle}},\ }\href@noop {} {\emph {\bibinfo {title} {Applied Hierarchical
  Modeling in Ecology: Analysis of distribution, abundance and species richness
  in R and BUGS: Volume 2: Dynamic and Advanced Models}}}\ (\bibinfo
  {publisher} {Academic Press},\ \bibinfo {year} {2020})\BibitemShut {NoStop}%
\bibitem [{\citenamefont {Nias}(1968)}]{nias1968clone}%
  \BibitemOpen
  \bibfield  {author} {\bibinfo {author} {\bibfnamefont {A.}~\bibnamefont
  {Nias}},\ }\bibfield  {title} {\bibinfo {title} {Clone size analysis: a
  parameter in the study of cell population kinetics},\ }\href@noop {}
  {\bibfield  {journal} {\bibinfo  {journal} {Cell Proliferation}\ }\textbf
  {\bibinfo {volume} {1}},\ \bibinfo {pages} {153} (\bibinfo {year}
  {1968})}\BibitemShut {NoStop}%
\bibitem [{\citenamefont {Rulands}\ \emph {et~al.}(2018)\citenamefont
  {Rulands}, \citenamefont {Lescroart}, \citenamefont {Chabab}, \citenamefont
  {Hindley}, \citenamefont {Prior}, \citenamefont {Sznurkowska}, \citenamefont
  {Huch}, \citenamefont {Philpott}, \citenamefont {Blanpain},\ and\
  \citenamefont {Simons}}]{rulands2018universality}%
  \BibitemOpen
  \bibfield  {author} {\bibinfo {author} {\bibfnamefont {S.}~\bibnamefont
  {Rulands}}, \bibinfo {author} {\bibfnamefont {F.}~\bibnamefont {Lescroart}},
  \bibinfo {author} {\bibfnamefont {S.}~\bibnamefont {Chabab}}, \bibinfo
  {author} {\bibfnamefont {C.~J.}\ \bibnamefont {Hindley}}, \bibinfo {author}
  {\bibfnamefont {N.}~\bibnamefont {Prior}}, \bibinfo {author} {\bibfnamefont
  {M.~K.}\ \bibnamefont {Sznurkowska}}, \bibinfo {author} {\bibfnamefont
  {M.}~\bibnamefont {Huch}}, \bibinfo {author} {\bibfnamefont {A.}~\bibnamefont
  {Philpott}}, \bibinfo {author} {\bibfnamefont {C.}~\bibnamefont {Blanpain}},\
  and\ \bibinfo {author} {\bibfnamefont {B.~D.}\ \bibnamefont {Simons}},\
  }\bibfield  {title} {\bibinfo {title} {Universality of clone dynamics during
  tissue development},\ }\href@noop {} {\bibfield  {journal} {\bibinfo
  {journal} {Nature physics}\ }\textbf {\bibinfo {volume} {14}},\ \bibinfo
  {pages} {469} (\bibinfo {year} {2018})}\BibitemShut {NoStop}%
\bibitem [{\citenamefont {de~Greef}\ \emph {et~al.}(2020)\citenamefont
  {de~Greef}, \citenamefont {Oakes}, \citenamefont {Gerritsen}, \citenamefont
  {Ismail}, \citenamefont {Heather}, \citenamefont {Hermsen}, \citenamefont
  {Chain},\ and\ \citenamefont {de~Boer}}]{de2020naive}%
  \BibitemOpen
  \bibfield  {author} {\bibinfo {author} {\bibfnamefont {P.~C.}\ \bibnamefont
  {de~Greef}}, \bibinfo {author} {\bibfnamefont {T.}~\bibnamefont {Oakes}},
  \bibinfo {author} {\bibfnamefont {B.}~\bibnamefont {Gerritsen}}, \bibinfo
  {author} {\bibfnamefont {M.}~\bibnamefont {Ismail}}, \bibinfo {author}
  {\bibfnamefont {J.~M.}\ \bibnamefont {Heather}}, \bibinfo {author}
  {\bibfnamefont {R.}~\bibnamefont {Hermsen}}, \bibinfo {author} {\bibfnamefont
  {B.}~\bibnamefont {Chain}},\ and\ \bibinfo {author} {\bibfnamefont {R.~J.}\
  \bibnamefont {de~Boer}},\ }\bibfield  {title} {\bibinfo {title} {The naive
  t-cell receptor repertoire has an extremely broad distribution of clone
  sizes},\ }\href@noop {} {\bibfield  {journal} {\bibinfo  {journal} {Elife}\
  }\textbf {\bibinfo {volume} {9}},\ \bibinfo {pages} {e49900} (\bibinfo {year}
  {2020})}\BibitemShut {NoStop}%
\bibitem [{\citenamefont {McGill}\ \emph {et~al.}(2007)\citenamefont {McGill},
  \citenamefont {Etienne}, \citenamefont {Gray}, \citenamefont {Alonso},
  \citenamefont {Anderson}, \citenamefont {Benecha}, \citenamefont {Dornelas},
  \citenamefont {Enquist}, \citenamefont {Green}, \citenamefont {He} \emph
  {et~al.}}]{mcgill2007species}%
  \BibitemOpen
  \bibfield  {author} {\bibinfo {author} {\bibfnamefont {B.~J.}\ \bibnamefont
  {McGill}}, \bibinfo {author} {\bibfnamefont {R.~S.}\ \bibnamefont {Etienne}},
  \bibinfo {author} {\bibfnamefont {J.~S.}\ \bibnamefont {Gray}}, \bibinfo
  {author} {\bibfnamefont {D.}~\bibnamefont {Alonso}}, \bibinfo {author}
  {\bibfnamefont {M.~J.}\ \bibnamefont {Anderson}}, \bibinfo {author}
  {\bibfnamefont {H.~K.}\ \bibnamefont {Benecha}}, \bibinfo {author}
  {\bibfnamefont {M.}~\bibnamefont {Dornelas}}, \bibinfo {author}
  {\bibfnamefont {B.~J.}\ \bibnamefont {Enquist}}, \bibinfo {author}
  {\bibfnamefont {J.~L.}\ \bibnamefont {Green}}, \bibinfo {author}
  {\bibfnamefont {F.}~\bibnamefont {He}}, \emph {et~al.},\ }\bibfield  {title}
  {\bibinfo {title} {Species abundance distributions: moving beyond single
  prediction theories to integration within an ecological framework},\
  }\href@noop {} {\bibfield  {journal} {\bibinfo  {journal} {Ecology letters}\
  }\textbf {\bibinfo {volume} {10}},\ \bibinfo {pages} {995} (\bibinfo {year}
  {2007})}\BibitemShut {NoStop}%
\bibitem [{\citenamefont {Matthews}\ and\ \citenamefont
  {Whittaker}(2015)}]{matthews2015species}%
  \BibitemOpen
  \bibfield  {author} {\bibinfo {author} {\bibfnamefont {T.~J.}\ \bibnamefont
  {Matthews}}\ and\ \bibinfo {author} {\bibfnamefont {R.~J.}\ \bibnamefont
  {Whittaker}},\ }\bibfield  {title} {\bibinfo {title} {On the species
  abundance distribution in applied ecology and biodiversity management},\
  }\href@noop {} {\bibfield  {journal} {\bibinfo  {journal} {Journal of Applied
  Ecology}\ }\textbf {\bibinfo {volume} {52}},\ \bibinfo {pages} {443}
  (\bibinfo {year} {2015})}\BibitemShut {NoStop}%
\bibitem [{\citenamefont {Rahbek}\ and\ \citenamefont
  {Graves}(2001)}]{rahbek2001multiscale}%
  \BibitemOpen
  \bibfield  {author} {\bibinfo {author} {\bibfnamefont {C.}~\bibnamefont
  {Rahbek}}\ and\ \bibinfo {author} {\bibfnamefont {G.~R.}\ \bibnamefont
  {Graves}},\ }\bibfield  {title} {\bibinfo {title} {Multiscale assessment of
  patterns of avian species richness},\ }\href@noop {} {\bibfield  {journal}
  {\bibinfo  {journal} {Proceedings of the National Academy of Sciences}\
  }\textbf {\bibinfo {volume} {98}},\ \bibinfo {pages} {4534} (\bibinfo {year}
  {2001})}\BibitemShut {NoStop}%
\bibitem [{\citenamefont {Hong}\ \emph {et~al.}(2006)\citenamefont {Hong},
  \citenamefont {Bunge}, \citenamefont {Jeon},\ and\ \citenamefont
  {Epstein}}]{hong2006predicting}%
  \BibitemOpen
  \bibfield  {author} {\bibinfo {author} {\bibfnamefont {S.-H.}\ \bibnamefont
  {Hong}}, \bibinfo {author} {\bibfnamefont {J.}~\bibnamefont {Bunge}},
  \bibinfo {author} {\bibfnamefont {S.-O.}\ \bibnamefont {Jeon}},\ and\
  \bibinfo {author} {\bibfnamefont {S.~S.}\ \bibnamefont {Epstein}},\
  }\bibfield  {title} {\bibinfo {title} {Predicting microbial species
  richness},\ }\href@noop {} {\bibfield  {journal} {\bibinfo  {journal}
  {Proceedings of the National Academy of Sciences}\ }\textbf {\bibinfo
  {volume} {103}},\ \bibinfo {pages} {117} (\bibinfo {year}
  {2006})}\BibitemShut {NoStop}%
\bibitem [{\citenamefont {Adler}\ \emph {et~al.}(2011)\citenamefont {Adler},
  \citenamefont {Seabloom}, \citenamefont {Borer}, \citenamefont {Hillebrand},
  \citenamefont {Hautier}, \citenamefont {Hector}, \citenamefont {Harpole},
  \citenamefont {O’Halloran}, \citenamefont {Grace}, \citenamefont {Anderson}
  \emph {et~al.}}]{adler2011productivity}%
  \BibitemOpen
  \bibfield  {author} {\bibinfo {author} {\bibfnamefont {P.~B.}\ \bibnamefont
  {Adler}}, \bibinfo {author} {\bibfnamefont {E.~W.}\ \bibnamefont {Seabloom}},
  \bibinfo {author} {\bibfnamefont {E.~T.}\ \bibnamefont {Borer}}, \bibinfo
  {author} {\bibfnamefont {H.}~\bibnamefont {Hillebrand}}, \bibinfo {author}
  {\bibfnamefont {Y.}~\bibnamefont {Hautier}}, \bibinfo {author} {\bibfnamefont
  {A.}~\bibnamefont {Hector}}, \bibinfo {author} {\bibfnamefont {W.~S.}\
  \bibnamefont {Harpole}}, \bibinfo {author} {\bibfnamefont {L.~R.}\
  \bibnamefont {O’Halloran}}, \bibinfo {author} {\bibfnamefont {J.~B.}\
  \bibnamefont {Grace}}, \bibinfo {author} {\bibfnamefont {T.~M.}\ \bibnamefont
  {Anderson}}, \emph {et~al.},\ }\bibfield  {title} {\bibinfo {title}
  {Productivity is a poor predictor of plant species richness},\ }\href@noop {}
  {\bibfield  {journal} {\bibinfo  {journal} {Science}\ }\textbf {\bibinfo
  {volume} {333}},\ \bibinfo {pages} {1750} (\bibinfo {year}
  {2011})}\BibitemShut {NoStop}%
\bibitem [{\citenamefont {Valencia}\ \emph {et~al.}(2020)\citenamefont
  {Valencia}, \citenamefont {de~Bello}, \citenamefont {Galland}, \citenamefont
  {Adler}, \citenamefont {Lep{\v{s}}}, \citenamefont {Anna}, \citenamefont {van
  Klink}, \citenamefont {Carmona}, \citenamefont {Danihelka}, \citenamefont
  {Dengler} \emph {et~al.}}]{valencia2020synchrony}%
  \BibitemOpen
  \bibfield  {author} {\bibinfo {author} {\bibfnamefont {E.}~\bibnamefont
  {Valencia}}, \bibinfo {author} {\bibfnamefont {F.}~\bibnamefont {de~Bello}},
  \bibinfo {author} {\bibfnamefont {T.}~\bibnamefont {Galland}}, \bibinfo
  {author} {\bibfnamefont {P.~B.}\ \bibnamefont {Adler}}, \bibinfo {author}
  {\bibfnamefont {J.}~\bibnamefont {Lep{\v{s}}}}, \bibinfo {author}
  {\bibfnamefont {E.}~\bibnamefont {Anna}}, \bibinfo {author} {\bibfnamefont
  {R.}~\bibnamefont {van Klink}}, \bibinfo {author} {\bibfnamefont {C.~P.}\
  \bibnamefont {Carmona}}, \bibinfo {author} {\bibfnamefont {J.}~\bibnamefont
  {Danihelka}}, \bibinfo {author} {\bibfnamefont {J.}~\bibnamefont {Dengler}},
  \emph {et~al.},\ }\bibfield  {title} {\bibinfo {title} {Synchrony matters
  more than species richness in plant community stability at a global scale},\
  }\href@noop {} {\bibfield  {journal} {\bibinfo  {journal} {Proceedings of the
  National Academy of Sciences}\ }\textbf {\bibinfo {volume} {117}},\ \bibinfo
  {pages} {24345} (\bibinfo {year} {2020})}\BibitemShut {NoStop}%
\bibitem [{\citenamefont {Pimm}(1984)}]{pimm1984complexity}%
  \BibitemOpen
  \bibfield  {author} {\bibinfo {author} {\bibfnamefont {S.~L.}\ \bibnamefont
  {Pimm}},\ }\bibfield  {title} {\bibinfo {title} {The complexity and stability
  of ecosystems},\ }\href@noop {} {\bibfield  {journal} {\bibinfo  {journal}
  {Nature}\ }\textbf {\bibinfo {volume} {307}},\ \bibinfo {pages} {321}
  (\bibinfo {year} {1984})}\BibitemShut {NoStop}%
\bibitem [{\citenamefont {Ives}\ \emph {et~al.}(2000)\citenamefont {Ives},
  \citenamefont {Klug},\ and\ \citenamefont {Gross}}]{ives2000stability}%
  \BibitemOpen
  \bibfield  {author} {\bibinfo {author} {\bibfnamefont {A.~R.}\ \bibnamefont
  {Ives}}, \bibinfo {author} {\bibfnamefont {J.~L.}\ \bibnamefont {Klug}},\
  and\ \bibinfo {author} {\bibfnamefont {K.}~\bibnamefont {Gross}},\ }\bibfield
   {title} {\bibinfo {title} {Stability and species richness in complex
  communities},\ }\href@noop {} {\bibfield  {journal} {\bibinfo  {journal}
  {Ecology Letters}\ }\textbf {\bibinfo {volume} {3}},\ \bibinfo {pages} {399}
  (\bibinfo {year} {2000})}\BibitemShut {NoStop}%
\bibitem [{\citenamefont {Jousset}\ \emph {et~al.}(2011)\citenamefont
  {Jousset}, \citenamefont {Schulz}, \citenamefont {Scheu},\ and\ \citenamefont
  {Eisenhauer}}]{jousset2011intraspecific}%
  \BibitemOpen
  \bibfield  {author} {\bibinfo {author} {\bibfnamefont {A.}~\bibnamefont
  {Jousset}}, \bibinfo {author} {\bibfnamefont {W.}~\bibnamefont {Schulz}},
  \bibinfo {author} {\bibfnamefont {S.}~\bibnamefont {Scheu}},\ and\ \bibinfo
  {author} {\bibfnamefont {N.}~\bibnamefont {Eisenhauer}},\ }\bibfield  {title}
  {\bibinfo {title} {Intraspecific genotypic richness and relatedness predict
  the invasibility of microbial communities},\ }\href@noop {} {\bibfield
  {journal} {\bibinfo  {journal} {The ISME journal}\ }\textbf {\bibinfo
  {volume} {5}},\ \bibinfo {pages} {1108} (\bibinfo {year} {2011})}\BibitemShut
  {NoStop}%
\bibitem [{\citenamefont {Mallon}\ \emph {et~al.}(2015)\citenamefont {Mallon},
  \citenamefont {Van~Elsas},\ and\ \citenamefont
  {Salles}}]{mallon2015microbial}%
  \BibitemOpen
  \bibfield  {author} {\bibinfo {author} {\bibfnamefont {C.~A.}\ \bibnamefont
  {Mallon}}, \bibinfo {author} {\bibfnamefont {J.~D.}\ \bibnamefont
  {Van~Elsas}},\ and\ \bibinfo {author} {\bibfnamefont {J.~F.}\ \bibnamefont
  {Salles}},\ }\bibfield  {title} {\bibinfo {title} {Microbial invasions: the
  process, patterns, and mechanisms},\ }\href@noop {} {\bibfield  {journal}
  {\bibinfo  {journal} {Trends in microbiology}\ }\textbf {\bibinfo {volume}
  {23}},\ \bibinfo {pages} {719} (\bibinfo {year} {2015})}\BibitemShut
  {NoStop}%
\bibitem [{\citenamefont {Capit{\'a}n}\ \emph {et~al.}(2017)\citenamefont
  {Capit{\'a}n}, \citenamefont {Cuenda},\ and\ \citenamefont
  {Alonso}}]{capitan2017stochastic}%
  \BibitemOpen
  \bibfield  {author} {\bibinfo {author} {\bibfnamefont {J.~A.}\ \bibnamefont
  {Capit{\'a}n}}, \bibinfo {author} {\bibfnamefont {S.}~\bibnamefont
  {Cuenda}},\ and\ \bibinfo {author} {\bibfnamefont {D.}~\bibnamefont
  {Alonso}},\ }\bibfield  {title} {\bibinfo {title} {Stochastic competitive
  exclusion leads to a cascade of species extinctions},\ }\href@noop {}
  {\bibfield  {journal} {\bibinfo  {journal} {Journal of Theoretical Biology}\
  }\textbf {\bibinfo {volume} {419}},\ \bibinfo {pages} {137} (\bibinfo {year}
  {2017})}\BibitemShut {NoStop}%
\bibitem [{\citenamefont {Lynch}\ and\ \citenamefont
  {Neufeld}(2015)}]{lynch2015ecology}%
  \BibitemOpen
  \bibfield  {author} {\bibinfo {author} {\bibfnamefont {M.~D.}\ \bibnamefont
  {Lynch}}\ and\ \bibinfo {author} {\bibfnamefont {J.~D.}\ \bibnamefont
  {Neufeld}},\ }\bibfield  {title} {\bibinfo {title} {Ecology and exploration
  of the rare biosphere},\ }\href@noop {} {\bibfield  {journal} {\bibinfo
  {journal} {Nature Reviews Microbiology}\ }\textbf {\bibinfo {volume} {13}},\
  \bibinfo {pages} {217} (\bibinfo {year} {2015})}\BibitemShut {NoStop}%
\bibitem [{\citenamefont {Leidinger}\ and\ \citenamefont
  {Cabral}(2017)}]{leidinger2017biodiversity}%
  \BibitemOpen
  \bibfield  {author} {\bibinfo {author} {\bibfnamefont {L.}~\bibnamefont
  {Leidinger}}\ and\ \bibinfo {author} {\bibfnamefont {J.~S.}\ \bibnamefont
  {Cabral}},\ }\bibfield  {title} {\bibinfo {title} {Biodiversity dynamics on
  islands: Explicitly accounting for causality in mechanistic models},\
  }\href@noop {} {\bibfield  {journal} {\bibinfo  {journal} {Diversity}\
  }\textbf {\bibinfo {volume} {9}},\ \bibinfo {pages} {30} (\bibinfo {year}
  {2017})}\BibitemShut {NoStop}%
\bibitem [{\citenamefont {Magurran}(2013)}]{magurran2013measuring}%
  \BibitemOpen
  \bibfield  {author} {\bibinfo {author} {\bibfnamefont {A.~E.}\ \bibnamefont
  {Magurran}},\ }\href@noop {} {\emph {\bibinfo {title} {Measuring biological
  diversity}}}\ (\bibinfo  {publisher} {John Wiley \& Sons},\ \bibinfo {year}
  {2013})\BibitemShut {NoStop}%
\bibitem [{\citenamefont {Hardin}(1960)}]{hardin1960competitive}%
  \BibitemOpen
  \bibfield  {author} {\bibinfo {author} {\bibfnamefont {G.}~\bibnamefont
  {Hardin}},\ }\bibfield  {title} {\bibinfo {title} {The competitive exclusion
  principle},\ }\href@noop {} {\bibfield  {journal} {\bibinfo  {journal}
  {Science}\ }\textbf {\bibinfo {volume} {131}},\ \bibinfo {pages} {1292}
  (\bibinfo {year} {1960})}\BibitemShut {NoStop}%
\bibitem [{\citenamefont {MacArthur}\ and\ \citenamefont
  {Levins}(1967)}]{macarthur1967limiting}%
  \BibitemOpen
  \bibfield  {author} {\bibinfo {author} {\bibfnamefont {R.}~\bibnamefont
  {MacArthur}}\ and\ \bibinfo {author} {\bibfnamefont {R.}~\bibnamefont
  {Levins}},\ }\bibfield  {title} {\bibinfo {title} {The limiting similarity,
  convergence, and divergence of coexisting species},\ }\href@noop {}
  {\bibfield  {journal} {\bibinfo  {journal} {The american naturalist}\
  }\textbf {\bibinfo {volume} {101}},\ \bibinfo {pages} {377} (\bibinfo {year}
  {1967})}\BibitemShut {NoStop}%
\bibitem [{\citenamefont {Mac~Arthur}(1969)}]{MacArthur1969species}%
  \BibitemOpen
  \bibfield  {author} {\bibinfo {author} {\bibfnamefont {R.}~\bibnamefont
  {Mac~Arthur}},\ }\bibfield  {title} {\bibinfo {title} {Species packing, and
  what competition minimizes},\ }\href@noop {} {\bibfield  {journal} {\bibinfo
  {journal} {Proceedings of the National Academy of Sciences}\ }\textbf
  {\bibinfo {volume} {64}},\ \bibinfo {pages} {1369} (\bibinfo {year}
  {1969})}\BibitemShut {NoStop}%
\bibitem [{\citenamefont {Gause}(2019)}]{gause2019struggle}%
  \BibitemOpen
  \bibfield  {author} {\bibinfo {author} {\bibfnamefont {G.~F.}\ \bibnamefont
  {Gause}},\ }\href@noop {} {\emph {\bibinfo {title} {The Struggle for
  Existence: A Classic of Mathematical Biology and Ecology}}}\ (\bibinfo
  {publisher} {Courier Dover Publications},\ \bibinfo {year}
  {2019})\BibitemShut {NoStop}%
\bibitem [{\citenamefont {Scheffer}\ and\ \citenamefont {van
  Nes}(2006)}]{scheffer2006self}%
  \BibitemOpen
  \bibfield  {author} {\bibinfo {author} {\bibfnamefont {M.}~\bibnamefont
  {Scheffer}}\ and\ \bibinfo {author} {\bibfnamefont {E.~H.}\ \bibnamefont {van
  Nes}},\ }\bibfield  {title} {\bibinfo {title} {Self-organized similarity, the
  evolutionary emergence of groups of similar species},\ }\href@noop {}
  {\bibfield  {journal} {\bibinfo  {journal} {Proceedings of the National
  Academy of Sciences}\ }\textbf {\bibinfo {volume} {103}},\ \bibinfo {pages}
  {6230} (\bibinfo {year} {2006})}\BibitemShut {NoStop}%
\bibitem [{\citenamefont {Vergnon}\ \emph {et~al.}(2012)\citenamefont
  {Vergnon}, \citenamefont {Van~Nes},\ and\ \citenamefont
  {Scheffer}}]{vergnon2012emergent}%
  \BibitemOpen
  \bibfield  {author} {\bibinfo {author} {\bibfnamefont {R.}~\bibnamefont
  {Vergnon}}, \bibinfo {author} {\bibfnamefont {E.~H.}\ \bibnamefont
  {Van~Nes}},\ and\ \bibinfo {author} {\bibfnamefont {M.}~\bibnamefont
  {Scheffer}},\ }\bibfield  {title} {\bibinfo {title} {Emergent neutrality
  leads to multimodal species abundance distributions},\ }\href@noop {}
  {\bibfield  {journal} {\bibinfo  {journal} {Nature communications}\ }\textbf
  {\bibinfo {volume} {3}},\ \bibinfo {pages} {1} (\bibinfo {year}
  {2012})}\BibitemShut {NoStop}%
\bibitem [{\citenamefont {Kessler}\ and\ \citenamefont
  {Shnerb}(2015)}]{kessler2015generalized}%
  \BibitemOpen
  \bibfield  {author} {\bibinfo {author} {\bibfnamefont {D.~A.}\ \bibnamefont
  {Kessler}}\ and\ \bibinfo {author} {\bibfnamefont {N.~M.}\ \bibnamefont
  {Shnerb}},\ }\bibfield  {title} {\bibinfo {title} {Generalized model of
  island biodiversity},\ }\href@noop {} {\bibfield  {journal} {\bibinfo
  {journal} {Physical Review E}\ }\textbf {\bibinfo {volume} {91}},\ \bibinfo
  {pages} {042705} (\bibinfo {year} {2015})}\BibitemShut {NoStop}%
\bibitem [{\citenamefont {Bunin}(2016)}]{bunin2016interaction}%
  \BibitemOpen
  \bibfield  {author} {\bibinfo {author} {\bibfnamefont {G.}~\bibnamefont
  {Bunin}},\ }\bibfield  {title} {\bibinfo {title} {Interaction patterns and
  diversity in assembled ecological communities}} (\bibinfo {year} {2016}),\
  \bibinfo {note} {arXiv [Preprint]
  https://arxiv.org/abs/1607.04734}\BibitemShut {NoStop}%
\bibitem [{\citenamefont {Roy}\ \emph {et~al.}(2020)\citenamefont {Roy},
  \citenamefont {Barbier}, \citenamefont {Biroli},\ and\ \citenamefont
  {Bunin}}]{roy2020complex}%
  \BibitemOpen
  \bibfield  {author} {\bibinfo {author} {\bibfnamefont {F.}~\bibnamefont
  {Roy}}, \bibinfo {author} {\bibfnamefont {M.}~\bibnamefont {Barbier}},
  \bibinfo {author} {\bibfnamefont {G.}~\bibnamefont {Biroli}},\ and\ \bibinfo
  {author} {\bibfnamefont {G.}~\bibnamefont {Bunin}},\ }\bibfield  {title}
  {\bibinfo {title} {Complex interactions can create persistent fluctuations in
  high-diversity ecosystems},\ }\href@noop {} {\bibfield  {journal} {\bibinfo
  {journal} {PLoS computational biology}\ }\textbf {\bibinfo {volume} {16}},\
  \bibinfo {pages} {e1007827} (\bibinfo {year} {2020})}\BibitemShut {NoStop}%
\bibitem [{\citenamefont {Fisher}\ and\ \citenamefont
  {Mehta}(2014)}]{fisher2014transition}%
  \BibitemOpen
  \bibfield  {author} {\bibinfo {author} {\bibfnamefont {C.~K.}\ \bibnamefont
  {Fisher}}\ and\ \bibinfo {author} {\bibfnamefont {P.}~\bibnamefont {Mehta}},\
  }\bibfield  {title} {\bibinfo {title} {The transition between the niche and
  neutral regimes in ecology},\ }\href@noop {} {\bibfield  {journal} {\bibinfo
  {journal} {Proceedings of the National Academy of Sciences}\ }\textbf
  {\bibinfo {volume} {111}},\ \bibinfo {pages} {13111} (\bibinfo {year}
  {2014})}\BibitemShut {NoStop}%
\bibitem [{\citenamefont {Verberk}(2011)}]{verberk2011explaining}%
  \BibitemOpen
  \bibfield  {author} {\bibinfo {author} {\bibfnamefont {W.}~\bibnamefont
  {Verberk}},\ }\bibfield  {title} {\bibinfo {title} {Explaining general
  patterns in species abundance and distributions},\ }\href@noop {} {\bibfield
  {journal} {\bibinfo  {journal} {Nature Education Knowledge}\ }\textbf
  {\bibinfo {volume} {3}},\ \bibinfo {pages} {38} (\bibinfo {year}
  {2011})}\BibitemShut {NoStop}%
\bibitem [{\citenamefont {Fowler}\ and\ \citenamefont
  {Ruokolainen}(2013)}]{fowler2013colonization}%
  \BibitemOpen
  \bibfield  {author} {\bibinfo {author} {\bibfnamefont {M.~S.}\ \bibnamefont
  {Fowler}}\ and\ \bibinfo {author} {\bibfnamefont {L.}~\bibnamefont
  {Ruokolainen}},\ }\bibfield  {title} {\bibinfo {title} {Colonization,
  covariance and colour: Environmental and ecological drivers of
  diversity--stability relationships},\ }\href@noop {} {\bibfield  {journal}
  {\bibinfo  {journal} {Journal of theoretical biology}\ }\textbf {\bibinfo
  {volume} {324}},\ \bibinfo {pages} {32} (\bibinfo {year} {2013})}\BibitemShut
  {NoStop}%
\bibitem [{\citenamefont {Barab{\'a}s}\ \emph {et~al.}(2016)\citenamefont
  {Barab{\'a}s}, \citenamefont {J.~Michalska-Smith},\ and\ \citenamefont
  {Allesina}}]{barabas2016effect}%
  \BibitemOpen
  \bibfield  {author} {\bibinfo {author} {\bibfnamefont {G.}~\bibnamefont
  {Barab{\'a}s}}, \bibinfo {author} {\bibfnamefont {M.}~\bibnamefont
  {J.~Michalska-Smith}},\ and\ \bibinfo {author} {\bibfnamefont
  {S.}~\bibnamefont {Allesina}},\ }\bibfield  {title} {\bibinfo {title} {The
  effect of intra-and interspecific competition on coexistence in multispecies
  communities},\ }\href@noop {} {\bibfield  {journal} {\bibinfo  {journal} {The
  American Naturalist}\ }\textbf {\bibinfo {volume} {188}},\ \bibinfo {pages}
  {E1} (\bibinfo {year} {2016})}\BibitemShut {NoStop}%
\bibitem [{\citenamefont {Evans}\ \emph {et~al.}(2013)\citenamefont {Evans},
  \citenamefont {Ralph}, \citenamefont {Schreiber},\ and\ \citenamefont
  {Sen}}]{evans2013stochastic}%
  \BibitemOpen
  \bibfield  {author} {\bibinfo {author} {\bibfnamefont {S.~N.}\ \bibnamefont
  {Evans}}, \bibinfo {author} {\bibfnamefont {P.~L.}\ \bibnamefont {Ralph}},
  \bibinfo {author} {\bibfnamefont {S.~J.}\ \bibnamefont {Schreiber}},\ and\
  \bibinfo {author} {\bibfnamefont {A.}~\bibnamefont {Sen}},\ }\bibfield
  {title} {\bibinfo {title} {Stochastic population growth in spatially
  heterogeneous environments},\ }\href@noop {} {\bibfield  {journal} {\bibinfo
  {journal} {Journal of mathematical biology}\ }\textbf {\bibinfo {volume}
  {66}},\ \bibinfo {pages} {423} (\bibinfo {year} {2013})}\BibitemShut
  {NoStop}%
\bibitem [{\citenamefont {Tejo}\ \emph {et~al.}(2021)\citenamefont {Tejo},
  \citenamefont {Qui{\~n}inao}, \citenamefont {Rebolledo},\ and\ \citenamefont
  {Marquet}}]{tejo2021coexistence}%
  \BibitemOpen
  \bibfield  {author} {\bibinfo {author} {\bibfnamefont {M.}~\bibnamefont
  {Tejo}}, \bibinfo {author} {\bibfnamefont {C.}~\bibnamefont {Qui{\~n}inao}},
  \bibinfo {author} {\bibfnamefont {R.}~\bibnamefont {Rebolledo}},\ and\
  \bibinfo {author} {\bibfnamefont {P.~A.}\ \bibnamefont {Marquet}},\
  }\bibfield  {title} {\bibinfo {title} {Coexistence, dispersal and spatial
  structure in metacommunities: a stochastic model approach},\ }\href@noop {}
  {\bibfield  {journal} {\bibinfo  {journal} {Theoretical Ecology}\ }\textbf
  {\bibinfo {volume} {14}},\ \bibinfo {pages} {279} (\bibinfo {year}
  {2021})}\BibitemShut {NoStop}%
\bibitem [{\citenamefont {Alonso}\ \emph {et~al.}(2006)\citenamefont {Alonso},
  \citenamefont {Etienne},\ and\ \citenamefont {McKane}}]{alonso2006merits}%
  \BibitemOpen
  \bibfield  {author} {\bibinfo {author} {\bibfnamefont {D.}~\bibnamefont
  {Alonso}}, \bibinfo {author} {\bibfnamefont {R.~S.}\ \bibnamefont
  {Etienne}},\ and\ \bibinfo {author} {\bibfnamefont {A.~J.}\ \bibnamefont
  {McKane}},\ }\bibfield  {title} {\bibinfo {title} {The merits of neutral
  theory},\ }\href@noop {} {\bibfield  {journal} {\bibinfo  {journal} {Trends
  in ecology \& evolution}\ }\textbf {\bibinfo {volume} {21}},\ \bibinfo
  {pages} {451} (\bibinfo {year} {2006})}\BibitemShut {NoStop}%
\bibitem [{\citenamefont {Haegeman}\ and\ \citenamefont
  {Loreau}(2011)}]{haegeman2011mathematical}%
  \BibitemOpen
  \bibfield  {author} {\bibinfo {author} {\bibfnamefont {B.}~\bibnamefont
  {Haegeman}}\ and\ \bibinfo {author} {\bibfnamefont {M.}~\bibnamefont
  {Loreau}},\ }\bibfield  {title} {\bibinfo {title} {A mathematical synthesis
  of niche and neutral theories in community ecology},\ }\href@noop {}
  {\bibfield  {journal} {\bibinfo  {journal} {Journal of theoretical biology}\
  }\textbf {\bibinfo {volume} {269}},\ \bibinfo {pages} {150} (\bibinfo {year}
  {2011})}\BibitemShut {NoStop}%
\bibitem [{\citenamefont {Baxter}\ \emph {et~al.}(2007)\citenamefont {Baxter},
  \citenamefont {Blythe},\ and\ \citenamefont {McKane}}]{baxter2007exact}%
  \BibitemOpen
  \bibfield  {author} {\bibinfo {author} {\bibfnamefont {G.~J.}\ \bibnamefont
  {Baxter}}, \bibinfo {author} {\bibfnamefont {R.~A.}\ \bibnamefont {Blythe}},\
  and\ \bibinfo {author} {\bibfnamefont {A.~J.}\ \bibnamefont {McKane}},\
  }\bibfield  {title} {\bibinfo {title} {Exact solution of the multi-allelic
  diffusion model},\ }\href@noop {} {\bibfield  {journal} {\bibinfo  {journal}
  {Mathematical bioSciences}\ }\textbf {\bibinfo {volume} {209}},\ \bibinfo
  {pages} {124} (\bibinfo {year} {2007})}\BibitemShut {NoStop}%
\bibitem [{\citenamefont {Xu}\ and\ \citenamefont
  {Chou}(2018)}]{xu2018immigration}%
  \BibitemOpen
  \bibfield  {author} {\bibinfo {author} {\bibfnamefont {S.}~\bibnamefont
  {Xu}}\ and\ \bibinfo {author} {\bibfnamefont {T.}~\bibnamefont {Chou}},\
  }\bibfield  {title} {\bibinfo {title} {Immigration-induced phase transition
  in a regulated multispecies birth-death process},\ }\href@noop {} {\bibfield
  {journal} {\bibinfo  {journal} {Journal of Physics A: Mathematical and
  Theoretical}\ }\textbf {\bibinfo {volume} {51}},\ \bibinfo {pages} {425602}
  (\bibinfo {year} {2018})}\BibitemShut {NoStop}%
\bibitem [{\citenamefont {Capit{\'a}n}\ \emph {et~al.}(2020)\citenamefont
  {Capit{\'a}n}, \citenamefont {Cuenda},\ and\ \citenamefont
  {Alonso}}]{capitan2020competitive}%
  \BibitemOpen
  \bibfield  {author} {\bibinfo {author} {\bibfnamefont {J.~A.}\ \bibnamefont
  {Capit{\'a}n}}, \bibinfo {author} {\bibfnamefont {S.}~\bibnamefont
  {Cuenda}},\ and\ \bibinfo {author} {\bibfnamefont {D.}~\bibnamefont
  {Alonso}},\ }\bibfield  {title} {\bibinfo {title} {Competitive dominance in
  plant communities: Modeling approaches and theoretical predictions},\
  }\href@noop {} {\bibfield  {journal} {\bibinfo  {journal} {Journal of
  Theoretical Biology}\ }\textbf {\bibinfo {volume} {502}},\ \bibinfo {pages}
  {110349} (\bibinfo {year} {2020})}\BibitemShut {NoStop}%
\bibitem [{\citenamefont {Black}\ and\ \citenamefont
  {McKane}(2012)}]{black2012stochastic}%
  \BibitemOpen
  \bibfield  {author} {\bibinfo {author} {\bibfnamefont {A.~J.}\ \bibnamefont
  {Black}}\ and\ \bibinfo {author} {\bibfnamefont {A.~J.}\ \bibnamefont
  {McKane}},\ }\bibfield  {title} {\bibinfo {title} {Stochastic formulation of
  ecological models and their applications},\ }\href@noop {} {\bibfield
  {journal} {\bibinfo  {journal} {Trends in ecology \& evolution}\ }\textbf
  {\bibinfo {volume} {27}},\ \bibinfo {pages} {337} (\bibinfo {year}
  {2012})}\BibitemShut {NoStop}%
\bibitem [{\citenamefont {Chesson}(2018)}]{chesson2018updates}%
  \BibitemOpen
  \bibfield  {author} {\bibinfo {author} {\bibfnamefont {P.}~\bibnamefont
  {Chesson}},\ }\bibfield  {title} {\bibinfo {title} {Updates on mechanisms of
  maintenance of species diversity},\ }\href@noop {} {\bibfield  {journal}
  {\bibinfo  {journal} {Journal of ecology}\ }\textbf {\bibinfo {volume}
  {106}},\ \bibinfo {pages} {1773} (\bibinfo {year} {2018})}\BibitemShut
  {NoStop}%
\bibitem [{\citenamefont {MacArthur}(1970)}]{macArthur1970species}%
  \BibitemOpen
  \bibfield  {author} {\bibinfo {author} {\bibfnamefont {R.}~\bibnamefont
  {MacArthur}},\ }\bibfield  {title} {\bibinfo {title} {Species packing and
  competitive equilibrium for many species},\ }\href@noop {} {\bibfield
  {journal} {\bibinfo  {journal} {Theoretical population biology}\ }\textbf
  {\bibinfo {volume} {1}},\ \bibinfo {pages} {1} (\bibinfo {year}
  {1970})}\BibitemShut {NoStop}%
\bibitem [{\citenamefont {Chesson}(1990)}]{chesson1990macarthur}%
  \BibitemOpen
  \bibfield  {author} {\bibinfo {author} {\bibfnamefont {P.}~\bibnamefont
  {Chesson}},\ }\bibfield  {title} {\bibinfo {title} {Macarthur's
  consumer-resource model},\ }\href@noop {} {\bibfield  {journal} {\bibinfo
  {journal} {Theoretical Population Biology}\ }\textbf {\bibinfo {volume}
  {37}},\ \bibinfo {pages} {26} (\bibinfo {year} {1990})}\BibitemShut {NoStop}%
\bibitem [{\citenamefont {O’Dwyer}(2018)}]{o2018whence}%
  \BibitemOpen
  \bibfield  {author} {\bibinfo {author} {\bibfnamefont {J.~P.}\ \bibnamefont
  {O’Dwyer}},\ }\bibfield  {title} {\bibinfo {title} {Whence
  lotka-volterra?},\ }\href@noop {} {\bibfield  {journal} {\bibinfo  {journal}
  {Theoretical Ecology}\ }\textbf {\bibinfo {volume} {11}},\ \bibinfo {pages}
  {441} (\bibinfo {year} {2018})}\BibitemShut {NoStop}%
\bibitem [{\citenamefont {Allesina}\ and\ \citenamefont
  {Tang}(2012)}]{allesina2012stability}%
  \BibitemOpen
  \bibfield  {author} {\bibinfo {author} {\bibfnamefont {S.}~\bibnamefont
  {Allesina}}\ and\ \bibinfo {author} {\bibfnamefont {S.}~\bibnamefont
  {Tang}},\ }\bibfield  {title} {\bibinfo {title} {Stability criteria for
  complex ecosystems},\ }\href@noop {} {\bibfield  {journal} {\bibinfo
  {journal} {Nature}\ }\textbf {\bibinfo {volume} {483}},\ \bibinfo {pages}
  {205} (\bibinfo {year} {2012})}\BibitemShut {NoStop}%
\bibitem [{\citenamefont {Gardiner}\ \emph {et~al.}(1985)\citenamefont
  {Gardiner} \emph {et~al.}}]{gardiner1985handbook}%
  \BibitemOpen
  \bibfield  {author} {\bibinfo {author} {\bibfnamefont {C.~W.}\ \bibnamefont
  {Gardiner}} \emph {et~al.},\ }\href@noop {} {\emph {\bibinfo {title}
  {Handbook of stochastic methods}}},\ Vol.~\bibinfo {volume} {3}\ (\bibinfo
  {publisher} {springer Berlin},\ \bibinfo {year} {1985})\BibitemShut {NoStop}%
\bibitem [{\citenamefont {Grimmett}\ and\ \citenamefont
  {Stirzaker}(2001)}]{grimmett2001probability}%
  \BibitemOpen
  \bibfield  {author} {\bibinfo {author} {\bibfnamefont {G.}~\bibnamefont
  {Grimmett}}\ and\ \bibinfo {author} {\bibfnamefont {D.}~\bibnamefont
  {Stirzaker}},\ }\href@noop {} {\emph {\bibinfo {title} {Probability and
  Random Processes}}}\ (\bibinfo  {publisher} {Oxford University Press},\
  \bibinfo {year} {2001})\BibitemShut {NoStop}%
\bibitem [{\citenamefont {Schnakenberg}(1976)}]{schnakenberg1976network}%
  \BibitemOpen
  \bibfield  {author} {\bibinfo {author} {\bibfnamefont {J.}~\bibnamefont
  {Schnakenberg}},\ }\bibfield  {title} {\bibinfo {title} {Network theory of
  microscopic and macroscopic behavior of master equation systems},\
  }\href@noop {} {\bibfield  {journal} {\bibinfo  {journal} {Reviews of Modern
  physics}\ }\textbf {\bibinfo {volume} {48}},\ \bibinfo {pages} {571}
  (\bibinfo {year} {1976})}\BibitemShut {NoStop}%
\bibitem [{\citenamefont {Meyn}\ and\ \citenamefont
  {Tweedie}(1993)}]{meyn1993stability}%
  \BibitemOpen
  \bibfield  {author} {\bibinfo {author} {\bibfnamefont {S.~P.}\ \bibnamefont
  {Meyn}}\ and\ \bibinfo {author} {\bibfnamefont {R.~L.}\ \bibnamefont
  {Tweedie}},\ }\bibfield  {title} {\bibinfo {title} {Stability of markovian
  processes iii: Foster--lyapunov criteria for continuous-time processes},\
  }\href@noop {} {\bibfield  {journal} {\bibinfo  {journal} {Advances in
  Applied Probability}\ }\textbf {\bibinfo {volume} {25}},\ \bibinfo {pages}
  {518} (\bibinfo {year} {1993})}\BibitemShut {NoStop}%
\bibitem [{\citenamefont {Gupta}\ \emph {et~al.}(2014)\citenamefont {Gupta},
  \citenamefont {Briat},\ and\ \citenamefont {Khammash}}]{gupta2014scalable}%
  \BibitemOpen
  \bibfield  {author} {\bibinfo {author} {\bibfnamefont {A.}~\bibnamefont
  {Gupta}}, \bibinfo {author} {\bibfnamefont {C.}~\bibnamefont {Briat}},\ and\
  \bibinfo {author} {\bibfnamefont {M.}~\bibnamefont {Khammash}},\ }\bibfield
  {title} {\bibinfo {title} {A scalable computational framework for
  establishing long-term behavior of stochastic reaction networks},\
  }\href@noop {} {\bibfield  {journal} {\bibinfo  {journal} {PLoS computational
  biology}\ }\textbf {\bibinfo {volume} {10}},\ \bibinfo {pages} {e1003669}
  (\bibinfo {year} {2014})}\BibitemShut {NoStop}%
\bibitem [{\citenamefont {Marquet}\ \emph {et~al.}(2020)\citenamefont
  {Marquet}, \citenamefont {Tejo},\ and\ \citenamefont
  {Rebolledo}}]{dobson2020unsolved}%
  \BibitemOpen
  \bibfield  {author} {\bibinfo {author} {\bibfnamefont {P.~A.}\ \bibnamefont
  {Marquet}}, \bibinfo {author} {\bibfnamefont {M.}~\bibnamefont {Tejo}},\ and\
  \bibinfo {author} {\bibfnamefont {R.}~\bibnamefont {Rebolledo}},\ }\bibfield
  {title} {\bibinfo {title} {What is the species richness distribution?},\ }in\
  \href@noop {} {\emph {\bibinfo {booktitle} {Unsolved problems in ecology}}},\
  \bibinfo {editor} {edited by\ \bibinfo {editor} {\bibfnamefont
  {A.}~\bibnamefont {Dobson}}, \bibinfo {editor} {\bibfnamefont {R.~D.}\
  \bibnamefont {Holt}},\ and\ \bibinfo {editor} {\bibfnamefont
  {D.}~\bibnamefont {Tilman}}}\ (\bibinfo  {publisher} {Princeton University
  Press},\ \bibinfo {address} {USA},\ \bibinfo {year} {2020})\BibitemShut
  {NoStop}%
\bibitem [{\citenamefont {MacArthur}\ and\ \citenamefont
  {Wilson}(1967)}]{macarthur1967theory}%
  \BibitemOpen
  \bibfield  {author} {\bibinfo {author} {\bibfnamefont {R.~H.}\ \bibnamefont
  {MacArthur}}\ and\ \bibinfo {author} {\bibfnamefont {E.~O.}\ \bibnamefont
  {Wilson}},\ }\href@noop {} {\emph {\bibinfo {title} {The Theory of Island
  Biogeography: By Robert H. MacArthur and Edward O. Wilson}}}\ (\bibinfo
  {publisher} {Princeton University press},\ \bibinfo {year}
  {1967})\BibitemShut {NoStop}%
\bibitem [{\citenamefont {Goyal}\ \emph {et~al.}(2015)\citenamefont {Goyal},
  \citenamefont {Kim}, \citenamefont {Chen},\ and\ \citenamefont
  {Chou}}]{goyal2015mechanisms}%
  \BibitemOpen
  \bibfield  {author} {\bibinfo {author} {\bibfnamefont {S.}~\bibnamefont
  {Goyal}}, \bibinfo {author} {\bibfnamefont {S.}~\bibnamefont {Kim}}, \bibinfo
  {author} {\bibfnamefont {I.~S.}\ \bibnamefont {Chen}},\ and\ \bibinfo
  {author} {\bibfnamefont {T.}~\bibnamefont {Chou}},\ }\bibfield  {title}
  {\bibinfo {title} {Mechanisms of blood homeostasis: lineage tracking and a
  neutral model of cell populations in rhesus macaques},\ }\href@noop {}
  {\bibfield  {journal} {\bibinfo  {journal} {BMC biology}\ }\textbf {\bibinfo
  {volume} {13}},\ \bibinfo {pages} {1} (\bibinfo {year} {2015})}\BibitemShut
  {NoStop}%
\bibitem [{\citenamefont {Hu}\ \emph {et~al.}(2021)\citenamefont {Hu},
  \citenamefont {Amor}, \citenamefont {Barbier}, \citenamefont {Bunin},\ and\
  \citenamefont {Gore}}]{gore2021}%
  \BibitemOpen
  \bibfield  {author} {\bibinfo {author} {\bibfnamefont {J.}~\bibnamefont
  {Hu}}, \bibinfo {author} {\bibfnamefont {D.~R.}\ \bibnamefont {Amor}},
  \bibinfo {author} {\bibfnamefont {M.}~\bibnamefont {Barbier}}, \bibinfo
  {author} {\bibfnamefont {G.}~\bibnamefont {Bunin}},\ and\ \bibinfo {author}
  {\bibfnamefont {J.}~\bibnamefont {Gore}},\ }\bibfield  {title} {\bibinfo
  {title} {Emergent phases of ecological diversity and dynamics mapped in
  microcosms},\ }\Eprint
  {https://arxiv.org/abs/https://www.biorxiv.org/content/early/2021/10/29/2021.10.28.466339.full.pdf}
  {https://www.biorxiv.org/content/early/2021/10/29/2021.10.28.466339.full.pdf}
   (\bibinfo {year} {2021}),\ \bibinfo {note} {biorXiv [Preprint]
  https://doi.org/10.1101/2021.10.28.466339}\BibitemShut {NoStop}%
\bibitem [{\citenamefont {Iyer-Biswas}\ and\ \citenamefont
  {Zilman}(2016)}]{iyer2016first}%
  \BibitemOpen
  \bibfield  {author} {\bibinfo {author} {\bibfnamefont {S.}~\bibnamefont
  {Iyer-Biswas}}\ and\ \bibinfo {author} {\bibfnamefont {A.}~\bibnamefont
  {Zilman}},\ }\bibfield  {title} {\bibinfo {title} {First-passage processes in
  cellular biology},\ }\href@noop {} {\bibfield  {journal} {\bibinfo  {journal}
  {Advances in chemical physics}\ }\textbf {\bibinfo {volume} {160}},\ \bibinfo
  {pages} {261} (\bibinfo {year} {2016})}\BibitemShut {NoStop}%
\bibitem [{\citenamefont {Redner}(2001)}]{redner2001guide}%
  \BibitemOpen
  \bibfield  {author} {\bibinfo {author} {\bibfnamefont {S.}~\bibnamefont
  {Redner}},\ }\href@noop {} {\emph {\bibinfo {title} {A guide to first-passage
  processes}}}\ (\bibinfo  {publisher} {Cambridge university press},\ \bibinfo
  {year} {2001})\BibitemShut {NoStop}%
\bibitem [{\citenamefont {May}(1972)}]{may1972will}%
  \BibitemOpen
  \bibfield  {author} {\bibinfo {author} {\bibfnamefont {R.~M.}\ \bibnamefont
  {May}},\ }\bibfield  {title} {\bibinfo {title} {Will a large complex system
  be stable?},\ }\href@noop {} {\bibfield  {journal} {\bibinfo  {journal}
  {Nature}\ }\textbf {\bibinfo {volume} {238}},\ \bibinfo {pages} {413}
  (\bibinfo {year} {1972})}\BibitemShut {NoStop}%
\bibitem [{\citenamefont {Allesina}\ and\ \citenamefont
  {Pascual}(2008)}]{allesina2008network}%
  \BibitemOpen
  \bibfield  {author} {\bibinfo {author} {\bibfnamefont {S.}~\bibnamefont
  {Allesina}}\ and\ \bibinfo {author} {\bibfnamefont {M.}~\bibnamefont
  {Pascual}},\ }\bibfield  {title} {\bibinfo {title} {Network structure,
  predator--prey modules, and stability in large food webs},\ }\href@noop {}
  {\bibfield  {journal} {\bibinfo  {journal} {Theoretical Ecology}\ }\textbf
  {\bibinfo {volume} {1}},\ \bibinfo {pages} {55} (\bibinfo {year}
  {2008})}\BibitemShut {NoStop}%
\bibitem [{\citenamefont {Bunin}(2017)}]{bunin2017ecological}%
  \BibitemOpen
  \bibfield  {author} {\bibinfo {author} {\bibfnamefont {G.}~\bibnamefont
  {Bunin}},\ }\bibfield  {title} {\bibinfo {title} {Ecological communities with
  lotka-volterra dynamics},\ }\href@noop {} {\bibfield  {journal} {\bibinfo
  {journal} {Physical Review E}\ }\textbf {\bibinfo {volume} {95}},\ \bibinfo
  {pages} {042414} (\bibinfo {year} {2017})}\BibitemShut {NoStop}%
\bibitem [{\citenamefont {Weigelt}\ \emph {et~al.}(2010)\citenamefont
  {Weigelt}, \citenamefont {Marquard}, \citenamefont {Temperton}, \citenamefont
  {Roscher}, \citenamefont {Scherber}, \citenamefont {Mwangi}, \citenamefont
  {Von~Felten}, \citenamefont {Buchmann}, \citenamefont {Schmid}, \citenamefont
  {Schulze} \emph {et~al.}}]{weigelt2010jena}%
  \BibitemOpen
  \bibfield  {author} {\bibinfo {author} {\bibfnamefont {A.}~\bibnamefont
  {Weigelt}}, \bibinfo {author} {\bibfnamefont {E.}~\bibnamefont {Marquard}},
  \bibinfo {author} {\bibfnamefont {V.~M.}\ \bibnamefont {Temperton}}, \bibinfo
  {author} {\bibfnamefont {C.}~\bibnamefont {Roscher}}, \bibinfo {author}
  {\bibfnamefont {C.}~\bibnamefont {Scherber}}, \bibinfo {author}
  {\bibfnamefont {P.~N.}\ \bibnamefont {Mwangi}}, \bibinfo {author}
  {\bibfnamefont {S.}~\bibnamefont {Von~Felten}}, \bibinfo {author}
  {\bibfnamefont {N.}~\bibnamefont {Buchmann}}, \bibinfo {author}
  {\bibfnamefont {B.}~\bibnamefont {Schmid}}, \bibinfo {author} {\bibfnamefont
  {E.-D.}\ \bibnamefont {Schulze}}, \emph {et~al.},\ }\bibfield  {title}
  {\bibinfo {title} {The jena experiment: six years of data from a grassland
  biodiversity experiment},\ }\href@noop {} {\bibfield  {journal} {\bibinfo
  {journal} {Ecology}\ }\textbf {\bibinfo {volume} {91}},\ \bibinfo {pages}
  {930} (\bibinfo {year} {2010})}\BibitemShut {NoStop}%
\bibitem [{\citenamefont {Dur{\'a}n}\ \emph {et~al.}(2021)\citenamefont
  {Dur{\'a}n}, \citenamefont {Hern{\'a}ndez}, \citenamefont {Suesca},
  \citenamefont {Acevedo}, \citenamefont {Acosta}, \citenamefont {Forero},
  \citenamefont {Rozo},\ and\ \citenamefont
  {Pedraza}}]{duran2021slipstreaming}%
  \BibitemOpen
  \bibfield  {author} {\bibinfo {author} {\bibfnamefont {D.~C.}\ \bibnamefont
  {Dur{\'a}n}}, \bibinfo {author} {\bibfnamefont {C.~A.}\ \bibnamefont
  {Hern{\'a}ndez}}, \bibinfo {author} {\bibfnamefont {E.}~\bibnamefont
  {Suesca}}, \bibinfo {author} {\bibfnamefont {R.}~\bibnamefont {Acevedo}},
  \bibinfo {author} {\bibfnamefont {I.~M.}\ \bibnamefont {Acosta}}, \bibinfo
  {author} {\bibfnamefont {D.~A.}\ \bibnamefont {Forero}}, \bibinfo {author}
  {\bibfnamefont {F.~E.}\ \bibnamefont {Rozo}},\ and\ \bibinfo {author}
  {\bibfnamefont {J.~M.}\ \bibnamefont {Pedraza}},\ }\bibfield  {title}
  {\bibinfo {title} {Slipstreaming mother machine: A microfluidic device for
  single-cell dynamic imaging of yeast},\ }\href@noop {} {\bibfield  {journal}
  {\bibinfo  {journal} {Micromachines}\ }\textbf {\bibinfo {volume} {12}},\
  \bibinfo {pages} {4} (\bibinfo {year} {2021})}\BibitemShut {NoStop}%
\bibitem [{\citenamefont {Baca{\"e}r}(2011)}]{bacaer2011yule}%
  \BibitemOpen
  \bibfield  {author} {\bibinfo {author} {\bibfnamefont {N.}~\bibnamefont
  {Baca{\"e}r}},\ }\bibfield  {title} {\bibinfo {title} {Yule and evolution
  (1924)},\ }in\ \href@noop {} {\emph {\bibinfo {booktitle} {A Short History of
  Mathematical Population Dynamics}}}\ (\bibinfo  {publisher} {Springer},\
  \bibinfo {year} {2011})\ pp.\ \bibinfo {pages} {81--88}\BibitemShut {NoStop}%
\bibitem [{\citenamefont {Jeraldo}\ \emph {et~al.}(2012)\citenamefont
  {Jeraldo}, \citenamefont {Sipos}, \citenamefont {Chia}, \citenamefont
  {Brulc}, \citenamefont {Dhillon}, \citenamefont {Konkel}, \citenamefont
  {Larson}, \citenamefont {Nelson}, \citenamefont {Qu}, \citenamefont {Schook}
  \emph {et~al.}}]{jeraldo2012quantification}%
  \BibitemOpen
  \bibfield  {author} {\bibinfo {author} {\bibfnamefont {P.}~\bibnamefont
  {Jeraldo}}, \bibinfo {author} {\bibfnamefont {M.}~\bibnamefont {Sipos}},
  \bibinfo {author} {\bibfnamefont {N.}~\bibnamefont {Chia}}, \bibinfo {author}
  {\bibfnamefont {J.~M.}\ \bibnamefont {Brulc}}, \bibinfo {author}
  {\bibfnamefont {A.~S.}\ \bibnamefont {Dhillon}}, \bibinfo {author}
  {\bibfnamefont {M.~E.}\ \bibnamefont {Konkel}}, \bibinfo {author}
  {\bibfnamefont {C.~L.}\ \bibnamefont {Larson}}, \bibinfo {author}
  {\bibfnamefont {K.~E.}\ \bibnamefont {Nelson}}, \bibinfo {author}
  {\bibfnamefont {A.}~\bibnamefont {Qu}}, \bibinfo {author} {\bibfnamefont
  {L.~B.}\ \bibnamefont {Schook}}, \emph {et~al.},\ }\bibfield  {title}
  {\bibinfo {title} {Quantification of the relative roles of niche and neutral
  processes in structuring gastrointestinal microbiomes},\ }\href@noop {}
  {\bibfield  {journal} {\bibinfo  {journal} {Proceedings of the National
  Academy of Sciences}\ }\textbf {\bibinfo {volume} {109}},\ \bibinfo {pages}
  {9692} (\bibinfo {year} {2012})}\BibitemShut {NoStop}%
\bibitem [{\citenamefont {Dornelas}\ and\ \citenamefont
  {Connolly}(2008)}]{dornelas2008multiple}%
  \BibitemOpen
  \bibfield  {author} {\bibinfo {author} {\bibfnamefont {M.}~\bibnamefont
  {Dornelas}}\ and\ \bibinfo {author} {\bibfnamefont {S.~R.}\ \bibnamefont
  {Connolly}},\ }\bibfield  {title} {\bibinfo {title} {Multiple modes in a
  coral species abundance distribution},\ }\href@noop {} {\bibfield  {journal}
  {\bibinfo  {journal} {Ecology Letters}\ }\textbf {\bibinfo {volume} {11}},\
  \bibinfo {pages} {1008} (\bibinfo {year} {2008})}\BibitemShut {NoStop}%
\bibitem [{\citenamefont {Callaghan}\ \emph {et~al.}(2021)\citenamefont
  {Callaghan}, \citenamefont {Nakagawa},\ and\ \citenamefont
  {Cornwell}}]{callaghan2021global}%
  \BibitemOpen
  \bibfield  {author} {\bibinfo {author} {\bibfnamefont {C.~T.}\ \bibnamefont
  {Callaghan}}, \bibinfo {author} {\bibfnamefont {S.}~\bibnamefont
  {Nakagawa}},\ and\ \bibinfo {author} {\bibfnamefont {W.~K.}\ \bibnamefont
  {Cornwell}},\ }\bibfield  {title} {\bibinfo {title} {Global abundance
  estimates for 9,700 bird species},\ }\href@noop {} {\bibfield  {journal}
  {\bibinfo  {journal} {Proceedings of the National Academy of Sciences}\
  }\textbf {\bibinfo {volume} {118}} (\bibinfo {year} {2021})}\BibitemShut
  {NoStop}%
\bibitem [{\citenamefont {Ser-Giacomi}\ \emph {et~al.}(2018)\citenamefont
  {Ser-Giacomi}, \citenamefont {Zinger}, \citenamefont {Malviya}, \citenamefont
  {De~Vargas}, \citenamefont {Karsenti}, \citenamefont {Bowler},\ and\
  \citenamefont {De~Monte}}]{ser2018ubiquitous}%
  \BibitemOpen
  \bibfield  {author} {\bibinfo {author} {\bibfnamefont {E.}~\bibnamefont
  {Ser-Giacomi}}, \bibinfo {author} {\bibfnamefont {L.}~\bibnamefont {Zinger}},
  \bibinfo {author} {\bibfnamefont {S.}~\bibnamefont {Malviya}}, \bibinfo
  {author} {\bibfnamefont {C.}~\bibnamefont {De~Vargas}}, \bibinfo {author}
  {\bibfnamefont {E.}~\bibnamefont {Karsenti}}, \bibinfo {author}
  {\bibfnamefont {C.}~\bibnamefont {Bowler}},\ and\ \bibinfo {author}
  {\bibfnamefont {S.}~\bibnamefont {De~Monte}},\ }\bibfield  {title} {\bibinfo
  {title} {Ubiquitous abundance distribution of non-dominant plankton across
  the global ocean},\ }\href@noop {} {\bibfield  {journal} {\bibinfo  {journal}
  {Nature ecology \& evolution}\ }\textbf {\bibinfo {volume} {2}},\ \bibinfo
  {pages} {1243} (\bibinfo {year} {2018})}\BibitemShut {NoStop}%
\bibitem [{\citenamefont {Matthews}\ \emph {et~al.}(2014)\citenamefont
  {Matthews}, \citenamefont {Borges},\ and\ \citenamefont
  {Whittaker}}]{matthews2014multimodal}%
  \BibitemOpen
  \bibfield  {author} {\bibinfo {author} {\bibfnamefont {T.~J.}\ \bibnamefont
  {Matthews}}, \bibinfo {author} {\bibfnamefont {P.~A.}\ \bibnamefont
  {Borges}},\ and\ \bibinfo {author} {\bibfnamefont {R.~J.}\ \bibnamefont
  {Whittaker}},\ }\bibfield  {title} {\bibinfo {title} {Multimodal species
  abundance distributions: a deconstruction approach reveals the processes
  behind the pattern},\ }\href@noop {} {\bibfield  {journal} {\bibinfo
  {journal} {Oikos}\ }\textbf {\bibinfo {volume} {123}},\ \bibinfo {pages}
  {533} (\bibinfo {year} {2014})}\BibitemShut {NoStop}%
\bibitem [{\citenamefont {Oakes}\ \emph {et~al.}(2017)\citenamefont {Oakes},
  \citenamefont {Heather}, \citenamefont {Best}, \citenamefont {Byng-Maddick},
  \citenamefont {Husovsky}, \citenamefont {Ismail}, \citenamefont {Joshi},
  \citenamefont {Maxwell}, \citenamefont {Noursadeghi}, \citenamefont {Riddell}
  \emph {et~al.}}]{oakes2017quantitative}%
  \BibitemOpen
  \bibfield  {author} {\bibinfo {author} {\bibfnamefont {T.}~\bibnamefont
  {Oakes}}, \bibinfo {author} {\bibfnamefont {J.~M.}\ \bibnamefont {Heather}},
  \bibinfo {author} {\bibfnamefont {K.}~\bibnamefont {Best}}, \bibinfo {author}
  {\bibfnamefont {R.}~\bibnamefont {Byng-Maddick}}, \bibinfo {author}
  {\bibfnamefont {C.}~\bibnamefont {Husovsky}}, \bibinfo {author}
  {\bibfnamefont {M.}~\bibnamefont {Ismail}}, \bibinfo {author} {\bibfnamefont
  {K.}~\bibnamefont {Joshi}}, \bibinfo {author} {\bibfnamefont
  {G.}~\bibnamefont {Maxwell}}, \bibinfo {author} {\bibfnamefont
  {M.}~\bibnamefont {Noursadeghi}}, \bibinfo {author} {\bibfnamefont
  {N.}~\bibnamefont {Riddell}}, \emph {et~al.},\ }\bibfield  {title} {\bibinfo
  {title} {Quantitative characterization of the t cell receptor repertoire of
  na{\"\i}ve and memory subsets using an integrated experimental and
  computational pipeline which is robust, economical, and versatile},\
  }\href@noop {} {\bibfield  {journal} {\bibinfo  {journal} {Frontiers in
  immunology}\ }\textbf {\bibinfo {volume} {8}},\ \bibinfo {pages} {1267}
  (\bibinfo {year} {2017})}\BibitemShut {NoStop}%
\bibitem [{\citenamefont {Descheemaeker}\ and\ \citenamefont
  {de~Buyl}(2020)}]{descheemaeker2020stochastic}%
  \BibitemOpen
  \bibfield  {author} {\bibinfo {author} {\bibfnamefont {L.}~\bibnamefont
  {Descheemaeker}}\ and\ \bibinfo {author} {\bibfnamefont {S.}~\bibnamefont
  {de~Buyl}},\ }\bibfield  {title} {\bibinfo {title} {Stochastic logistic
  models reproduce experimental time series of microbial communities},\
  }\href@noop {} {\bibfield  {journal} {\bibinfo  {journal} {Elife}\ }\textbf
  {\bibinfo {volume} {9}},\ \bibinfo {pages} {e55650} (\bibinfo {year}
  {2020})}\BibitemShut {NoStop}%
\bibitem [{\citenamefont {Li}\ and\ \citenamefont
  {Chesson}(2018)}]{li2018seed}%
  \BibitemOpen
  \bibfield  {author} {\bibinfo {author} {\bibfnamefont {Y.~M.}\ \bibnamefont
  {Li}}\ and\ \bibinfo {author} {\bibfnamefont {P.}~\bibnamefont {Chesson}},\
  }\bibfield  {title} {\bibinfo {title} {Seed demographic comparisons reveal
  spatial and temporal niche differentiation between native and invasive
  species in a community of desert winter annual plants},\ }\href@noop {}
  {\bibfield  {journal} {\bibinfo  {journal} {Evolutionary Ecology Research}\
  }\textbf {\bibinfo {volume} {19}},\ \bibinfo {pages} {71} (\bibinfo {year}
  {2018})}\BibitemShut {NoStop}%
\bibitem [{\citenamefont {Chesson}\ and\ \citenamefont
  {Kuang}(2008)}]{chesson2008interaction}%
  \BibitemOpen
  \bibfield  {author} {\bibinfo {author} {\bibfnamefont {P.}~\bibnamefont
  {Chesson}}\ and\ \bibinfo {author} {\bibfnamefont {J.~J.}\ \bibnamefont
  {Kuang}},\ }\bibfield  {title} {\bibinfo {title} {The interaction between
  predation and competition},\ }\href@noop {} {\bibfield  {journal} {\bibinfo
  {journal} {Nature}\ }\textbf {\bibinfo {volume} {456}},\ \bibinfo {pages}
  {235} (\bibinfo {year} {2008})}\BibitemShut {NoStop}%
\bibitem [{\citenamefont {HilleRisLambers}\ \emph {et~al.}(2012)\citenamefont
  {HilleRisLambers}, \citenamefont {Adler}, \citenamefont {Harpole},
  \citenamefont {Levine}, \citenamefont {Mayfield} \emph
  {et~al.}}]{hillerislambers2012rethinking}%
  \BibitemOpen
  \bibfield  {author} {\bibinfo {author} {\bibfnamefont {J.}~\bibnamefont
  {HilleRisLambers}}, \bibinfo {author} {\bibfnamefont {P.~B.}\ \bibnamefont
  {Adler}}, \bibinfo {author} {\bibfnamefont {W.~S.}\ \bibnamefont {Harpole}},
  \bibinfo {author} {\bibfnamefont {J.~M.}\ \bibnamefont {Levine}}, \bibinfo
  {author} {\bibfnamefont {M.~M.}\ \bibnamefont {Mayfield}}, \emph {et~al.},\
  }\bibfield  {title} {\bibinfo {title} {Rethinking community assembly through
  the lens of coexistence theory},\ }\href@noop {} {\bibfield  {journal}
  {\bibinfo  {journal} {Annual Review of Ecology, Evolution and Systematics}\
  }\textbf {\bibinfo {volume} {43}},\ \bibinfo {pages} {2012} (\bibinfo {year}
  {2012})}\BibitemShut {NoStop}%
\end{thebibliography}%


\begin{thebibliography}{26}%
\makeatletter
\providecommand \@ifxundefined [1]{%
 \@ifx{#1\undefined}
}%
\providecommand \@ifnum [1]{%
 \ifnum #1\expandafter \@firstoftwo
 \else \expandafter \@secondoftwo
 \fi
}%
\providecommand \@ifx [1]{%
 \ifx #1\expandafter \@firstoftwo
 \else \expandafter \@secondoftwo
 \fi
}%
\providecommand \natexlab [1]{#1}%
\providecommand \enquote  [1]{``#1''}%
\providecommand \bibnamefont  [1]{#1}%
\providecommand \bibfnamefont [1]{#1}%
\providecommand \citenamefont [1]{#1}%
\providecommand \href@noop [0]{\@secondoftwo}%
\providecommand \href [0]{\begingroup \@sanitize@url \@href}%
\providecommand \@href[1]{\@@startlink{#1}\@@href}%
\providecommand \@@href[1]{\endgroup#1\@@endlink}%
\providecommand \@sanitize@url [0]{\catcode `\\12\catcode `\$12\catcode
  `\&12\catcode `\#12\catcode `\^12\catcode `\_12\catcode `\%12\relax}%
\providecommand \@@startlink[1]{}%
\providecommand \@@endlink[0]{}%
\providecommand \url  [0]{\begingroup\@sanitize@url \@url }%
\providecommand \@url [1]{\endgroup\@href {#1}{\urlprefix }}%
\providecommand \urlprefix  [0]{URL }%
\providecommand \Eprint [0]{\href }%
\providecommand \doibase [0]{https://doi.org/}%
\providecommand \selectlanguage [0]{\@gobble}%
\providecommand \bibinfo  [0]{\@secondoftwo}%
\providecommand \bibfield  [0]{\@secondoftwo}%
\providecommand \translation [1]{[#1]}%
\providecommand \BibitemOpen [0]{}%
\providecommand \bibitemStop [0]{}%
\providecommand \bibitemNoStop [0]{.\EOS\space}%
\providecommand \EOS [0]{\spacefactor3000\relax}%
\providecommand \BibitemShut  [1]{\csname bibitem#1\endcsname}%
\let\auto@bib@innerbib\@empty
\bibitem [{\citenamefont {Gardiner}\ \emph {et~al.}(1985)\citenamefont
  {Gardiner} \emph {et~al.}}]{gardiner1985handbook}%
  \BibitemOpen
  \bibfield  {author} {\bibinfo {author} {\bibfnamefont {C.~W.}\ \bibnamefont
  {Gardiner}} \emph {et~al.},\ }\href@noop {} {\emph {\bibinfo {title}
  {Handbook of stochastic methods}}},\ Vol.~\bibinfo {volume} {3}\ (\bibinfo
  {publisher} {springer Berlin},\ \bibinfo {year} {1985})\BibitemShut {NoStop}%
\bibitem [{\citenamefont {Fisher}\ and\ \citenamefont
  {Mehta}(2014)}]{fisher2014transition}%
  \BibitemOpen
  \bibfield  {author} {\bibinfo {author} {\bibfnamefont {C.~K.}\ \bibnamefont
  {Fisher}}\ and\ \bibinfo {author} {\bibfnamefont {P.}~\bibnamefont {Mehta}},\
  }\bibfield  {title} {\bibinfo {title} {The transition between the niche and
  neutral regimes in ecology},\ }\href@noop {} {\bibfield  {journal} {\bibinfo
  {journal} {Proceedings of the National Academy of Sciences}\ }\textbf
  {\bibinfo {volume} {111}},\ \bibinfo {pages} {13111} (\bibinfo {year}
  {2014})}\BibitemShut {NoStop}%
\bibitem [{\citenamefont {Lynch}\ and\ \citenamefont
  {Neufeld}(2015)}]{lynch2015ecology}%
  \BibitemOpen
  \bibfield  {author} {\bibinfo {author} {\bibfnamefont {M.~D.}\ \bibnamefont
  {Lynch}}\ and\ \bibinfo {author} {\bibfnamefont {J.~D.}\ \bibnamefont
  {Neufeld}},\ }\bibfield  {title} {\bibinfo {title} {Ecology and exploration
  of the rare biosphere},\ }\href@noop {} {\bibfield  {journal} {\bibinfo
  {journal} {Nature Reviews Microbiology}\ }\textbf {\bibinfo {volume} {13}},\
  \bibinfo {pages} {217} (\bibinfo {year} {2015})}\BibitemShut {NoStop}%
\bibitem [{\citenamefont {Verberk}(2011)}]{verberk2011explaining}%
  \BibitemOpen
  \bibfield  {author} {\bibinfo {author} {\bibfnamefont {W.}~\bibnamefont
  {Verberk}},\ }\bibfield  {title} {\bibinfo {title} {Explaining general
  patterns in species abundance and distributions},\ }\href@noop {} {\bibfield
  {journal} {\bibinfo  {journal} {Nature Education Knowledge}\ }\textbf
  {\bibinfo {volume} {3}},\ \bibinfo {pages} {38} (\bibinfo {year}
  {2011})}\BibitemShut {NoStop}%
\bibitem [{\citenamefont {Fowler}\ and\ \citenamefont
  {Ruokolainen}(2013)}]{fowler2013colonization}%
  \BibitemOpen
  \bibfield  {author} {\bibinfo {author} {\bibfnamefont {M.~S.}\ \bibnamefont
  {Fowler}}\ and\ \bibinfo {author} {\bibfnamefont {L.}~\bibnamefont
  {Ruokolainen}},\ }\bibfield  {title} {\bibinfo {title} {Colonization,
  covariance and colour: Environmental and ecological drivers of
  diversity--stability relationships},\ }\href@noop {} {\bibfield  {journal}
  {\bibinfo  {journal} {Journal of theoretical biology}\ }\textbf {\bibinfo
  {volume} {324}},\ \bibinfo {pages} {32} (\bibinfo {year} {2013})}\BibitemShut
  {NoStop}%
\bibitem [{\citenamefont {Barab{\'a}s}\ \emph {et~al.}(2016)\citenamefont
  {Barab{\'a}s}, \citenamefont {J.~Michalska-Smith},\ and\ \citenamefont
  {Allesina}}]{barabas2016effect}%
  \BibitemOpen
  \bibfield  {author} {\bibinfo {author} {\bibfnamefont {G.}~\bibnamefont
  {Barab{\'a}s}}, \bibinfo {author} {\bibfnamefont {M.}~\bibnamefont
  {J.~Michalska-Smith}},\ and\ \bibinfo {author} {\bibfnamefont
  {S.}~\bibnamefont {Allesina}},\ }\bibfield  {title} {\bibinfo {title} {The
  effect of intra-and interspecific competition on coexistence in multispecies
  communities},\ }\href@noop {} {\bibfield  {journal} {\bibinfo  {journal} {The
  American Naturalist}\ }\textbf {\bibinfo {volume} {188}},\ \bibinfo {pages}
  {E1} (\bibinfo {year} {2016})}\BibitemShut {NoStop}%
\bibitem [{\citenamefont {Marquet}\ \emph {et~al.}(2017)\citenamefont
  {Marquet}, \citenamefont {Espinoza}, \citenamefont {Abades}, \citenamefont
  {Ganz},\ and\ \citenamefont {Rebolledo}}]{marquet2017proportional}%
  \BibitemOpen
  \bibfield  {author} {\bibinfo {author} {\bibfnamefont {P.~A.}\ \bibnamefont
  {Marquet}}, \bibinfo {author} {\bibfnamefont {G.}~\bibnamefont {Espinoza}},
  \bibinfo {author} {\bibfnamefont {S.~R.}\ \bibnamefont {Abades}}, \bibinfo
  {author} {\bibfnamefont {A.}~\bibnamefont {Ganz}},\ and\ \bibinfo {author}
  {\bibfnamefont {R.}~\bibnamefont {Rebolledo}},\ }\bibfield  {title} {\bibinfo
  {title} {On the proportional abundance of species: Integrating population
  genetics and community ecology},\ }\href@noop {} {\bibfield  {journal}
  {\bibinfo  {journal} {Scientific reports}\ }\textbf {\bibinfo {volume} {7}},\
  \bibinfo {pages} {1} (\bibinfo {year} {2017})}\BibitemShut {NoStop}%
\bibitem [{\citenamefont {Van~Kampen}(1992)}]{van1992stochastic}%
  \BibitemOpen
  \bibfield  {author} {\bibinfo {author} {\bibfnamefont {N.~G.}\ \bibnamefont
  {Van~Kampen}},\ }\href@noop {} {\emph {\bibinfo {title} {Stochastic processes
  in physics and chemistry}}},\ Vol.~\bibinfo {volume} {1}\ (\bibinfo
  {publisher} {Elsevier},\ \bibinfo {year} {1992})\BibitemShut {NoStop}%
\bibitem [{\citenamefont {Grimmett}\ and\ \citenamefont
  {Stirzaker}(2001)}]{grimmett2001probability}%
  \BibitemOpen
  \bibfield  {author} {\bibinfo {author} {\bibfnamefont {G.}~\bibnamefont
  {Grimmett}}\ and\ \bibinfo {author} {\bibfnamefont {D.}~\bibnamefont
  {Stirzaker}},\ }\href@noop {} {\emph {\bibinfo {title} {Probability and
  Random Processes}}}\ (\bibinfo  {publisher} {Oxford University Press},\
  \bibinfo {year} {2001})\BibitemShut {NoStop}%
\bibitem [{\citenamefont {Schnakenberg}(1976)}]{schnakenberg1976network}%
  \BibitemOpen
  \bibfield  {author} {\bibinfo {author} {\bibfnamefont {J.}~\bibnamefont
  {Schnakenberg}},\ }\bibfield  {title} {\bibinfo {title} {Network theory of
  microscopic and macroscopic behavior of master equation systems},\
  }\href@noop {} {\bibfield  {journal} {\bibinfo  {journal} {Reviews of Modern
  physics}\ }\textbf {\bibinfo {volume} {48}},\ \bibinfo {pages} {571}
  (\bibinfo {year} {1976})}\BibitemShut {NoStop}%
\bibitem [{\citenamefont {Gupta}\ \emph {et~al.}(2014)\citenamefont {Gupta},
  \citenamefont {Briat},\ and\ \citenamefont {Khammash}}]{gupta2014scalable}%
  \BibitemOpen
  \bibfield  {author} {\bibinfo {author} {\bibfnamefont {A.}~\bibnamefont
  {Gupta}}, \bibinfo {author} {\bibfnamefont {C.}~\bibnamefont {Briat}},\ and\
  \bibinfo {author} {\bibfnamefont {M.}~\bibnamefont {Khammash}},\ }\bibfield
  {title} {\bibinfo {title} {A scalable computational framework for
  establishing long-term behavior of stochastic reaction networks},\
  }\href@noop {} {\bibfield  {journal} {\bibinfo  {journal} {PLoS computational
  biology}\ }\textbf {\bibinfo {volume} {10}},\ \bibinfo {pages} {e1003669}
  (\bibinfo {year} {2014})}\BibitemShut {NoStop}%
\bibitem [{\citenamefont {Meyn}\ and\ \citenamefont
  {Tweedie}(1993)}]{meyn1993stability}%
  \BibitemOpen
  \bibfield  {author} {\bibinfo {author} {\bibfnamefont {S.~P.}\ \bibnamefont
  {Meyn}}\ and\ \bibinfo {author} {\bibfnamefont {R.~L.}\ \bibnamefont
  {Tweedie}},\ }\bibfield  {title} {\bibinfo {title} {Stability of markovian
  processes iii: Foster--lyapunov criteria for continuous-time processes},\
  }\href@noop {} {\bibfield  {journal} {\bibinfo  {journal} {Advances in
  Applied Probability}\ }\textbf {\bibinfo {volume} {25}},\ \bibinfo {pages}
  {518} (\bibinfo {year} {1993})}\BibitemShut {NoStop}%
\bibitem [{\citenamefont {Bartoszynski}\ and\ \citenamefont
  {Niewiadomska-Bugaj}(2020)}]{bartoszynski2020probability}%
  \BibitemOpen
  \bibfield  {author} {\bibinfo {author} {\bibfnamefont {R.}~\bibnamefont
  {Bartoszynski}}\ and\ \bibinfo {author} {\bibfnamefont {M.}~\bibnamefont
  {Niewiadomska-Bugaj}},\ }\href@noop {} {\emph {\bibinfo {title} {Probability
  and statistical inference}}}\ (\bibinfo  {publisher} {John Wiley \& Sons},\
  \bibinfo {year} {2020})\BibitemShut {NoStop}%
\bibitem [{\citenamefont {Dimitrova}\ \emph {et~al.}(2020)\citenamefont
  {Dimitrova}, \citenamefont {Kaishev},\ and\ \citenamefont
  {Tan}}]{dimitrova2020computing}%
  \BibitemOpen
  \bibfield  {author} {\bibinfo {author} {\bibfnamefont {D.~S.}\ \bibnamefont
  {Dimitrova}}, \bibinfo {author} {\bibfnamefont {V.~K.}\ \bibnamefont
  {Kaishev}},\ and\ \bibinfo {author} {\bibfnamefont {S.}~\bibnamefont {Tan}},\
  }\bibfield  {title} {\bibinfo {title} {Computing the kolmogorov-smirnov
  distribution when the underlying cdf is purely discrete, mixed, or
  continuous},\ }\href@noop {} {\bibfield  {journal} {\bibinfo  {journal}
  {Journal of Statistical Software}\ }\textbf {\bibinfo {volume} {95}},\
  \bibinfo {pages} {1} (\bibinfo {year} {2020})}\BibitemShut {NoStop}%
\bibitem [{\citenamefont {Carr}\ \emph {et~al.}(2019)\citenamefont {Carr},
  \citenamefont {Diener}, \citenamefont {Baliga},\ and\ \citenamefont
  {Gibbons}}]{carr2019use}%
  \BibitemOpen
  \bibfield  {author} {\bibinfo {author} {\bibfnamefont {A.}~\bibnamefont
  {Carr}}, \bibinfo {author} {\bibfnamefont {C.}~\bibnamefont {Diener}},
  \bibinfo {author} {\bibfnamefont {N.~S.}\ \bibnamefont {Baliga}},\ and\
  \bibinfo {author} {\bibfnamefont {S.~M.}\ \bibnamefont {Gibbons}},\
  }\bibfield  {title} {\bibinfo {title} {Use and abuse of correlation analyses
  in microbial ecology},\ }\href@noop {} {\bibfield  {journal} {\bibinfo
  {journal} {The ISME journal}\ }\textbf {\bibinfo {volume} {13}},\ \bibinfo
  {pages} {2647} (\bibinfo {year} {2019})}\BibitemShut {NoStop}%
\bibitem [{\citenamefont {Chance}\ \emph {et~al.}(2019)\citenamefont {Chance},
  \citenamefont {McCollum}, \citenamefont {Street}, \citenamefont
  {Strickland},\ and\ \citenamefont {Lashley}}]{chance2019native}%
  \BibitemOpen
  \bibfield  {author} {\bibinfo {author} {\bibfnamefont {D.~P.}\ \bibnamefont
  {Chance}}, \bibinfo {author} {\bibfnamefont {J.~R.}\ \bibnamefont
  {McCollum}}, \bibinfo {author} {\bibfnamefont {G.~M.}\ \bibnamefont
  {Street}}, \bibinfo {author} {\bibfnamefont {B.~K.}\ \bibnamefont
  {Strickland}},\ and\ \bibinfo {author} {\bibfnamefont {M.~A.}\ \bibnamefont
  {Lashley}},\ }\bibfield  {title} {\bibinfo {title} {Native species abundance
  buffers non-native plant invasibility following intermediate forest
  management disturbances},\ }\href@noop {} {\bibfield  {journal} {\bibinfo
  {journal} {Forest Science}\ }\textbf {\bibinfo {volume} {65}},\ \bibinfo
  {pages} {336} (\bibinfo {year} {2019})}\BibitemShut {NoStop}%
\bibitem [{\citenamefont {Xu}\ and\ \citenamefont
  {Chou}(2018)}]{xu2018immigration}%
  \BibitemOpen
  \bibfield  {author} {\bibinfo {author} {\bibfnamefont {S.}~\bibnamefont
  {Xu}}\ and\ \bibinfo {author} {\bibfnamefont {T.}~\bibnamefont {Chou}},\
  }\bibfield  {title} {\bibinfo {title} {Immigration-induced phase transition
  in a regulated multispecies birth-death process},\ }\href@noop {} {\bibfield
  {journal} {\bibinfo  {journal} {Journal of Physics A: Mathematical and
  Theoretical}\ }\textbf {\bibinfo {volume} {51}},\ \bibinfo {pages} {425602}
  (\bibinfo {year} {2018})}\BibitemShut {NoStop}%
\bibitem [{\citenamefont {McKane}\ \emph {et~al.}(2004)\citenamefont {McKane},
  \citenamefont {Alonso},\ and\ \citenamefont {Sol{\'e}}}]{mckane2004analytic}%
  \BibitemOpen
  \bibfield  {author} {\bibinfo {author} {\bibfnamefont {A.~J.}\ \bibnamefont
  {McKane}}, \bibinfo {author} {\bibfnamefont {D.}~\bibnamefont {Alonso}},\
  and\ \bibinfo {author} {\bibfnamefont {R.~V.}\ \bibnamefont {Sol{\'e}}},\
  }\bibfield  {title} {\bibinfo {title} {Analytic solution of hubbell's model
  of local community dynamics},\ }\href@noop {} {\bibfield  {journal} {\bibinfo
   {journal} {Theoretical Population Biology}\ }\textbf {\bibinfo {volume}
  {65}},\ \bibinfo {pages} {67} (\bibinfo {year} {2004})}\BibitemShut {NoStop}%
\bibitem [{\citenamefont {Baxter}\ \emph {et~al.}(2007)\citenamefont {Baxter},
  \citenamefont {Blythe},\ and\ \citenamefont {McKane}}]{baxter2007exact}%
  \BibitemOpen
  \bibfield  {author} {\bibinfo {author} {\bibfnamefont {G.~J.}\ \bibnamefont
  {Baxter}}, \bibinfo {author} {\bibfnamefont {R.~A.}\ \bibnamefont {Blythe}},\
  and\ \bibinfo {author} {\bibfnamefont {A.~J.}\ \bibnamefont {McKane}},\
  }\bibfield  {title} {\bibinfo {title} {Exact solution of the multi-allelic
  diffusion model},\ }\href@noop {} {\bibfield  {journal} {\bibinfo  {journal}
  {Mathematical bioSciences}\ }\textbf {\bibinfo {volume} {209}},\ \bibinfo
  {pages} {124} (\bibinfo {year} {2007})}\BibitemShut {NoStop}%
\bibitem [{\citenamefont {Marquet}\ \emph {et~al.}(2020)\citenamefont
  {Marquet}, \citenamefont {Tejo},\ and\ \citenamefont
  {Rebolledo}}]{dobson2020unsolved}%
  \BibitemOpen
  \bibfield  {author} {\bibinfo {author} {\bibfnamefont {P.~A.}\ \bibnamefont
  {Marquet}}, \bibinfo {author} {\bibfnamefont {M.}~\bibnamefont {Tejo}},\ and\
  \bibinfo {author} {\bibfnamefont {R.}~\bibnamefont {Rebolledo}},\ }\bibfield
  {title} {\bibinfo {title} {What is the species richness distribution?},\ }in\
  \href@noop {} {\emph {\bibinfo {booktitle} {Unsolved problems in ecology}}},\
  \bibinfo {editor} {edited by\ \bibinfo {editor} {\bibfnamefont
  {A.}~\bibnamefont {Dobson}}, \bibinfo {editor} {\bibfnamefont {R.~D.}\
  \bibnamefont {Holt}},\ and\ \bibinfo {editor} {\bibfnamefont
  {D.}~\bibnamefont {Tilman}}}\ (\bibinfo  {publisher} {Princeton University
  Press},\ \bibinfo {address} {USA},\ \bibinfo {year} {2020})\BibitemShut
  {NoStop}%
\bibitem [{\citenamefont {Haegeman}\ and\ \citenamefont
  {Loreau}(2011)}]{haegeman2011mathematical}%
  \BibitemOpen
  \bibfield  {author} {\bibinfo {author} {\bibfnamefont {B.}~\bibnamefont
  {Haegeman}}\ and\ \bibinfo {author} {\bibfnamefont {M.}~\bibnamefont
  {Loreau}},\ }\bibfield  {title} {\bibinfo {title} {A mathematical synthesis
  of niche and neutral theories in community ecology},\ }\href@noop {}
  {\bibfield  {journal} {\bibinfo  {journal} {Journal of theoretical biology}\
  }\textbf {\bibinfo {volume} {269}},\ \bibinfo {pages} {150} (\bibinfo {year}
  {2011})}\BibitemShut {NoStop}%
\bibitem [{\citenamefont {Ser-Giacomi}\ \emph {et~al.}(2018)\citenamefont
  {Ser-Giacomi}, \citenamefont {Zinger}, \citenamefont {Malviya}, \citenamefont
  {De~Vargas}, \citenamefont {Karsenti}, \citenamefont {Bowler},\ and\
  \citenamefont {De~Monte}}]{ser2018ubiquitous}%
  \BibitemOpen
  \bibfield  {author} {\bibinfo {author} {\bibfnamefont {E.}~\bibnamefont
  {Ser-Giacomi}}, \bibinfo {author} {\bibfnamefont {L.}~\bibnamefont {Zinger}},
  \bibinfo {author} {\bibfnamefont {S.}~\bibnamefont {Malviya}}, \bibinfo
  {author} {\bibfnamefont {C.}~\bibnamefont {De~Vargas}}, \bibinfo {author}
  {\bibfnamefont {E.}~\bibnamefont {Karsenti}}, \bibinfo {author}
  {\bibfnamefont {C.}~\bibnamefont {Bowler}},\ and\ \bibinfo {author}
  {\bibfnamefont {S.}~\bibnamefont {De~Monte}},\ }\bibfield  {title} {\bibinfo
  {title} {Ubiquitous abundance distribution of non-dominant plankton across
  the global ocean},\ }\href@noop {} {\bibfield  {journal} {\bibinfo  {journal}
  {Nature ecology \& evolution}\ }\textbf {\bibinfo {volume} {2}},\ \bibinfo
  {pages} {1243} (\bibinfo {year} {2018})}\BibitemShut {NoStop}%
\bibitem [{\citenamefont {Mora}\ and\ \citenamefont
  {Walczak}(2018)}]{mora2018quantifying}%
  \BibitemOpen
  \bibfield  {author} {\bibinfo {author} {\bibfnamefont {T.}~\bibnamefont
  {Mora}}\ and\ \bibinfo {author} {\bibfnamefont {A.~M.}\ \bibnamefont
  {Walczak}},\ }\bibfield  {title} {\bibinfo {title} {Quantifying lymphocyte
  receptor diversity},\ }in\ \href@noop {} {\emph {\bibinfo {booktitle}
  {Systems Immunology}}}\ (\bibinfo  {publisher} {CRC Press},\ \bibinfo {year}
  {2018})\ pp.\ \bibinfo {pages} {183--198}\BibitemShut {NoStop}%
\bibitem [{\citenamefont {Hubbell}(1979)}]{hubbell1979tree}%
  \BibitemOpen
  \bibfield  {author} {\bibinfo {author} {\bibfnamefont {S.~P.}\ \bibnamefont
  {Hubbell}},\ }\bibfield  {title} {\bibinfo {title} {Tree dispersion,
  abundance, and diversity in a tropical dry forest},\ }\href@noop {}
  {\bibfield  {journal} {\bibinfo  {journal} {Science}\ }\textbf {\bibinfo
  {volume} {203}},\ \bibinfo {pages} {1299} (\bibinfo {year}
  {1979})}\BibitemShut {NoStop}%
\bibitem [{\citenamefont {Descheemaeker}\ and\ \citenamefont
  {de~Buyl}(2020)}]{descheemaeker2020stochastic}%
  \BibitemOpen
  \bibfield  {author} {\bibinfo {author} {\bibfnamefont {L.}~\bibnamefont
  {Descheemaeker}}\ and\ \bibinfo {author} {\bibfnamefont {S.}~\bibnamefont
  {de~Buyl}},\ }\bibfield  {title} {\bibinfo {title} {Stochastic logistic
  models reproduce experimental time series of microbial communities},\
  }\href@noop {} {\bibfield  {journal} {\bibinfo  {journal} {Elife}\ }\textbf
  {\bibinfo {volume} {9}},\ \bibinfo {pages} {e55650} (\bibinfo {year}
  {2020})}\BibitemShut {NoStop}%
\bibitem [{\citenamefont {Jeraldo}\ \emph {et~al.}(2012)\citenamefont
  {Jeraldo}, \citenamefont {Sipos}, \citenamefont {Chia}, \citenamefont
  {Brulc}, \citenamefont {Dhillon}, \citenamefont {Konkel}, \citenamefont
  {Larson}, \citenamefont {Nelson}, \citenamefont {Qu}, \citenamefont {Schook}
  \emph {et~al.}}]{jeraldo2012quantification}%
  \BibitemOpen
  \bibfield  {author} {\bibinfo {author} {\bibfnamefont {P.}~\bibnamefont
  {Jeraldo}}, \bibinfo {author} {\bibfnamefont {M.}~\bibnamefont {Sipos}},
  \bibinfo {author} {\bibfnamefont {N.}~\bibnamefont {Chia}}, \bibinfo {author}
  {\bibfnamefont {J.~M.}\ \bibnamefont {Brulc}}, \bibinfo {author}
  {\bibfnamefont {A.~S.}\ \bibnamefont {Dhillon}}, \bibinfo {author}
  {\bibfnamefont {M.~E.}\ \bibnamefont {Konkel}}, \bibinfo {author}
  {\bibfnamefont {C.~L.}\ \bibnamefont {Larson}}, \bibinfo {author}
  {\bibfnamefont {K.~E.}\ \bibnamefont {Nelson}}, \bibinfo {author}
  {\bibfnamefont {A.}~\bibnamefont {Qu}}, \bibinfo {author} {\bibfnamefont
  {L.~B.}\ \bibnamefont {Schook}}, \emph {et~al.},\ }\bibfield  {title}
  {\bibinfo {title} {Quantification of the relative roles of niche and neutral
  processes in structuring gastrointestinal microbiomes},\ }\href@noop {}
  {\bibfield  {journal} {\bibinfo  {journal} {Proceedings of the National
  Academy of Sciences}\ }\textbf {\bibinfo {volume} {109}},\ \bibinfo {pages}
  {9692} (\bibinfo {year} {2012})}\BibitemShut {NoStop}%
\end{thebibliography}%

\end{document}


\title{Supplemental materials to the manuscript Phenomenology and Dynamics of Competitive Ecosystems Beyond the Niche-Neutral1
Regimes}
\maketitle

\section{Fokker-Planck and Stochastic Differential Equation (SDE) formulation}

{In this study, we treat} the species abundance $n_i$ as a discrete variable based on its biological nature. However, for large systems, it is often convenient to approximate the master equation in the continuum limit. 
We begin with the multi-dimensional  master equation in Eq.~2 in the main text,
\begin{eqnarray}
\label{eq:master-eq}
    \partial_tP(n_1,n_2\dots,n_S;t) &=& \sum_{i}\left\{ q_{i}^+(\vec{n}-\vec{e_i})P(\vec{n}-\vec{e_i};t)+q^-_{i}(\vec{n_i}+\vec{e_i}) P(\vec{n}+\vec{e_i};t)-\left[q_{i}^+(\vec{n})+q_{i}^-(\vec{n}) \right]P(\vec{n};t) \right\}
\end{eqnarray}
 where $q^+_{i}(\vec{n})$ and $q^-_{i}(\vec{n})$ represent the birth and death rate of species $i$ (respectively), which generally {depend on} the state vector $\vec{n}=(n_1,\dots, n_S)$. Here, $\vec{e}_i\equiv (0, \dots, 1, \dots , 0)$ (the one is located in the $i$-th component) and the transition rates are given by
\begin{align}
\label{eq:rates}
q_i^+(\vec{n})&=r^+ n_i +\mu,  \\
q_i^-(\vec{n})&=r^- n_i + \frac{r}{K} n_i \left(n_i +\sum_{j\neq i} \rho _{j,i} n_j\right) \nonumber
\end{align}
for each species $i\in \{1,2,\dots,S\}$.

Assuming  that the carrying capacity is large $K \gg 1$, we define $x_i$ to be the corresponding continuous limit of $n_i$, which can be rescaled by the characteristic system size, $y_i = x_i / K$.
This probability density of this continuous variable is  $p(\vec{y};t)=K P(\vec{n};t)$ and the re-scaled rates are $Q_i^{+/-}(\vec{y}) = q_i^{+/-}(\vec{n})/K$. Using Eq.~\ref{eq:rates}, we get 
\begin{align*}
    Q_i^{+}(\vec{y}) &= r^+ y_i + \mu / K \\
    Q_i^{+}(\vec{y}) &= r^- y_i + r y_i \left( y_i + \sum_{j\neq i}\rho_{j,i} y_j \right).
\end{align*}
Written in the continuous variables, the master equation from Eq.~\ref{eq:master-eq} becomes
\begin{equation}
    \label{eq:master-eq-cont}
    \partial_t  p(\vec{y};t) =
    K \sum_{i}\left\{Q^+_i(\vec{y}-\vec{e}_i)p(\vec{y}-\vec{e}_i,t)+ 
    Q^-_i (\vec{y}+\vec{e}_i)p(\vec{y}+\vec{e}_i,t) - \left[Q^+_i(\vec{y})+Q^-_i(\vec{y})\right]p(\vec{y},t)\vphantom{\vec{e}_i} \right\},
\end{equation}
where $\vec{e}_i$ is the change in abundance $\vec{y}$ due to an individual birth-death event, $\vec{e}_i=(\delta_{1i}/K, \delta_{2i}/K, ..., \delta_{Si}/K)$ where $\delta_{ij}$ is the Kronecker delta as in Eq.~\ref{eq:master-eq}. 
    The Fokker-Planck approximation is obtained via Taylor expansion of the right-hand side of Eq.~\ref{eq:master-eq-cont} to the second order in $1/K$:
\begin{multline*}
    Q^+_i(\vec{y}-\vec{e}_i)p(\vec{y}-\vec{e}_i;t)= Q^+_i(\vec{y})p(\vec{y};t)+\sum_j (-(\vec{e}_i)_j) \frac{\partial}{\partial y_j}\left( Q^+_i(\vec{y})p(\vec{y};t)\vphantom{\frac{1}{1}}\right)
    + \frac{1}{2!}\sum_j \sum_k (\vec{e}_i)_j(\vec{e}_i)_k \frac{\partial^2}{\partial y_j \partial y_k}\left( Q^+_i(\vec{y})p(\vec{y};t)\vphantom{\frac{1}{1}}\right)+...
\end{multline*}
\begin{multline*}
    Q^-_i(\vec{y}+\vec{e}_i)p(\vec{y}+\vec{e}_i;t)= Q^-_i(\vec{y})p(\vec{y},t)+\sum_j (\vec{e}_i)_j \frac{\partial}{\partial y_j}\left( Q^-_i(\vec{y})p(\vec{y};t)\vphantom{\frac{1}{1}}\right)
    + \frac{1}{2!}\sum_j \sum_k (\vec{e}_i)_j (\vec{e}_i)_k \frac{\partial^2}{\partial y_j \partial y_k}\left( Q^-_i(\vec{y})P(\vec{y};t)\vphantom{\frac{1}{1}}\right)+...
\end{multline*}
Finally, we obtain the Fokker-Planck equation (FPE) for the master equation Eq.~\ref{eq:master-eq}
\begin{equation}
\label{eq:fokker-planck}
    \partial_t  p(\vec{y};t) =
    -\sum_j  \frac{\partial}{\partial y_j}\left[\left( Q^+_i(\vec{y})-Q^-_i(\vec{y})\right)p(\vec{y};t)\vphantom{\frac{1}{1}}\right]
    + \frac{1}{2 K}\sum_{j,k}  \frac{\partial^2}{\partial y_j \partial y_k}\left[\left( Q^+_i(\vec{y})+Q^-_i(\vec{y})\right)p(\vec{y};t)\vphantom{\frac{1}{1}}\right] + \mathcal{O}(1/K^2).
\end{equation}
Using It\^o's prescription for SDEs, this FPE corresponds to the Stochastic Differential Equation (SDE)\cite{gardiner1985handbook}: 
\begin{equation}
    \label{eq:langevin}
    d y_i = \left( Q^+_i(\vec{y})-Q^-_i(\vec{y}) \right)dt + \sqrt{ \frac{Q^+_i(\vec{y})+Q^-_i(\vec{y})}{K} } dW_i
\end{equation}
where $W_i$ is a Wiener process. Multiplying both sides of this equation by the carrying capacity, and keeping in mind that $K dy_i = dx_i$, we get
\begin{equation}
    \label{eq:langevin-LV}
    d x_i = \left( q^+_i(\vec{x})-q^-_i(\vec{x}) \right)dt + \sqrt{ K \left( q^+_i(\vec{x})+q^-_i(\vec{x})\right) } dW_i.
\end{equation}
Note that the drift term in the SDE recovers the deterministic Lotka-Voltera equations [4] in the main text.

The continuum description of the population dynamics has been devised in the past in a number of works \cite{fisher2014transition,lynch2015ecology,verberk2011explaining,fowler2013colonization,barabas2016effect, marquet2017proportional}. In particular, a common choice for the diffusion term is to be proportional to the square root of the abundance~\cite{fisher2014transition}, such that the noise is independent of other species abundances, as in  
\begin{equation}
    \label{eq:langevin-LV-sqrt-noise}
    d x_i = \left( q^+_i(\vec{x})-q^-_i(\vec{x}) \right)dt + \sqrt{ K r x_i } dW_i.
\end{equation}

In another example, in the the neutral regime, under further approximations and assumptions, Marquet et al \cite{marquet2017proportional} derived an expression for the normalized species abundance $z=x_i/J$

\begin{equation}
 \label{eq:langevin-marquet}
    d z = \left( b(z)-d(z) \right)dt + \sqrt{2c(z)} dW_i,
\end{equation}
where $b,d,c$ are phenomenological functions. We note that with appropriate choice of birth and death rates, this equation can be viewed as a variant of one of our mean field approximations of Section  \textbf{Approximations of Species Abundance Distribution, Approximation Approach III: Estimating $\langle J|n\rangle$ using Mean-Field Approximation} {, see below}.

\begin{figure}[ht!]
    \centering
    \includegraphics[width=0.7\columnwidth]{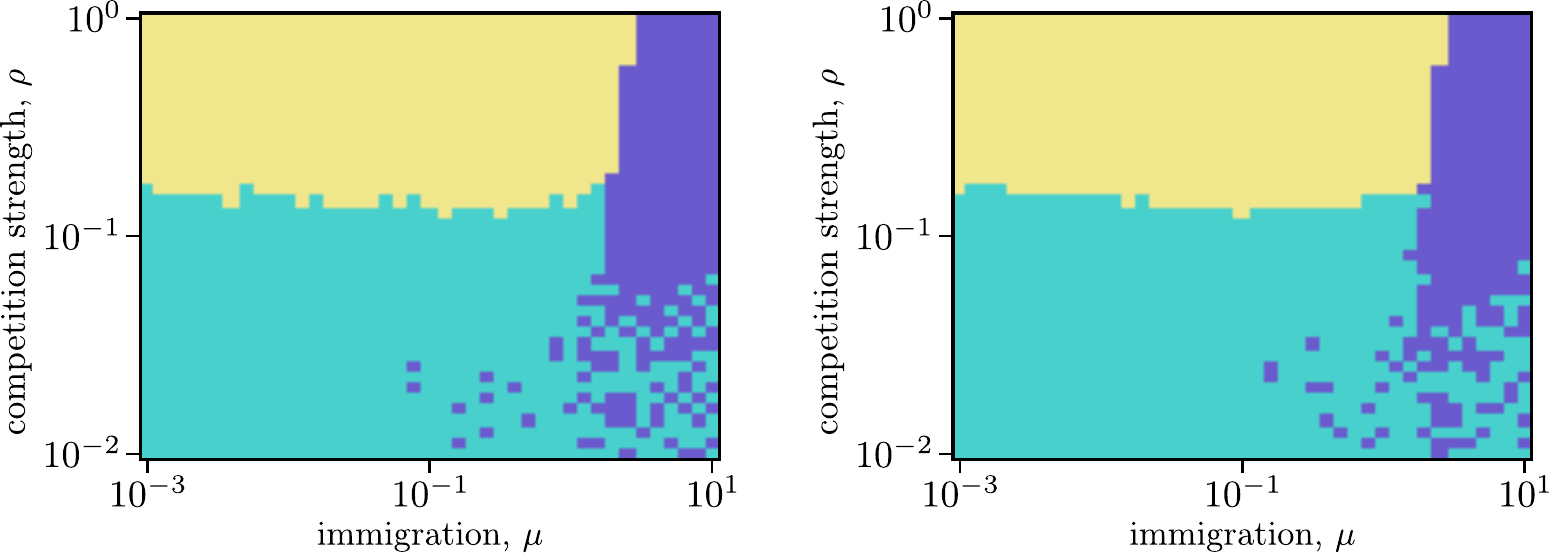}
    \caption{Langevin numerical simulations. Left: Modality regimes with noise Eq.~\ref{eq:langevin-LV} from the rates defined in the main text. Right: Modality regimes with $\sqrt{x}$ noise as in Eq.~\ref{eq:langevin-LV-sqrt-noise}. Colours correspond to the modality regime presented in the main text: yellow is the `rare-biosphere' regime, teal is the bimodal regime and purple is the immigration dominated unimodal regime.}
    \label{fig:langevin}
\end{figure}

However, it is important to note that the results of the Fokker-Planck/SDE approximations often substantially deviate from the exact predictions of the master equation presented above and given in the literature~\cite{van1992stochastic}. To assess the accuracy of the Fokker-Planck approximation, we simulated the corresponding SDE of Eq.~\ref{eq:langevin-LV} using an Euler integration method. 
We find that the Fokker-Planck approximation does not fully reproduce the complete phase space of the modality regimes for either noise specified in Eq.~\ref{eq:langevin-LV}, as shown in   Fig.~\ref{fig:langevin}.
In particular, in the Fokker-Planck approximation, the `rare-biosphere' regime at high competition strength ($\rho > 2\cdot10{^{-1}}$) persists even at low immigration rates, so that no bimodality is observed on the neutral manifold ($\rho = 1$). Additionally, the multimodal regime is absent in all Langevin results. Other SDE approximations, such as Eq.~\ref{eq:langevin-LV-sqrt-noise} also result in similar discrepancies with the exact results.

\section{Approximations of Species Abundance Distribution (SAD)}
In this section we find approximate solutions for species abundance distributions using various mean-field approaches. We also find the exact solution for SAD in some parameter regimes.

The set of differential equations in Eq.\ref{eq:master-eq} can be written in the matrix form as 
\begin{equation}
    \mathbf{\dot P = M P},
\end{equation}
where $\mathbf{P}$ is the vector whose entry $\mathbf{P}_{\alpha}$ is the probability of being in state $\alpha$ at time $t$. Note that the state space spans the countably infinite set ${\mathbb{Z}^{*}}^S$. Here, the matrix $\mathbf{M}$ is the generator of our Markov chain $\mathbf{P}(t)=e^{\mathbf{M}t}\cdot \mathbf{P}(0)$ \cite{grimmett2001probability}. In our case $ \mathbf{M}$ is irreducible (there is no subset of states disconnected from the rest) \cite{grimmett2001probability,schnakenberg1976network} and the matrix elements $\mathbf{M}_{\alpha,\beta}$ are the transition rates from state $\beta = (n_1^{\beta}, n_2^{\beta}, ..., n_S^{\beta})$ to state $\alpha = (n_1^{\alpha}, n_2^{\alpha}, ..., n_S^{\alpha})$. 

The only non-zero entries in the generator $\mathbf{M}_{\alpha,\beta}$ are the ones for which a single species abundance differs by one individual between states $\beta$ and $\alpha$, i.e. $n_k^{\beta} = n_k^{\alpha} \pm 1$.
The entries where the labeled states have $n_k^{\beta} = n_k^{\alpha}  - 1$ are the birth rates for $n_k^{\beta} $, whereas the entries whose labels correspond to $n_k^{\beta} = n_k^{\alpha}  + 1$ are the death rates for $n_k^\beta$.
It can be shown that a stationary solution exists for an irreducible, non-explosive continuous Markov chain when a solution exists to the matrix equation $\mathbf{0 = M P}$ \cite{grimmett2001probability,schnakenberg1976network}.

In general, it remains difficult to show that a stationary solution exists for an arbitrary, given Markov chain. However, our model belongs to a class of models that have been extensively studies in the context of stochastic chemical reactions.
For this class of stochastic processes, a unique  stationary solution exists under the following conditions \cite{gupta2014scalable}.

\begin{cond}
For a positive vector $\vec{\nu} \in \mathbb{R}^S$, there exists positive constants $c_1$ and $c_2$ such that for all $\vec{n} \in \mathcal{S}$
\begin{equation}
    \sum_{j=1}^{R} \lambda_j(\vec{n})\langle \vec{\nu}, \vec{\zeta_j} \rangle \leq c_1 - c_2 \langle \nu , \vec{n} \rangle
    \label{eq:drift}
\end{equation}
where $\lambda_j(\vec{n})$ are the rates out of state $\vec{n}$ and $\zeta_j$ is the stoichiometric vector for the corresponding rate.
\end{cond}

For the stochastic process in Eq.~\ref{eq:master-eq}, $\mathcal{S} = {\mathbb{Z}^*}^S$.
Every state has $R=2S$ rates associated to it; each of S species has a rate of birth and a rate of death.
For species $i$, the stoichiometric vector of the birth reaction is $\vec{\zeta}\equiv (0, \dots, 1, \dots , 0)$ (1 is the $i$-th component) and, similarly, the stoichiometric vector for the death reaction is $\vec{\zeta}\equiv (0, \dots, -1, \dots , 0)$.
The bracket notation $\langle \cdot , \cdot \rangle$ is the standard inner product on $\mathbb{R}^S$.

This condition, known as the drift condition, is a sufficient condition for the existence of the stationary distribution of a Markov chains according to the following proposition\cite{gupta2014scalable,meyn1993stability}.

\begin{prop}
Assume that the state space $\mathcal{S}$ of the Markov process is irreducible and \textbf{Condition 1} holds. 
Then this process is exponentially ergodic in the sense that there exists a unique
distribution $\mathbf{P}_s\in \mathcal{P}(\mathcal{S})$ along with constants $B, c > 0$ such that for any $\vec{n} \in \mathcal{S}$
\begin{equation*}
    \textrm{sup}_{A\subset \mathcal{S}} |\mathbf{P}(A)-\mathbf{P}_s(A)| \leq B e^{-ct} , \quad \forall t \geq  0
\end{equation*}
\end{prop}
In other words, the distribution converges to the stationary distribution exponentially fast.

Given that the Markov process is irreducible, we need to show that \textbf{Condition 1} holds.
We choose $\nu = (1,1,...,1)$ and $c_2=r$ (our basal growth rate).
We shall show that $c_1$ can be chosen to be a constant such that our system satisfies \textbf{Condition 1}.
It is fairly easy to show that, starting from Eq.~\ref{eq:drift},
\begin{equation}
    \begin{gathered}
    \sum_{j=1}^{R} \lambda_j(\vec{n})\langle \vec{\nu}, \vec{\zeta_j} \rangle \leq c_1 - c_2 \langle \nu , \vec{n} \rangle \nonumber\\
    \sum_i^S q_i^+(\vec{n}) - q_i^-(\vec{n}) + r n_i \leq c_1 \nonumber \\
    S\mu + \sum_i^S r n_i \left( 2 - \frac{n_i +\rho \sum_{j \neq i}n_j}{K}\right) \leq c_1.
    \label{eq:cond-solve}
    \end{gathered}
\end{equation}
The left hand side is the equation is bounded from above, for which a maximum is found at $n_i = K/[(1-\rho)+\rho S]$ $\forall i$.
Thus, setting $c_1$ larger than this maximum guarantees a stationary solutions exists by \textbf{Proposition 1}.

\subsection{Derivation of the SAD from global balance equation}
\label{App:ZeroFlux}

Consider the multi-dimensional  master equation given in Eq.~\ref{eq:master-eq}. To find the master equation for one species, say $n_1$, we sum over all other species on both sides of the equation; i.e. 
\begin{eqnarray}
  \partial_t P_1(n_1)&=&\sum_{n_2=0}^{\infty}\dots \sum_{n_S=0}^{\infty}  \partial_tP(n_1,n_2\dots,n_S) \\ \nonumber &=&  \sum_{n_2=0}^{\infty}\dots \sum_{n_S=0}^{\infty} \left\{\sum_{i}\left\{ q_{i}^+(\vec{n}-\vec{e_i})P(\vec{n}-\vec{e_i})+q^-_{i}(\vec{n}+\vec{e_i}) P(\vec{n}+\vec{e_i})-\left[q_{i}^+(\vec{n})+q_{i}(\vec{n}) \right]P(\vec{n}) \right\}\right\}.
  \end{eqnarray}
Since the birth rates at $n_i = -1$ and death rates at $n_i = 0$ go to zero, i.e. $q_{i}^+(n_1, \dots, -1, \dots, n_S)= 0$ and $q_{i}^-(n_1, \dots, 0, \dots, n_S)=0$, we get for every $n_i$:
 \begin{eqnarray}
      \sum_{n_i=0}^{\infty}  q_{i}^+(n_1, \dots n_i-1, \dots, n_S)P(n_1, \dots n_i-1, \dots, n_S) &=& \sum_{n_i=0}^{\infty}  q_{i}^+(n_1, \dots n_i, \dots, n_S)P(n_1, \dots n_i, \dots, n_S)
,  \\ \nonumber
 \sum_{n_i=0}^{\infty}  q_{i}^-(n_1, \dots n_i+1, \dots, n_S)P(n_1, \dots n_i+1, \dots, n_S)&=& \sum_{n_i=0}^{\infty}  q_{i}^-(n_1, \dots n_i, \dots, n_S)P(n_1, \dots n_i, \dots, n_S)
 \end{eqnarray}

This simplification allows cancellation of terms for all the species except $n_1$ and the above equation may be written as
\begin{eqnarray}
\label{eq:ME-intermediate}
  \partial_t P_1(n_1) = \sum_{n_2=0}^{\infty}\dots \sum_{n_S=0}^{\infty} \left\{ q_{1}^+(\vec{n}-\vec{e_1})P(\vec{n}-\vec{e_1})+q^-_{1}(\vec{n}+\vec{e_1}) P(\vec{n}+\vec{e_1})-\left[q_{1}^+(\vec{n})+q^-_{1}(\vec{n}) \right]P(\vec{n})\right\}.
\end{eqnarray}
For simplicity, we define $F^{+}(\vec{n})\equiv q_1^+(\vec{n})P(\vec{n})$ and $F^{-}(\vec{n})\equiv q_1^-(\vec{n})P(\vec{n})$, thus Eq.~\ref{eq:ME-intermediate}
can be written as 
\begin{eqnarray}
    \partial_t P_1(n_1) = \sum_{n_2=0}^{\infty}\dots \sum_{n_S=0}^{\infty} \left\{ F^{+}(n_1-1,n_2,\dots)-F^{+}(n_1,n_2,\dots)
     +F^{-}(n_1+1,n_2,\dots)-F^{-}(n_1,n_2,\dots) \right\}.
\end{eqnarray}
{To solve this equation, we use a z-transform ($n_1\rightarrow z$): for a function $G(n_1)$, the corresponding transformed function is ${\cal Z}[G(n_1)]\equiv \hat{G}(z)=\sum_{n_1=0}^{\infty} G(n_1) z^{-n_1} $. 
Using z-transforms we get}
\begin{eqnarray}
    \partial_t \hat{P_1}(z) = \sum_{n_2=0}^{\infty}\dots \sum_{n_S=0}^{\infty} \hat{F}^+(z,n_2,\dots)(1-z^{-1})+\hat{F}^-(z,n_2,n_3\dots)(1-z).
\end{eqnarray}
{Here, we have used a property of  z-transforms: ${\cal Z}[G(n)-G(n-1)]=(1-z^{-1})\hat{G}(z)$, and ${\cal Z}[G(n+1)-G(n)]=(1-z)\hat{G}(z)-z G(0)$. Now, to find the stationary solution, we set $\partial_t P(z)=0 $, and after re-organizing the equation we get} 
\begin{eqnarray}
     \sum_{n_2=0}^{\infty}\dots \sum_{n_S=0}^{\infty}\hat{F}^+(z,n_2,n_3,\dots)=\sum_{n_2=0}^{\infty}\dots \sum_{n_S=0}^{\infty} \hat{F}^-(z,n_2,n_3,\dots)\frac{1-z}{z^{-1}-1}=\sum_{n_2=0}^{\infty}\dots \sum_{n_S=0}^{\infty}\hat{F}^-(z,n_2,n_3\dots)z .
\end{eqnarray}
Then, we use the inverse z-transform ($z\rightarrow n_1$) to solve for $P_1(n)$, and find
\begin{eqnarray}
      \sum_{n_2=0}^{\infty}\dots \sum_{n_S=0}^{\infty}q^+_{1}(\vec{n})P(\vec{n})= \sum_{n_2=0}^{\infty}\dots \sum_{n_S=0}^{\infty}q^-_{1}(\vec{n}+\vec{e_1})P(\vec{n}+\vec{e_1}). 
\end{eqnarray}
We use the Bayes formula, $P(n_1,n_2,n_3,\dots,n_S)=P(n_2,n_3,\dots, n_S|n_1)P_1(n_1)$, and obtain
\begin{eqnarray}
     \langle q_{1}^+(\vec{n})|n_1\rangle_{n_2,n_3,\dots,n_S}P_1(n_1) = \langle q_{1}^-(\vec{n}+\vec{e_1}) |n_1+1\rangle_{n_2,\dots n_S} P_1(n_1+1), 
\end{eqnarray}
where $ \langle *|n_1\rangle_{n_2, \dots, n_S}\equiv\langle *|n_1\rangle \equiv\sum_{n_2=0}^{\infty}\dots \sum_{n_S=0}^{\infty} (*)P(n_2, \dots, n_S|n_1) $. 
Up to {this point, the derivation has not made use of any additional assumption or approximation, and the above equation} is general. Since all species are equivalent,  $P_i(n_i)=P_j(n_j)$ for all $i,j$ we can drop the indices and define  $P(n)\equiv P_i(n_i)$ for every $i$. 

For the rates in Eq.~\ref{eq:rates}, the averaging of the rates gives an average birth rate,  $\langle q_{1}^+(\vec{n})|n_1\rangle=\mu + r^+ n_1$  and an average death rate $\langle q_{1}^-(\vec{n})|n_1\rangle=n_1\left(r^-+r n_1/K + r \rho \sum_{j\neq 1 }\langle n_j |n_1\rangle /K \right)$; 
{note that although $q^+_{i}$ depends on $n_i$, $q^-_{i}$ depends on the abundances of all species.} The averaged rates are also equivalent for all species and we drop the indices and define: $q^+(n) \equiv  \langle q_{i}^+(\vec{n})|n_i\rangle$ and $q^-(n) \equiv \langle q_{i}^-(\vec{n})|n_i\rangle$.

Solving the recursive equation with this new notation, we obtain
\begin{eqnarray}
    \label{eq:exact-sol}
    P(n)&=&P(0)\prod_{n'=1}^{n}\frac{q^{+}(n'-1)}{ q^-(n')} \label{eq:exact_appendix}
    \\ \nonumber
    &=&P(0)\prod_{n'=1}^{n}\frac{r^+(n'+a)}{n'\left(r^-+r n'/K + r\rho \sum_{j\neq 1 }\langle n_j |n'\rangle /K \right)}= P(0)\frac{(r^+)^{n}(a)_{n}}{n!\prod_{n'=1}^{n}\left(r^-+r n'/K + r\rho \sum_{j\neq 1 }\langle n_j |n'\rangle /K \right)},
\end{eqnarray}
where
 $a=\mu/r^+$ and $(a)_{n} \equiv a(a+1)\dots (a+n-1)$, is the Pochhammer symbol.  Here, $P(0)$ is obtained from the normalization.
We emphasize that the above abundance distribution $P(n)$ in Eq.~\ref{eq:exact_appendix} is exact, and no approximations have been introduced so far.

The denominator in the exact solution includes  the interactions between $n_1$ and all other species in the term $\sum_{j\neq 1}\langle n_j |n_1 \rangle $. Since this term is intractable, we need to apply some approximations in order to provide an explicit expression to $P_1(n_1)$. In the following subsections we present the results of three different  approximation approaches and discuss their limitations. We note that although the approximations presented below recover the regimes in the  main text, each approximation breaks down for different sets of parameters, see Fig.~\ref{fig:quality}.  

\subsubsection{Approximation Approach I: Estimating $\sum_{j\neq i}\langle n_j|n_i\rangle$ using mean-field approximation}
{In the first approximation approach,} we ignore the correlations between species abundances and assume that $\langle n_j |n_i \rangle = \langle n_j \rangle $. Thus, 
\begin{equation}
    P(n)\approx P(0)\frac{(r^+)^{n}(a)_{n}}{n!\prod_{n'=1}^{n}\left(r^-+r n'/K + r\rho \sum_{j\neq 1 }\langle n_j\rangle /K \right)} = P(0)\frac{(r^+)^{n}(a)_{n}}{n!\prod_{n'=1}^{n}\left(r^-+r n'/K + r\rho (S-1)\langle n\rangle /K \right)},
    \label{eq:MF1}
\end{equation}
where the last equality arises from symmetry; $\langle n_j \rangle = \langle n_i \rangle = \langle n\rangle $ for every species $i,j$. By definition, $\langle n \rangle = \sum_{n=0}^{\infty }n P(n)$. Hence
\begin{equation}
    \langle n \rangle \approx P(0)\sum_{n=0}^{\infty}n\frac{(r^+)^{n}(a)_{n}}{n!\prod_{n'=1}^{n}\left(r^-+r n'/K + r\rho (S-1)\langle n\rangle /K \right)}
    \label{eq:MF1_closure}
\end{equation}
where $P(0)=1/{_1}F_1[a,b;c]$ is the normalization coefficient, and ${_1}F_1[a,b;c]$ is the Kummer confluent hypergeometric function, $a=\frac{\mu}{r^+}$, $b=\frac{r^- K + r \rho (S-1) \langle n \rangle }{r}+1$ and $c=\frac{r^+ K}{r}$. 
{Finally the SAD is found by} the substitution of the numerical solution of $\langle n\rangle$, obtained from Eq.~\ref{eq:MF1_closure}, into Eq.~\ref{eq:MF1}.

\subsubsection{Approximation Approach II: Estimating total population size $\langle J|n\rangle $ using convolutions}
\label{Approximation Approach II:}
From Eq.~\ref{eq:exact-sol} the exact solution for the SAD is 
\begin{eqnarray}
\label{eq:sol-ME-Jgivenn}
    P(n)= P(0)\frac{(r^+)^{n}(a)_{n}}{n!\prod_{n'=1}^{n}\left(r^-+r (1-\rho)n'/K + r\rho \langle J |n'\rangle /K \right)},
\end{eqnarray}
where $J=\sum_{i=1}^{S} n_i $ is the total population size. Here we again assume that the total number of individuals in the system, $J$, weakly depends on any particular individual species abundance $n$; in other words, $\langle J | n \rangle \approx \tilde{J}$ is a random variable independent of the abundance $n$. Hence, our marginal probability can be written as a conditional probability
\begin{equation} 
    P(n)=P(n|\langle J |n \rangle ) \approx P(n|\tilde{J})=P(0)
    \frac{(a)_{n} \Tilde{c}^{n}}{n ! (\Tilde{b}+1)_{n} } 
\end{equation}
with $a=\frac{\mu}{r^+}$, $\tilde{b}= \frac{r^-K+r\rho \tilde{J}}{r(1-\rho)}$, and $\tilde{c}=\frac{r^+ K}{r(1-\rho)}$.
Moreover, we assume that the species abundances are mutually independent, ${P}(n_1,\dots n_S) \approx \prod_i P_i(n_i)$ such that the PDF of $\sum_i n_i$ now reads 
\begin{equation}
    {\rm Prob}\left(\left.\sum_i n_i\right|\tilde{J}\right)=\underbrace{P_1(n_1|\tilde{J})*P_2(n_2|\tilde{J})* \dots * P_S(n_S|\tilde{J})}_{S {\rm \ times}}
\end{equation}
where $A*B$ represents the convolution of $A$ with $B$. ${\rm Prob}\left(\sum_i n_i|\tilde{J}\right)$ is the {probability distribution function  for} $\sum_i n_i$ individuals where we assume that a single species PDF is $P(n|\tilde{J})$ with a given $\tilde{J}$. 
To capture the fact that $\tilde{J}$  {is closely related to the sum of random abundances (the total number of individuals, $J$)}, we consider
\begin{equation}
    P(\tilde{J})\approx \frac{{\rm Prob}\left(\left.\sum_i n_i=\tilde{J}\right|\tilde{J}\right)}{\sum_{\tilde{J}} {\rm Prob}\left(\left.\sum_i n_i=\tilde{J}\right|\tilde{J}\right)},
\end{equation}
where $P(\tilde{J})$ is the approximated distribution of $\tilde{J}$ (and may be used as an approximation for $P(J)$). 
Then
\begin{equation}
    P(n) = \sum_{\tilde{J}}P(n|\tilde{J}) P(\tilde{J})
\end{equation}
is the approximated PDF for each abundance - the SAD. 

Note that when $S$ is large, we find
\begin{equation}
    P\left(\left.\sum_i n_i\right|\tilde{J}\right) \sim  {\cal N}\left(S\langle n |\tilde{J} \rangle, S \cdot Var(n) \right),
\end{equation}
thus $P(\tilde{J}) \sim {\rm Prob}(\sum_i n_i =\tilde{J}|\tilde{J})$ reaches its maximum in the vicinity of $\tilde{J}$ which satisfies $\tilde{J}\approx S \langle n_i |\tilde{J} \rangle = \left\langle \sum_i n_i |\tilde{J} \right\rangle $. Furthermore, for the approximation $P(\tilde{J})\approx {\rm Prob}(\sum_i n_i =\tilde{J}|\tilde{J})$, the values of $\tilde{J}$ where $\tilde{J}\ll S\langle n_i |\tilde{J} \rangle  $ or $\tilde{J}\gg S\langle n_i |\tilde{J} \rangle $ are highly improbable, due to the Gaussian nature of $P(\sum_i n_i|\tilde{J})$ for large $S$.
 
\subsubsection{Approximation Approach III: Estimating total population size $\langle J|n\rangle $ using mean-field approximation } In a similar fashion to previous approximation approaches, we assume $\langle J|n\rangle \approx \langle J\rangle  $, thus
\begin{eqnarray}
    P(n) \approx P(0)\frac{(r^+)^{n}(a)_{n}}{n!\prod_{n'=1}^{n}\left(r^-+r (1-\rho)n'/K + r\rho \langle J \rangle /K \right)}=  P(0)\frac{(r^+)^{n}(a)_{n}}{n!\prod_{n'=1}^{n}\left(r^-+r (1-\rho)n'/K + r\rho S\langle n\rangle  /K \right)}.
\end{eqnarray}
Then, $\langle n \rangle$ is given by the numerical solution of
\begin{eqnarray}
    \langle n \rangle = \sum_{n=0}^{\infty} n P(0)\frac{(r^+)^{n}(a)_{n}}{n!\prod_{n'=1}^{n}\left(r^-+r (1-\rho)n'/K + r\rho S\langle n\rangle  /K \right)}.
\end{eqnarray}
where here the normalization factor is $P(0)=1/{_1}F_1[a,\tilde{b};\tilde{c}]$ with $a=\frac{\mu}{r^+}$, $\tilde{b}= \frac{r^-K+r\rho \langle J\rangle }{r(1-\rho)}$, and $\tilde{c}=\frac{r^+ K}{r(1-\rho)}$.

\subsection{Interspecies Correlations and Limitations of the Approximation Approaches}

For the approximation approaches described above, we ignore the correlations between the species abundances and assume the species abundance dynamics are mutually independent, which implies $P(\vec{n})=\prod_{i=1}^S P(n_i)$. 
Of course, this mutual independence is not exact except for $\rho=0$. For all other $\rho>0$ the dynamics of a species are coupled, through their death rates, to other species abundances.  Therefore, we expect deviations of our approximations from the true simulated SAD. In particular, we expect {to find that the interspecies correlation and the quality of approximation are interconnected}.

\subsubsection{KS and KL distances}

We quantify how closely our approximate distributions ($Q$'s) measure to the exact distribution ($P$) using two metrics. The first one, the Kolmogorov-Smirnov distance (KS), is defined as $KS(P,Q)\equiv\max \left|{\rm CDF}(P)-{\rm CDF}(Q)\right|$, where CDF represents the cumulative distribution function. The second metric we use is the symmetric Kullback–Leibler divergence (sometimes referred to as the Jensen-Shannon divergence), which is defined as $KL(P,Q)\equiv \sum_x (P-Q)\ln\left(P/Q\right)$. 

\begin{figure}[ht!]
    \centering
       \includegraphics[width=0.45\columnwidth,trim=100 00 100 200]{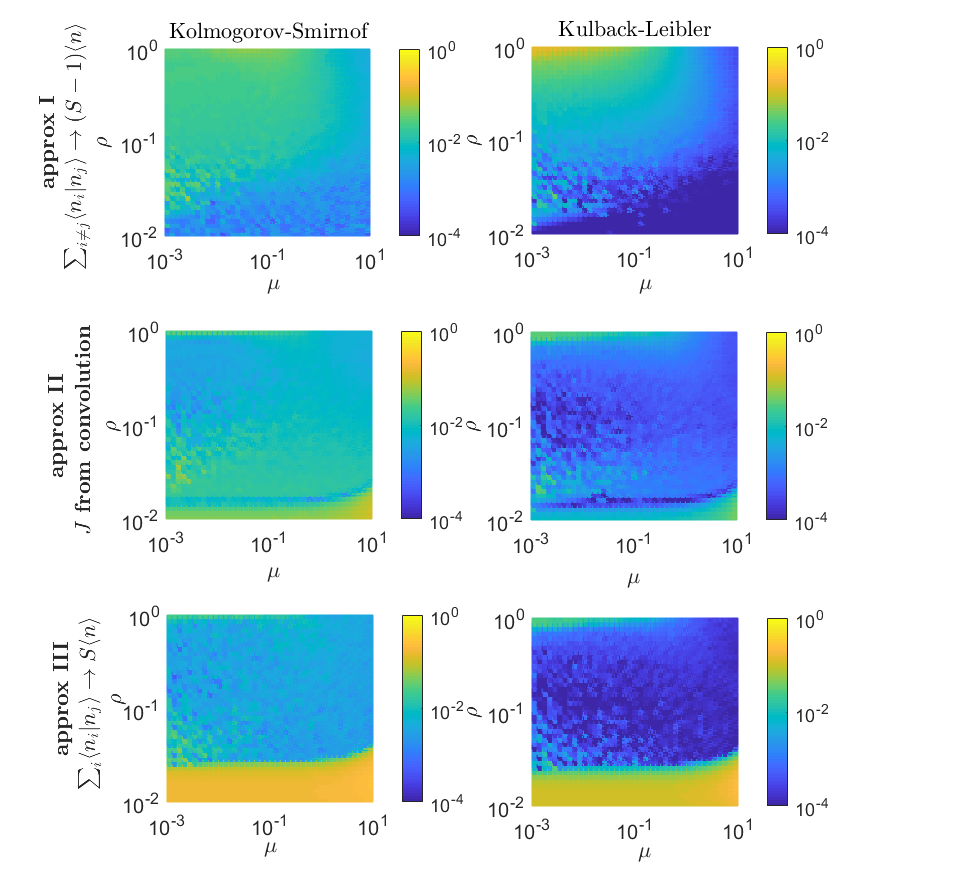}
    \includegraphics[width=0.45\columnwidth,trim=100 100 100 100]{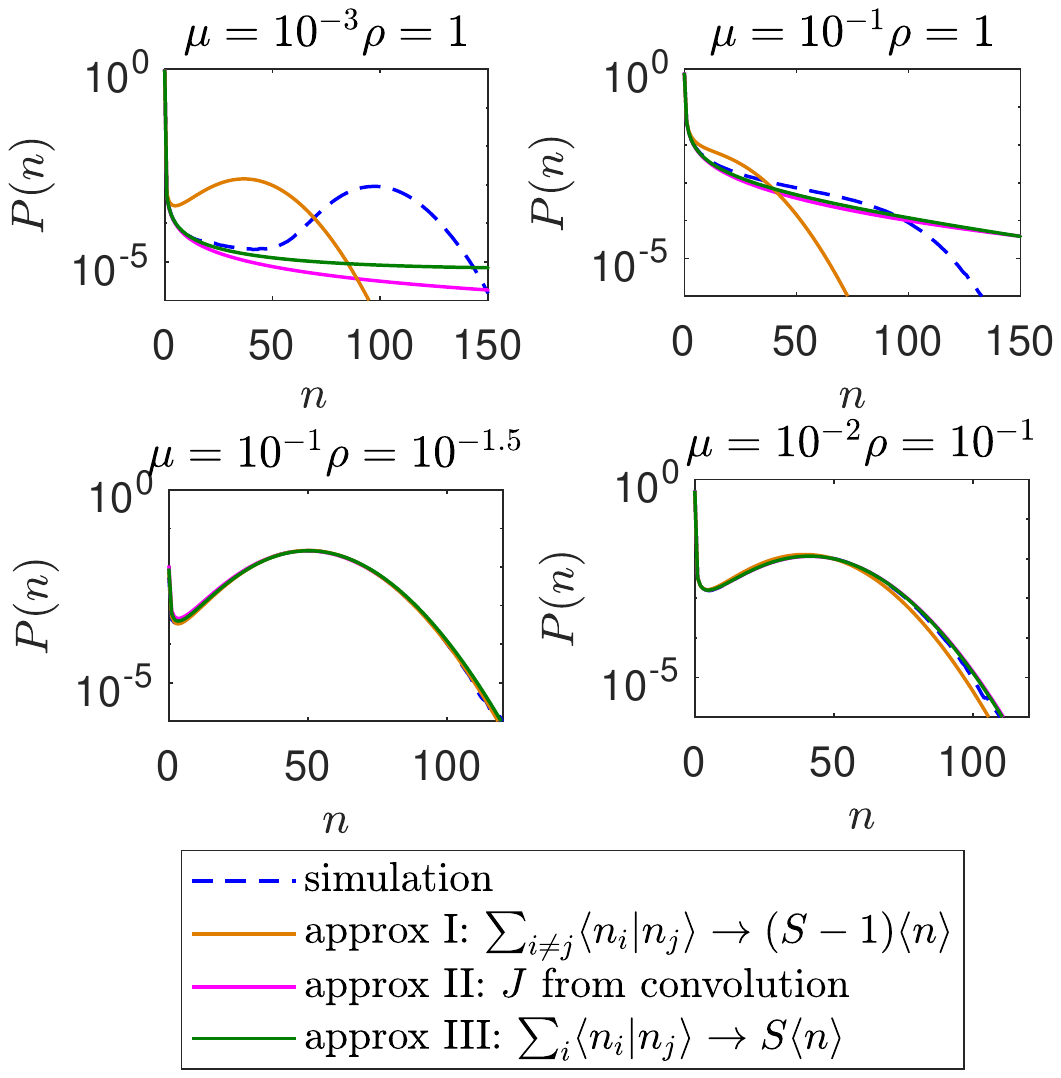}
    \caption{Quality of Approximation. { Left:  Kolmogorov-Smirnoff statistics (KS) and Kullback-Leibler divergence (KL) comparing the simulated SAD with our three suggested approximations (see SM above). These qualitatively represent regions of agreement (green, blue) or disagreement (yellow) with the approximations. As expected, the quality of the approximations follow the correlation results, where low correlation provides good agreement with the approximation. As noted before, approximation I is used in the main text to derive boundaries and analytics. Right: Four examples of the simulated and approximated SADs. Simulation SADs are represented with blue, dashed lines, whereas the three suggested approximations are given by solid lines (see legend).  Note that the KS distance might be low in certain regimes (i.e. green or blue regions), while KL exhibits high values (yellow), see for example $(\mu=10^{-3}, \rho=1)$. This is due to the fact the KS measures the difference between the distributions; as our distributions vary many orders of magnitude, the KS values are heavily determined by the maximum of the distribution.}} 
    \label{fig:quality}
\end{figure}

Intuitively, the KS metric captures the difference between an approximated and simulated distribution, {whereas} the KL divergence measures the ratio between the two distributions. 
Fig.~\ref{fig:quality} shows the KS and KL metrics comparing the three approximations presented above and the simulation results.

\begin{figure}[ht!]
    \centering
    \includegraphics[width=0.45\columnwidth,trim= 10 10 10 10, clip]{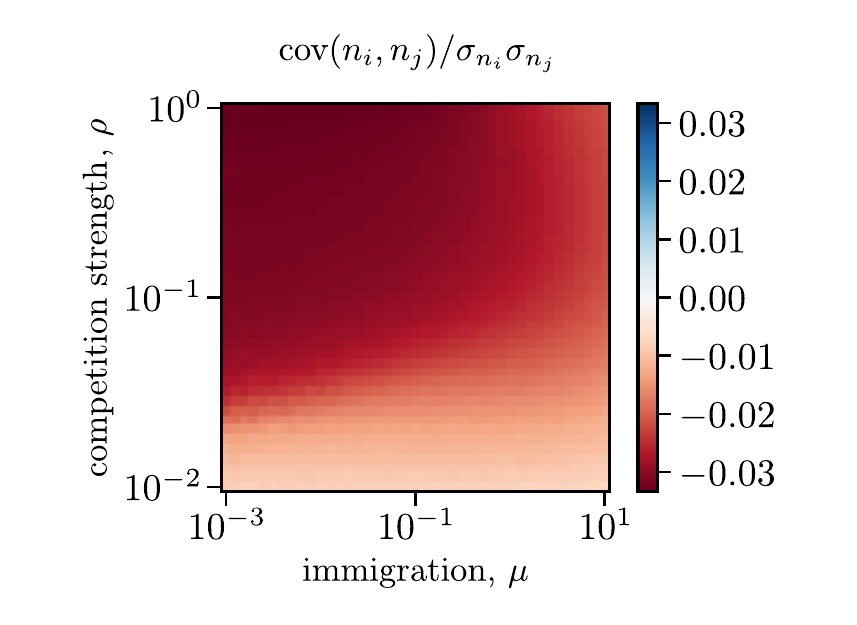}
    \includegraphics[width=0.45\columnwidth,trim= 10 10 10 10, clip]{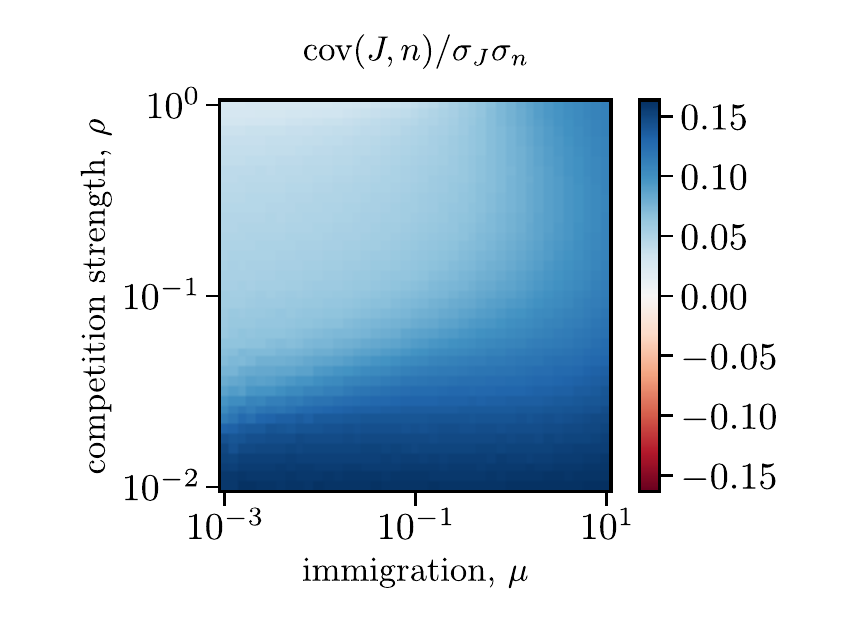}
    \caption{Abundance correlations. (Left panel) Pearson correlation coefficient between the abundances of any two different species. (Right panel) Pearson correlation coefficient between  the total population size $J=\sum n_j$ and an abundance of any species. Correlations were calculated from Gillespie simulation time course data with $6 \cdot 10^8$ time steps.
    }
    \label{fig:correlation}
\end{figure}

{
Both KS and KL distances {\em qualitatively} show regimes where the analytical approximations agree well with the simulated solutions (low distance scores; green and blue colors in the heatmap of Fig.~\ref{fig:quality}), as compared to regimes where the approximations do not agree with the simulations (represented with high values, yellow colors in Fig.~\ref{fig:quality}). Importantly, both KS and KL distances qualitatively reflect the regimes where correlations between the species' is large; this should be the case given that our approximations assume that the species are mutually uncorrelated. When this assumption of independence is fulfilled, there is good agreement between the analytical approximation and simulations (see Fig.~\ref{fig:quality} and the main text). For example, since Approximation III relies on a low correlation between $J$ and $n_i$, regions of low correlation show good agreement between the approximation and the simulation (see Fig.~\ref{fig:quality}, \ref{fig:correlation} and discussion in the section {\bf The abundances of different species are weakly anti-correlated} below).   \\
In Fig.~\ref{fig:quality} we present the KS distance. 
Importantly, we aim neither to confirm or deny the validity of our approximations using these metrics; as such, a specific threshold (p-value) for hypothesis testing is somewhat arbitrary to our purposes.
However, using $p=0.05$ for this KS test (as is commonly used in the literature, see e.g. \cite{bartoszynski2020probability, dimitrova2020computing}) we find that most regimes are well approximated except for at low competition strength in Approximations II and III as well as some regions close to $\rho=1$.}

In the main text, we choose to present the results for the regime boundaries obtained from the Approximation I above; even so, all approximations present reasonable agreement with the simulations. 
However, in some regimes, some approximations work better than the others: One approximation may better capture the location of a dominant abundance peak better, while other approximations show better agreement with the boundaries for various regimes. 
For example, only Approximation I captures the bi-modality at very low $\mu$ and $\rho=1$, where the other approximations seem to align slightly better with the simulated SAD.

{
\subsubsection{The abundances of different species are weakly anti-correlated}
\label{sec:Correlation}
So far, we have investigated the single species marginal abundance distribution $P(n)$, where our approximations assumed independence between the different species.  However, 
the species are not independent of each other due to the inter-species competitive interactions when $\rho>0$. It has been suggested that inter-species correlations reflect on the underlying community structure and the phase space \cite{carr2019use,chance2019native}. 
To investigate the connection between the population structure and the cross-species correlations, we calculated the cross-species abundances correlations, quantified via the Pearson correlation coefficient, as shown in Fig.~\ref{fig:correlation}A. The population  exhibits weak cross-species anti-correlation that increases with the competition strength $\rho$. This is expected given that the death rate of each species increases with the abundance of the other species and, consequently, these cross-species influences are more pronounced at high competition strengths. 
Conversely, higher immigration rates decrease the correlation between the species. 
Thus, the anti-correlation is most pronounced in the high competition and low immigration regime.
}

{
Furthermore, the impact of individual species on the total population size can be quantified by the correlation between the total population size $J$ and individual abundances $n_j$.
We found that the individual species abundances are positively correlated with the total abundance $J=\sum_i n_i$, which also fluctuates as the individuals of all species undergo birth and death events. 
Interestingly, as shown in Fig.~\ref{fig:correlation}B, the magnitude of this correlation $\text{cov}(J,n_i)/\sigma_n \sigma_J$ exhibits inverse 
trends compared to the inter-species anti-correlation: the correlation  $\text{cov}(J,n_i)/\sigma_n \sigma_J$ is weaker when the cross-species anti-correlation is stronger.
The magnitude of the correlation between the total population size $J$ and a species abundance $n$ exhibits similar behavior to the average richness (shown in main text Fig.~2A): $\text{cov}(J,n_i)/\sigma_n \sigma_J$ is high in the high immigration, low competition strength regime and is low otherwise.
This behaviour may be understood heuristically: 
whereas each species in a system with $S^*$ dominant species only contributes $\sim J/S^*$ to the total population size. 
}
{
Somewhat unexpectedly, neither the inter-species correlations nor the correlations between the species abundance and the total abundance distinguish between the different modality regimes but rather both increase with the richness. As expected, our mean-field approximation works best at very low ${\rm cov}(n_i,n_j)/\sigma_{n_i}\sigma_{n_j}$, whereas our mean-field deviates from the solution as anti-correlation gets stronger. 
}

\subsection{Asymptotic analysis}

{Species abundance distribution of the `rare-biosphere' are commonly described in the literature as being power-laws with exponential cutoff (see \cite{xu2018immigration}). In the following, we examine the tails of our approximate SADs and show that they exhibit similar behavior in some regimes.  } 

\subsubsection{{Neutral Dynamics; $\rho=1$}}
In the case where $\langle J|n_1\rangle \approx \langle J \rangle $, {which corresponds to our Approximation III (see text above), we obtain }
\begin{equation}
    P(n)= P(0) \left(\frac{r^+}{r^- + r \langle J \rangle /K}\right)^{n}\frac{ (a)_{n}}{n!} \approx  P(0) \left(\frac{r^+}{r^- + r \langle J \rangle /K}\right)^{n_1}\frac{ n_1^{a-1}}{\Gamma[a]}.
    \label{eq:neutral-asym}
\end{equation}
Note that this is valid only for the tail-end of the distributions when $ a\ll n$. {Notably, a power-law without the exponential cutoff is found when $\langle J\rangle =K$, as $r^+ = r^- + r$.}


\subsubsection{{General Competition Strength; $0 < \rho < 1$}}  
We can approximate the tail of the distribution for non-neutral models ($0 < \rho < 1$) using similar arguments as above, that is to say approximating $\langle J |n \rangle \approx \langle J \rangle $.
Using Eq.~\ref{eq:sol-ME-Jgivenn} and this approximation, we find that
\begin{equation}
    P(n) = P(0)\frac{(r^+)^{n}(a)_{n}}{n!\prod_{n'=1}^{n}(r^-+r(1-\rho)n'/K+\rho \langle J |n' \rangle /K)}\approx P(0)\frac{(a)_n \tilde{c}^n}{n!(\tilde{b}+1)_n}
    \label{eq:Tail_general_rho}
\end{equation}
where $a=\frac{\mu}{r^+}$, $\tilde{b}= \frac{r^-K+r\rho  \langle J \rangle }{r(1-\rho)}$, and $\tilde{c}=\frac{r^+ K}{r(1-\rho)}$.
Looking at the tail of the distribution ($n \gg a, \tilde{b}, \tilde{c}$) we find that 
\begin{equation}
    P(n) \xrightarrow{n \rightarrow \infty} P(0)\frac{\Gamma[\tilde{b}+1]}{\sqrt{2 \pi} \Gamma[a]} n^{a-\tilde{b}-\frac{3}{2}}(\tilde{c}/n)^{n}e^n
\end{equation} 
where we have used Stirling's approximation: $n! \approx \sqrt{2\pi n}n^n e^{-n}$.
In the high $n$ limit, we can further approximate $J\sim n$ such that $P(n)\sim n^{-n/(1-\rho)}e^n$.

In the case where $\rho \rightarrow 0$, there is no need for approximating $\langle J |n\rangle$, as it disappears in the exact solution.
We are left with the solution
\begin{equation}
    P(n) \xrightarrow{\rho\rightarrow 0}P(0) 
    \frac{(\mu/r^+)_{n} (r^+ K/r)^{n}}{n ! (r^-K/r+1)_{n} }
\end{equation}
which is exact.
The tail of this distribution goes as
\begin{equation}
    P(n) \xrightarrow{\rho\rightarrow 0,n\rightarrow \infty} P(0)\frac{\Gamma[\tilde{r^-K/r}+1]}{\sqrt{2 \pi} \Gamma[\mu/r^+]} n^{\frac{\mu}{r^+}-\frac{r^-K}{r}-\frac{3}{2}}(r^-K/rn)^{n}e^n.
\end{equation}

{We can check that we recover Eq.~\ref{eq:neutral-asym} in the limit when $\rho \rightarrow 1$ in Eq.~\ref{eq:Tail_general_rho}. However, both $\tilde{b}$ and $\tilde{c}$ in Eq.~\ref{eq:Tail_general_rho} go to $\infty$ as $\rho \rightarrow 1$.}
As such, we use the fact that $(x+1)_n \xrightarrow{ x \rightarrow\infty} x^{n}$ to write
\begin{equation}
     P(n) \xrightarrow{\rho\rightarrow 1}P(0) 
    \frac{(a)_{n} \{r^+ K/[r(1-\rho)]\}^{n}}{n ! \{(r^-K+rJ)/[r(1-\rho)]\}^{n} }= P(0) 
    \frac{(a)_{n} (r^+ K)^{n}}{n ! (r^-K+rJ)^{n} }  \xrightarrow{n \rightarrow \infty } P(0) \left( \frac{r^+K}{r^-K + rJ} \right)^n \frac{n^{a-1}}{\Gamma[a]}
\end{equation}
which agrees with what we found earlier for $\rho=1$ and constant $J$. 

\subsubsection{Comparing with Other Neutral SADs} This asymptotic behaviour may be compared to analytical solutions in Moran models for which $J$ is held constant.
These Moran type models are often solvable exactly, we choose to show their results for $J=S n_{det}$, where $n_{det}$ is the solution to our mean deterministic Lotka-Voltera equation. 
In~\cite{mckane2004analytic}, an analytical solution to the Hubbell model with immigration is found such that
\begin{equation}
    P(n)={J \choose n}\frac{\beta (n+p,n^*-n)}{\beta (p^*,n^*-J)} 
\end{equation}
where $p=1/S$, $n^*=(J-m)/(1-m) - p$, and $p^*=m p (J-1)/(1-m)$ .
In this model, $m$ is defined as the probability of immigration at any step.
This is different from our immigration rate, however we find a suitable transformation to be $m \approx \mu / \langle r^+_n + r^-_n \rangle$: the probability of immigration is the rate of immigration divided by the mean rate of birth and death reactions.
Note that the function $\beta(a,b)=\Gamma (a) \Gamma (b)/ \Gamma (a+b)$.

\begin{figure}[h!]
    \centering
    \includegraphics[width=0.7\columnwidth]{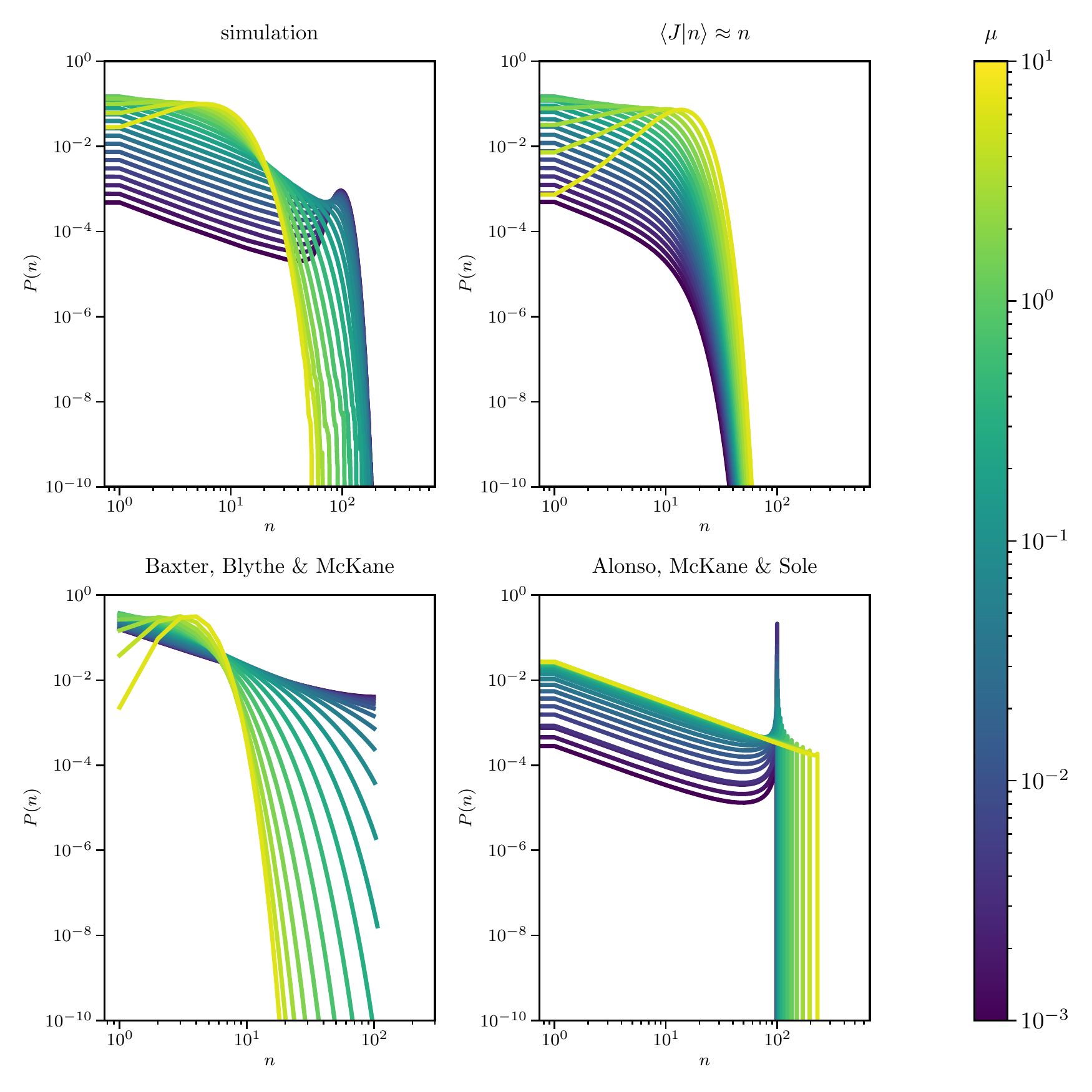}
   \caption{Asymptotic behaviour of various neutral models compared to simulations. (top panels) Using $\langle J | n \rangle \approx n $, our approximation does not recover a bimodality, however the analytical approximation clearly follows the simulation's power law with exponential cutoff. (bottom panels) Moran-like models in the literature of power-law SADs with exponential cutoff. Here, the total population size used is the total population size of the steady-state Lotka-Voltera equation, $J=S n_{det}$. The continuous Fokker-Planck diffusion model of Baxter, Blythe \& McKane\cite{baxter2007exact} shows the immigration dominated peak. However, the Hubbell community model solved by Alonso, McKane \& Sol\'{e}\cite{mckane2004analytic} shows a bimodality in the low immigation regime. Both have power laws with exponential cutoff in different regimes.}
    \label{fig:asymptotic}
\end{figure}

In~\cite{baxter2007exact}, a continuum Fokker-Planck equation is solved to evaluate a similar multi-allelic diffusion model abundance. 
However, in this formalism, immigration is replaced by mutations wherein $u_i$ is the rate of mutation of cell allele $i$.
Assuming all the mutation rates are equivalent, $u_i=u$.
The steady state joint probability distribution is 
\begin{equation}
    P(\vec{x})=\Gamma (2 S u ) \delta (1-\sum_i x_i)\prod^S_{i=0} \frac{x_i^{2u-1}}{\Gamma (2 u )}
\end{equation}
which may be integrated to find the SAD
\begin{equation}
    SAD(n) = \langle \sum_j \delta (x_j - n/J) \rangle_{P(\vec{x})} \approx \left( \frac{n}{J} \right)^{2 u - 1 } e^{- (2 u (S-1) -1 ) n / J }  
\end{equation}
Although mutations and immigration are not completely equivalent, mutations may take on a heuristic role similar to immigration that allows for no species to be truly extinct.
As such, we assume $u=\mu/r$.
Comparisons of these different asymptotic behaviours are found in Fig.~\ref{fig:asymptotic}.

\section{Richness Distribution}
As mentioned in previously, the number of (co-)exiting species is a stochastic variable we denote as $S^*$. In the following, we provide insights about its distribution. For clarity, we specify the distribution of the richness with its corresponding subscript, i.e. $P_{\rm rich, S}(S^*)$.
The richness distribution may be derived explicitly by integrating the joint distribution
 \begin{equation}
 \label{eq:binomial}
     P_{\rm rich, S}(S^*) = Prob\underbrace{\left(n_1, n_2,  \dots,  n_i, \dots , n_S \right)}_\textrm{ $n_i >0$ for exactly $S^*$ species} = \binom{S}{S^*} \sum_{n_1=1}^{\infty} \cdots \sum_{n_{S^*}=1}^{\infty} P(\underbrace{n_1,...,n_i,...,n_{S^*}}_{S^*},\underbrace{0,...,0}_{S-S^*})
 \end{equation}
where the second equality stems from the fact that all species are symmetric; the binomial coefficient appears from the permutations of relabeling species in the probability.

\subsection{The mean richness}
Here we derive the mean of the richness distribution by two different methods, which is possible  without calculating the whole distribution.

\subsubsection{Indicator variables}
For each species $i$, we define an indicator random variable $s_i$ which indicates whether a species is present ($s_i=1$) or absent ($s_i=0$).
Note that because the abundances of different species are correlated, these random variables are not independently distributed.
The marginal distribution of an indicator variable is simply $P_{s}(s_i=0)= P(0)$ and $P_{s}(s_i=1)= 1-P(0)$ .
Thus the mean of any indicator variable is
\begin{equation}
    \langle s_i \rangle = 0 \cdot P(0) + 1 \cdot (1 - P(0)) = ( 1 - P(0) ).
\end{equation}

The richness of the system is $S^* = \sum_{i=1}^S s_i$ and its mean is found as
\begin{equation}
    \langle S^* \rangle = \sum_{i=1}^S \langle s_i \rangle = S (1 - P(0)).
\end{equation}
leading to the equation in the main text. In a more formal derivation, the  mean richness can be written as
\begin{align}
    \langle S^* \rangle = \sum_{S^*=0}^S S^* P_{\rm rich, S}(S^*)
    & = \sum_{S^*=1}^S \frac{S!}{S^*! (S-S^*)!} S^* \sum_{n_1=1}^{\infty} \cdots \sum_{n_{S^*}=1}^{\infty} P(n_1,...,n_i,...,n_{S^*},0,...,0)\\
     & = S \sum_{S^*=1}^S \frac{(S-1)!}{(S^*-1)! ((S-1)-(S^*-1))!} \sum_{n_1=1}^{\infty} \cdots \sum_{n_{S^*}=1}^{\infty} P(n_1,...,n_i,...,n_{S^*},0,...,0)\\
     & = S \sum_{n_1=1}^{\infty}\left( P(n_1,0,...,0) + \sum_{S^*=2}^S \binom{S-1}{S^*-1} \sum_{n_2=1}^{\infty} \cdots \sum_{n_{S^*}=1}^{\infty} P(n_1,...,n_i,...,n_{S^*},0,...,0) \right). \label{eq:intermediate-rich}
\end{align}
As before, the symmetry of the species allows for a permutation of the species labels, such that 
\begin{equation}
    P(n_1,0,...,0) + \sum_{S^*=2}^S \binom{S-1}{S^*-1} \sum_{n_2=1}^{\infty} \cdots \sum_{n_{S^*}=1}^{\infty} P(n_1,...,n_i,...,n_{S^*},0,...,0) = \sum_{n_2=0}^{\infty} \cdots \sum_{n_{S}=0}^{\infty} P(n_1,n_2,...,n_S).
\end{equation}

Thus, we find that Eq.~\ref{eq:intermediate-rich} is equivalent to
\begin{align}
    \langle S^* \rangle = S \sum_{n_1=1}^{\infty} \left( \sum_{n_2=0}^{\infty}\dots \sum_{n_S=0}^{\infty} P(n_1,n_2,...,n_S) \right) = S \sum_{n_1=1}^{\infty} P(n_1) = S \left( 1 - P(0) \right)
\end{align}
We find that the mean richness can be expressed as a function of the total number of species and the marginal distribution of a species at zero.

{
\subsection{The Binomial Distribution Approximation}
In the limit of small $\rho\ll 1$, in the first approximation, the species can be assumed to be independent, $P(n_1,\dots,n_S) = \prod_{i=1}^S P(n_i)$, and the richness distribution becomes a binomial distribution
\begin{equation}
\label{eq:richness-approx}
     P_{\rm rich, S}(S^*) = \binom{S}{S^*} P(0)^{S-S^*}[1-P(0)]^{S^*},
 \end{equation}
where $P(0)S$ is the average number of temporarily extinct species, in agreement with \cite{dobson2020unsolved} (see Section 2 of the main text). 
}

{This result is supported by an alternative derivation based on our discussion of the first passage times of extinction or invasion in Section 3.C of the main text. The first passage times may be interpreted as the inverse of the rates of extinctions or successful invasions defining the effective rates of species richness transitions from species $S^* \rightarrow S^*+1$ (invasion) or $S^* \rightarrow S^*-1$ (exclusion).  
At stationary state, this can be encapsulated in an effective master equation for the richness distribution,
\begin{equation}
0 = q^+_{\rm rich, S}(S^*-1) P_{\rm rich, S}(S^*-1) + q^-_{\rm rich, S}(S^*+1) P_{\rm rich, S}(S^*+1) -\left[q^+_{\rm rich, S}(S^*) + q^-_{\rm rich, S}(S^*)\right]P_{\rm rich, S}(S^*),
\end{equation}
where $q^+_{\rm rich, S}(S^*)$ and $q^-_{\rm rich, S}(S^*)$ refer to the transition rates of species addition and removal respectively. This equation satisfies the flux balance condition for the richness distribution $P_{\rm rich, S}(S^*)$
\begin{equation}
    q^+_{\rm rich, S}(S^*)P_{\rm rich, S}(S^*) = q^-_{\rm rich, S}(S^*+1)P_{\rm rich, S}(S^*+1).
\end{equation}
Thus, the richness distribution is given by 
\begin{equation}
    P_{\rm rich, S}(S^*) = P_{\rm rich, S}(0) \prod_{s=1}^{S^*} \frac{q^+_{\rm rich, S}(s-1)}{q^-_{\rm rich, S}(s)}.
\end{equation}
In general the transition rates depend on the multi-dimensional state of the species composition, $(n_1, n_2, \cdots , n_S)$, for which we do not have an exact solution $P(\vec{n})$.
To derive an approximate expression for $P_{\rm rich, S}(S^*)$, we assume that the species are mutually independent. Consequently, a given species may enter the system (crossing from $n=0$ to $n=1$) with a rate $1/T(0\rightarrow 1)$ and leave the system at a rate $1/T(1\rightarrow 0)$, where both rates are independent of other species abundances.
The rate of increasing the species richness, $S^*$, is the sum of invasion rates for $S-S^*$ excluded species entering the system whereas the rate of decreasing the species richness is the sum of exclusion rates for $S^*$ co-existing species exiting the system
\begin{eqnarray}
    q^+_{\rm rich, S}(S^*) & \approx & \frac{S-S^*}{T(0 \rightarrow 1)}\\
    \label{eq:richness-rates1}
    q^-_{\rm rich, S}(S^*)& \approx& \frac{S^*}{T(1 \rightarrow 0)}.
    \label{eq:richness-rates2}
\end{eqnarray}
Under the above assumptions, the solution for the richness distribution is
\begin{equation}
    P_{\rm rich, S}(S^*) = P_{\rm rich, S}(0) \prod_{s=1}^{S^*} \frac{S-s+1}{s}\frac{T(1 \rightarrow 0)}{T(0 \rightarrow 1)} = \left(1+\frac{T(1 \rightarrow 0)}{T(0 \rightarrow 1)}\right)^{-S}\left(\frac{T(1 \rightarrow 0)}{T(0 \rightarrow 1)}\right)^{S^*}\frac{S!}{(S-S^*)!(S^*)!}.
\end{equation}
Since $T(0 \rightarrow 1)=1/\mu$ and $T(1 \rightarrow 0)=(1-P(0))/\mu P(0)$ (see Section 6)
 \begin{equation}
     P_{\rm rich, S}(S^*) = \binom{S}{S^*} P(0)^{S-S^*}[1-P(0)]^{S^*},
 \end{equation}
recovering Eq. \ref{eq:richness-approx}. This is not surprising as both approaches rely on the assumption of species independence. The approximate binomial distribution  recovers the correct mean ($\langle S^*\rangle =S(1-P(0))$) but not the variance nor the full shape of the distribution found in simulations. This binomial distribution of the richness was also derived in \cite{dobson2020unsolved}, where the authors similarly assumed linear transition rates.}

\begin{figure}[ht!]
    \centering

    \includegraphics[width=0.75\columnwidth]{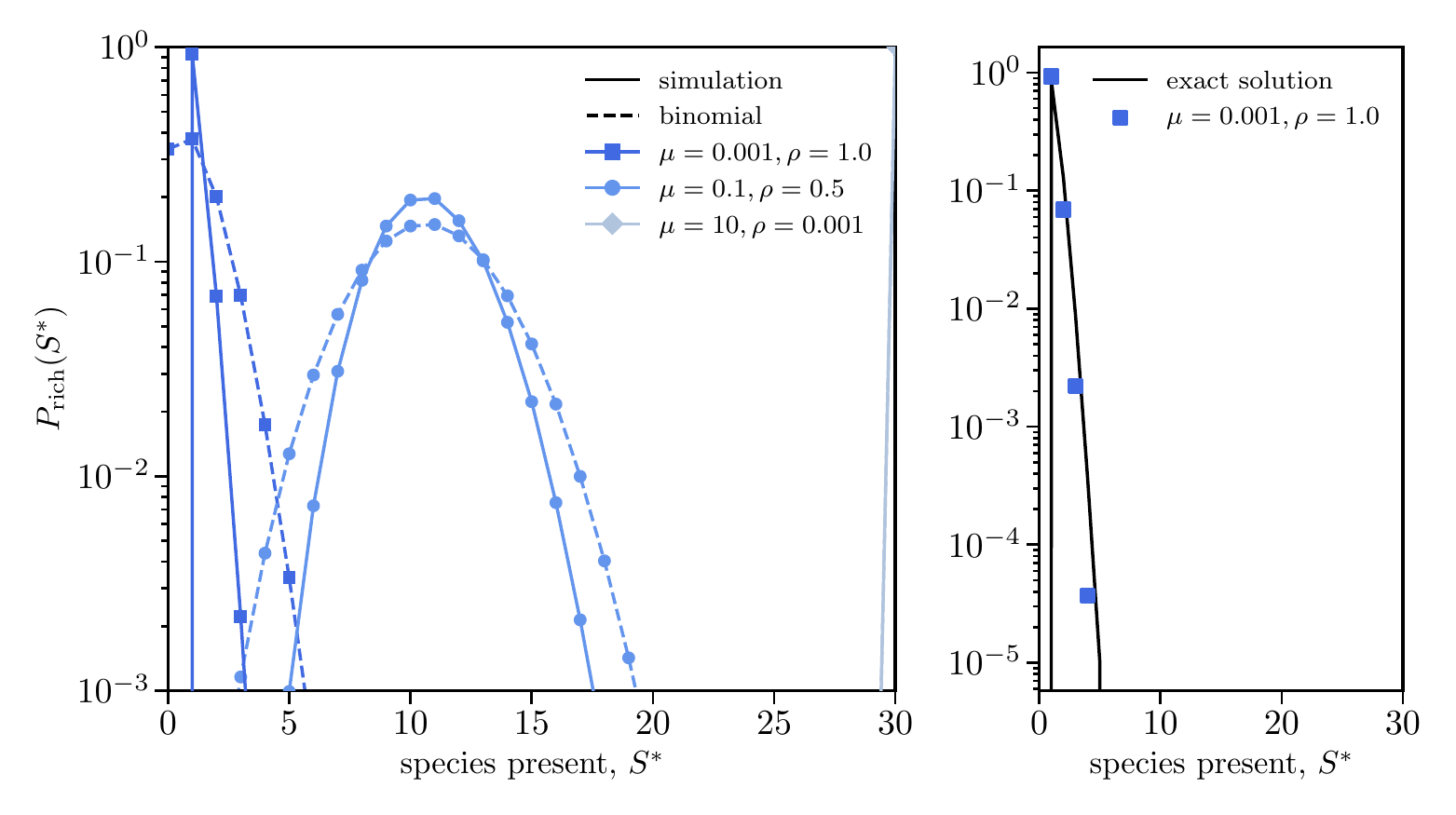}
    \caption{Sample richness distributions. Left panel: Simulation (full line) and binomial approximation (dashed line) richness distributions are shown. These sample distributions are taken from a set of parameters such that each richness regime is represented by one parameter: full coexistence ($\mu=10, \rho=0.01$), partial coexistence ($\mu=0.1, \rho=0.5$) and exclusion ($\mu=0.001, \rho=1.0$). For $\rho = 0.001$, the binomial approximation lines up well with the simulation; the binomial distribution is exact for $\rho = 0$. For larger competition strengths, the binomial distribution does not approximate well the simulated distribution.
    Right panel: Simulations (scatter points) plotted against the exact solution for $\rho=1$ (line) presented above. Although the binomial distribution is much wider than the simulated distribution for $\rho=1$ (see left panel), the exact solution given in Eq.~\ref{eq:neutral-rich} fits the narrow distribution.
    }
    \label{fig:richness}
\end{figure}

{
Fig. \ref{fig:richness} shows how the binomial approximation compares to the simulated richness distribution.
Outside of the regime where the binomial is exact ($\rho=0$), the binomial distribution tends to be wider than the simulated distribution. This is explained by the fact that the assumption that the species are mutually independent breaks down in the intermediate $\rho$ regime where each species abundance depends on the presented and abundance of other species.
This constrains the number of species present at any point in time and the species richness distribution is narrower around the mean richness than predicted by the independent binomial approximation.}

 \subsection{Exact Richness Distribution for $\rho=1$}
 {In this subsection we calculate the exact richness distribution for $\rho=1$ which is obtainable since the multidimensional probability for $\rho=1$ is known (Eq.~\ref{eq:exact_appendix} in Section 2): 
 \begin{equation}
     P(n_1, \dots, n_S) = \frac{1}{{_1F_1}[a S, \hat{b}, c]}\frac{c^{n_1+n_2 + \dots + n_S}}{(\hat{b})_{n_1+n_2+\dots n_S}} \prod_{i=1}^S
\frac{(a)_{n_i}}{n_i!} \end{equation}
with $a$ and $c$ defined as in Eq.~\ref{eq:exact_appendix}, and $\hat{b} \equiv \frac{K r^-}{r}+1$. This has also been derived in \cite{haegeman2011mathematical}. The richness distribution is obtained by integrating over the relevant area in the $S-$dimensional space. For example, in a two-species system, when $S=2$, the richness distribution is given by
\begin{eqnarray}
P_{\rm rich, 2}(S^*=0 ) &=& \frac{1}{{_1 F_1}[2a,\hat{b},c]} \\  
P_{\rm rich, 2}(S^*=1 ) &=&  \sum_{n_1=1}^{\infty}P(n_1, 0) + \sum_{n_2=1}^{\infty}P(0, n_2) = \frac{2}{{_1 F_1}[2a,\hat{b},c]}[{_1 F_1}[a,\hat{b},c] -1] \\ 
 P_{\rm rich, 2}(S^*=2 ) &=&  \sum_{n_1=1}^{\infty}\sum_{n_2=1}^{\infty}P(n_1, n_2) = 1-\frac{2}{{_1 F_1}[2a,\hat{b},c]}[{_1 F_1}[a,\hat{b},c] -1] - \frac{1}{{_1F_1}[2a,\hat{b},c]}
\end{eqnarray}
Similarly, one can show that for a given $S$ 
\begin{equation}
\label{eq:neutral-rich}
    P_{\rm rich, S}(S^*) = \binom{S}{S^*} \frac{{_1F_1}[S^* a ,\hat{b},c]+{\sum_{s=0} ^{S^*-1}}(-1)^{S^*-s}\binom{S^*}{s}{_1F_1}[s a ,\hat{b},c]}{{_1F_1}[aS,\hat{b},c]}
\end{equation}
}

\section{Defining Phase Boundaries}

\subsection{Boundaries for Richness Regimes} For the phase boundaries defined by richness, we use
the mean richness to be, $ \langle S^* \rangle = S \left(1-P(0)\right) 
$, where $P(0)$ is obtained numerically from the approximated SAD. Note that in the mean-field approximations $P(0)$ is explicitly given as a Kummer confluent hypergeometric function. Then, the transition between full richness, $SP(0) \rightarrow 0$, to partial coexistence, $S(1-P(0)) = S - 1$, can be defined as the arithmetic mean between the two boundaries as $S P(0) = 1/2$. Similarly, the transition boundary between partial coexistence and excluded regime is drawn where $SP(0)=S-3/2$, which lies in-between $S(1-P(0))=\langle S^* \rangle=1$ and $S(1-P(0))=\langle S^*\rangle =2$ .

\subsection{Derivation of $\tilde{n}$ } The boundaries defined by the modalities can be found numerically directly from the exact distribution, see Fig.~2 in the main text. However, we have found that closed expression for the boundaries may be derived from the mean-field approximation.     

The transition between  `rare-biosphere' to bimodal regimes, is found at the dividing line between the presence and absence of a local positive maximum in the SAD. 
The `rare-biosphere' regime is defined by $P(n)>P(n+1)$ for all $n$ since the SAD is monotonically decreasing.
In the bimodal regime, there {exists a $n>0$ such that} $P(n)<P(n+1)$.
Thus the boundary between the regimes occurs where, as function of parameters $\mu$ and $\rho$, there is $\tilde{n}$ that solves
  \begin{eqnarray}
      P(\tilde{n})=P(\tilde{n}+1).
    \end{eqnarray}
Using the $P(n)$ from the mean-field approximation, we find
    \begin{eqnarray}
      P(0)\frac{(a)_{\tilde{n}} c^{\tilde{n}}}{\tilde{n}! (b+1)_{\tilde{n}} }&=&P(0) \frac{(a)_{\tilde{n}+1} c^{\tilde{n}+1}}{(\tilde{n}+1)! (b+1)_{\tilde{n}+1} } 
      \\
      \frac{(b+1)_{\tilde{n}+1}(\tilde{n}+1)!}{\tilde{n}! (b+1)_{\tilde{n}} }&=& \frac{(a)_{\tilde{n}+1} c^{\tilde{n}+1}}{(a)_{\tilde{n}} c^{\tilde{n}} } 
      \\
      n(b+\tilde{n})&=& c (a+\tilde{n}-1)
      \\
      \tilde{n}&=& \frac{(c-b)\pm \sqrt{(c-b)^2+4(a-1)c}}{2}
  \end{eqnarray}
Substituting $a=\frac{\mu}{r^+}$, $b=\frac{r^- K + r \rho (S-1) \langle n \rangle }{r}+1$ and $c=\frac{r^+ K}{r}$ yields
\begin{eqnarray}
   \tilde{n}\approx
    \frac{K-\rho(S-1)\langle n\rangle}{2}\left\{1 \pm \sqrt{1+4\frac{(\mu-r^+) K}{r(K-\rho(S-1)\langle n\rangle)^2}}\right\}.
    \end{eqnarray}
To make further progress we need to replace $\langle n\rangle$. Here  using $S \langle n \rangle = \langle S^* \rangle \tilde{n}$, we find
\begin{equation}
    \tilde{n}=\frac{K}{2(1-\rho+\rho \langle S^* \rangle)}\left\{1+ \sqrt{1+\frac{4(\mu-r^+)(1-\rho+\rho \langle S^* \rangle)}{rK}}\right\} \approx \frac{K}{1-\rho+\rho \langle S^* \rangle}- \frac{\mu-r^+}{r}.
\end{equation}
   
\subsection{Boundaries for Modality Regimes}
In the `rare-biosphere' regime, the SAD is monotonically decreasing. Therefore, the `rare-biosphere' regime can be defined in terms of $\tilde{n}$ such that its solution is not physical, i.e. either an imaginary $\tilde{n}$, or a real but negative $\tilde{n}$. The transition line between real and imaginary $\tilde{n}$, i.e. where $\Im(\tilde{n})=0$, is given by  
\begin{eqnarray}
    {r\left[K-\rho  (S-1)\langle n\rangle\right]^2}=4{(r^+-\mu) K}.
\end{eqnarray}
The transition line between the negative to positive $\tilde{n}$, where $\tilde{n}=0$, is drawn where
\begin{equation}
    \frac{(K-\rho(S-1)  \langle n\rangle)^2}{4}=1+ \frac{4K(\mu-r^+)}{r(K-\rho(S-1)  \langle n\rangle)^2}.
\end{equation}
Therefore, the {`rare-biosphere'} regime is defined as the union of the regions defined by both two equations above.

The second {modality }boundary we derive is the border of uni-modality region with positive probable abundance. At zero abundance ($n=0$), the abundance distribution has either a local maximum $P(0)>P(1)$ corresponding to bimodality or the {`rare-biosphere'} regimes or it has a local minimum , i.e. $P(0)<P(1)$,  { which corresponds to} unimodality. Consequently, the boundary between these two possibilities is given by $P(0)=P(1)$, which can be cast as a {zero-flux equation,} yielding
\begin{equation}
    q^+(0) = \langle q^-(\vec{n})|1\rangle  \Longrightarrow  \mu = r^- +\frac{r}{K} \left[1-\rho + \rho \langle J\rangle\right]. 
\end{equation}
{This defines the boundary of the immigration dominated regime where the system transitions to non-monotonic unimodal distribution.}

\section{Dependence on Carrying Capacity $K$}

\begin{figure}[ht!]
    \centering
    \includegraphics[width=0.45\columnwidth]{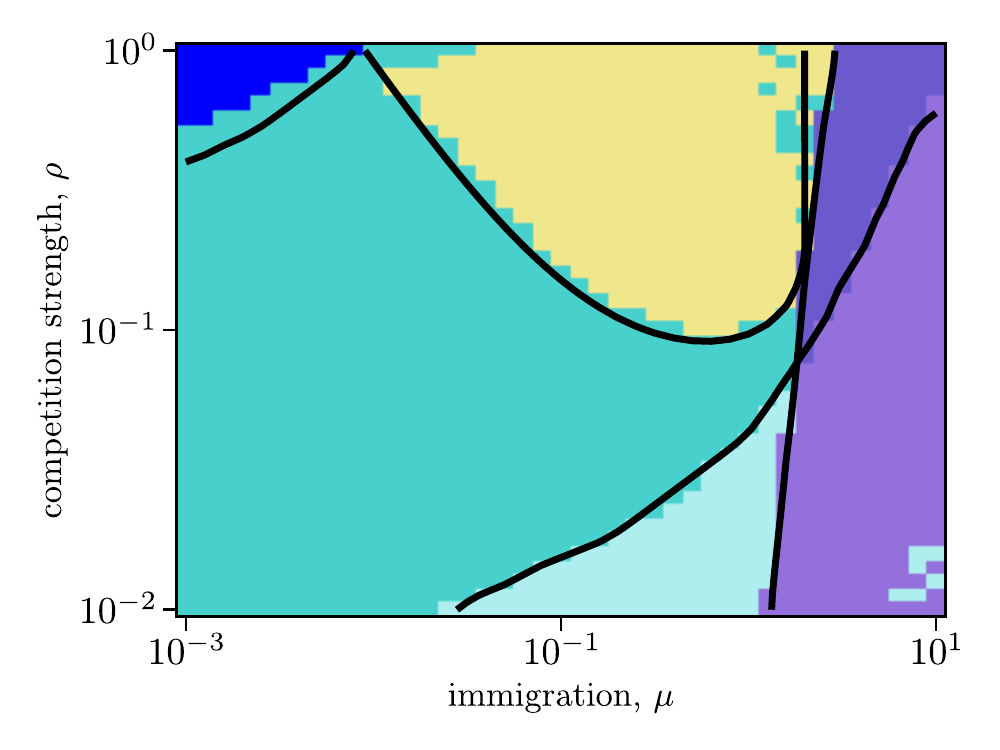}
    \includegraphics[width=0.45\columnwidth]{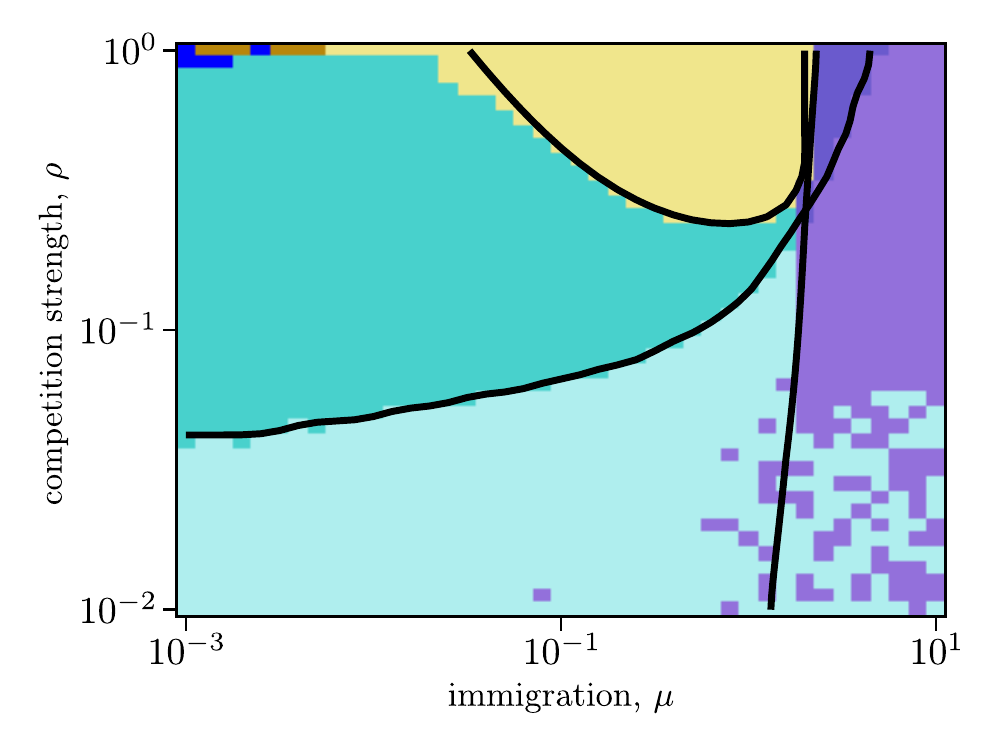}
    \caption{Varying carry capacity. Left panel: $K=50$. Right panel: $K=200$. Note that the the erroneous classification in certain regimes is due to numerical difficulty in determining optima of the SAD.}
    \label{fig:carry_capacity}
\end{figure}

{The model depends on a variety of parameters ($K$, $r^+$, $r^-$) which may also be varied.
We see that t}he regime boundaries depend on the carrying capacity, see Fig.~\ref{fig:carry_capacity}; as the carrying capacity increases, the {`rare-biosphere'} regime shrinks in size in the parameter space.
We also find that the full richness regime extends to regions of higher competition strength as $K$ increases. This suggests that interspecies competition decreases in the larger carrying capacity regimes.
Intuitively, the larger carrying capacity allows for more individual species to immigrate into the system and coexist.

\section{Kinetics}

\begin{figure}[ht!]
\begin{center}
\includegraphics[width=0.7\columnwidth]{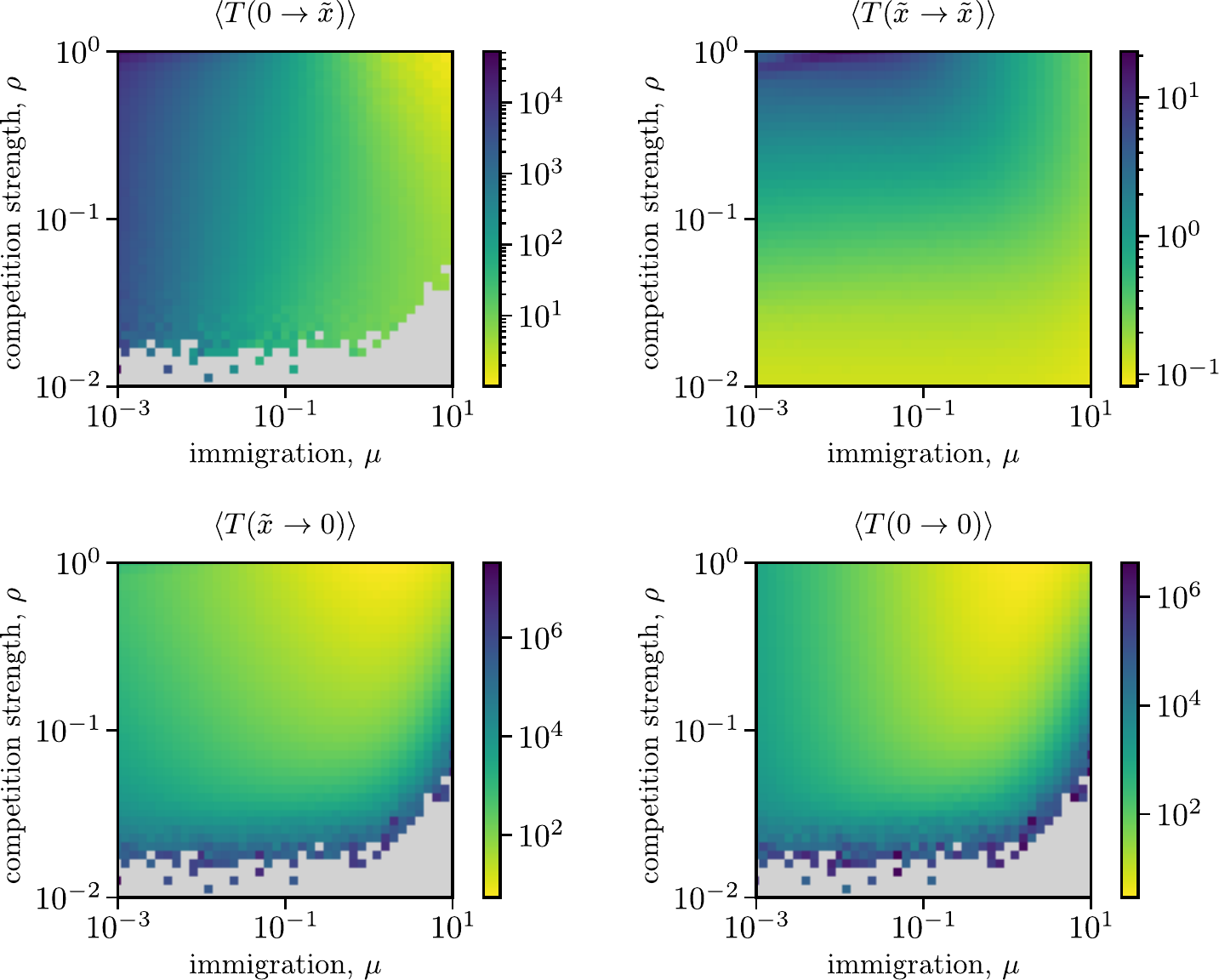}
\end{center}
\includegraphics[trim=140 260 140 230, width=0.24\columnwidth]{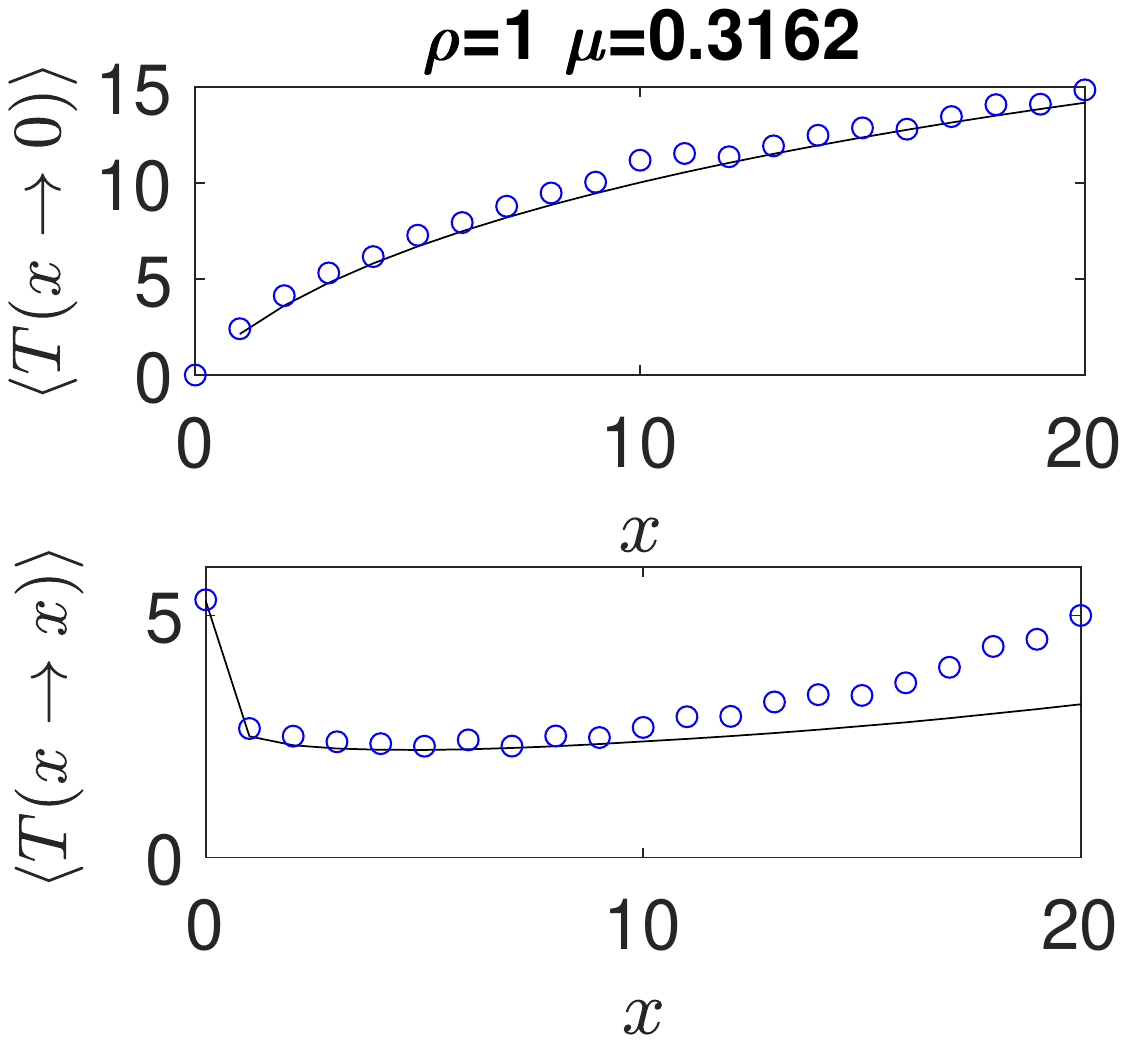} 
\includegraphics[trim=140 260 140 230, width=0.24\columnwidth]{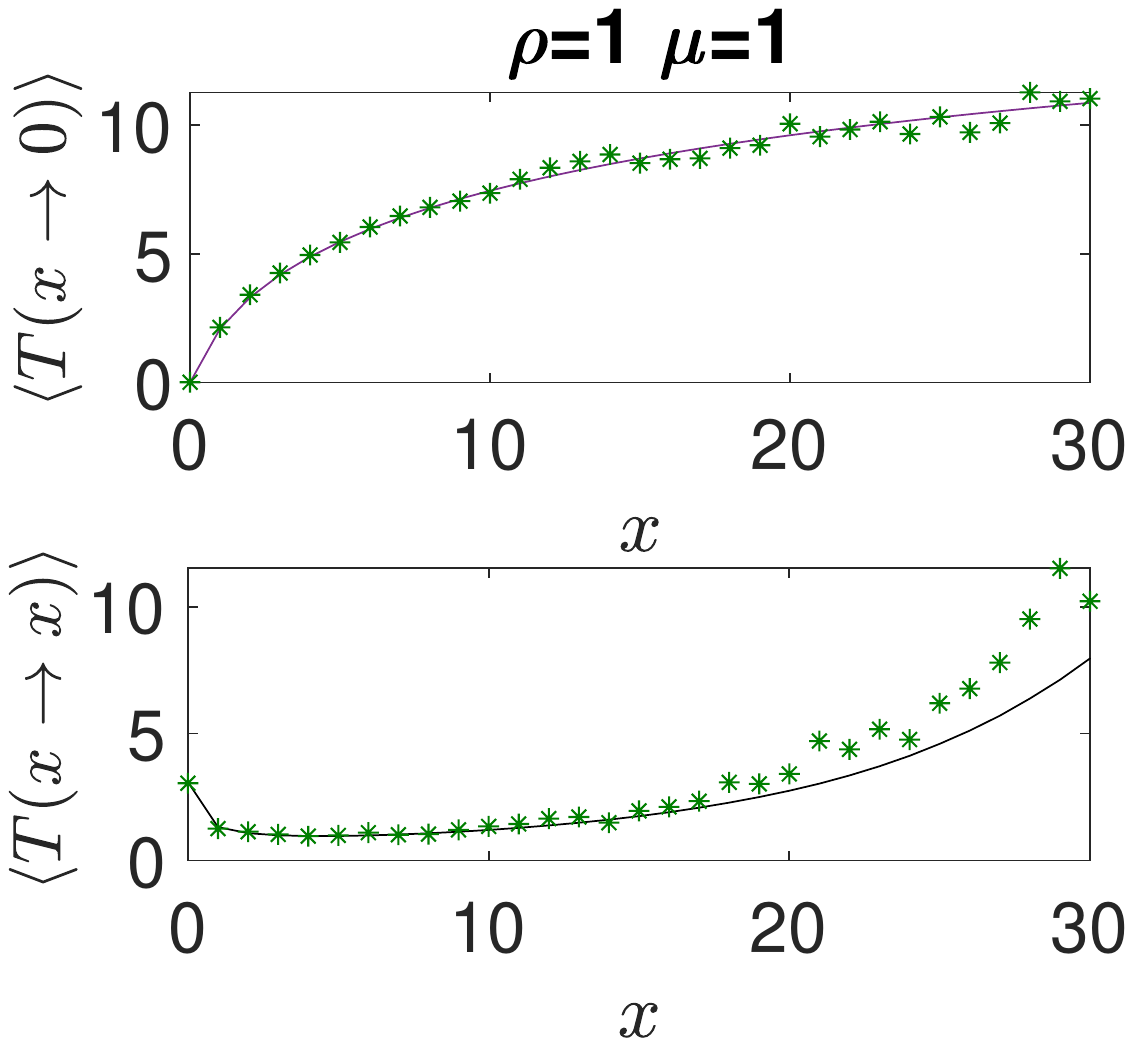} 
\includegraphics[trim=140 260 140 230, width=0.24\columnwidth]{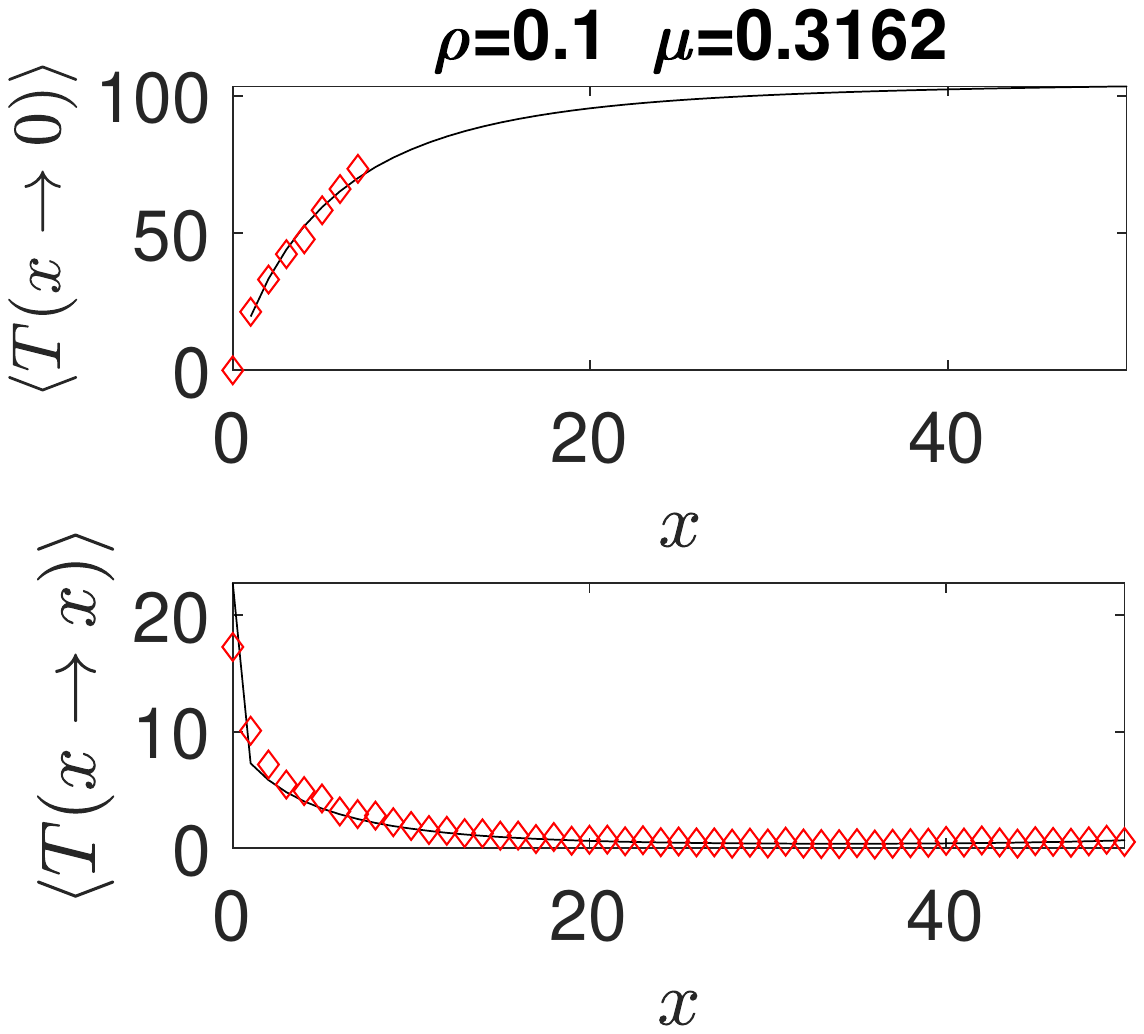} 
\includegraphics[trim=140 260 140 230, width=0.24\columnwidth]{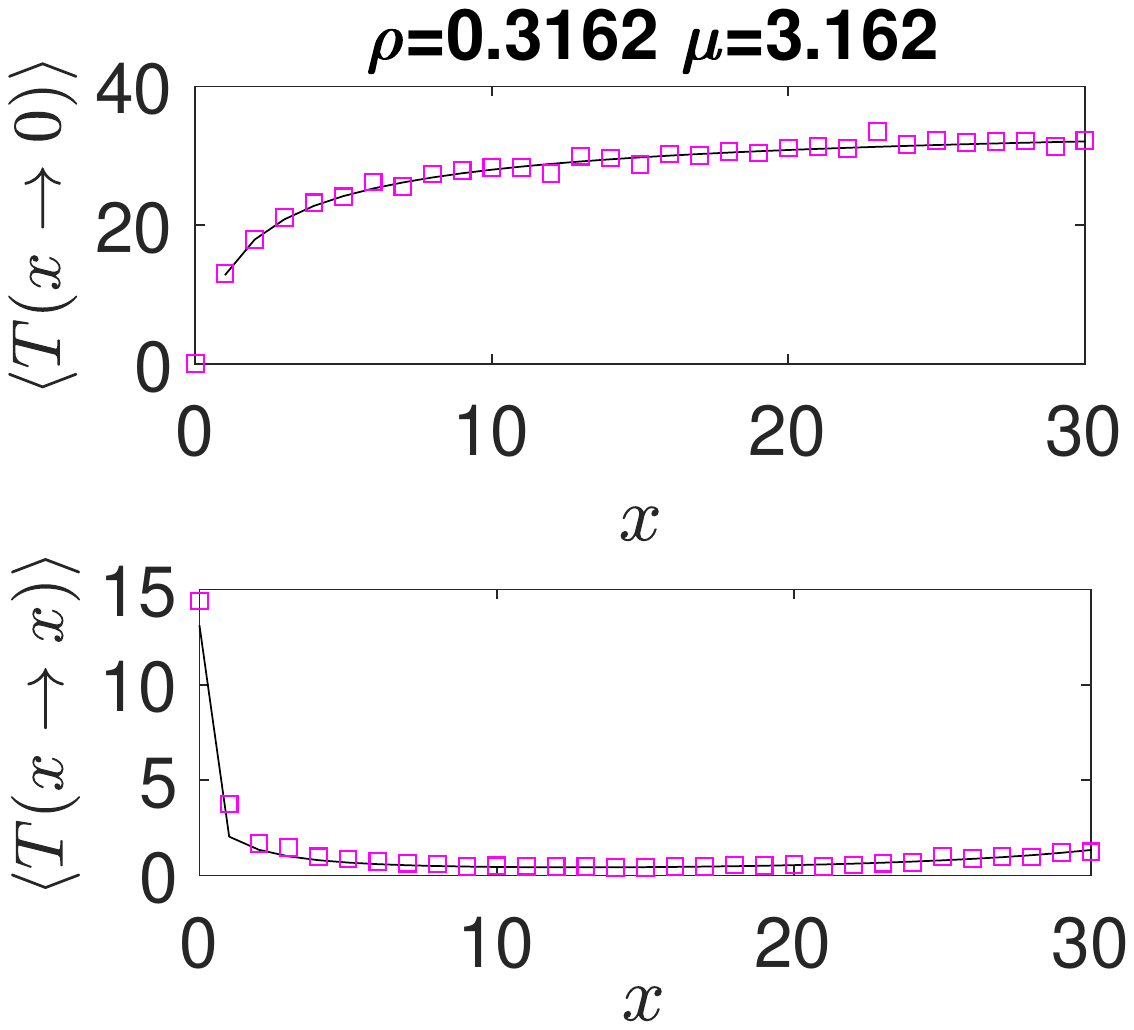}
\\
\includegraphics[trim=140 230 140 230, width=0.24\columnwidth]{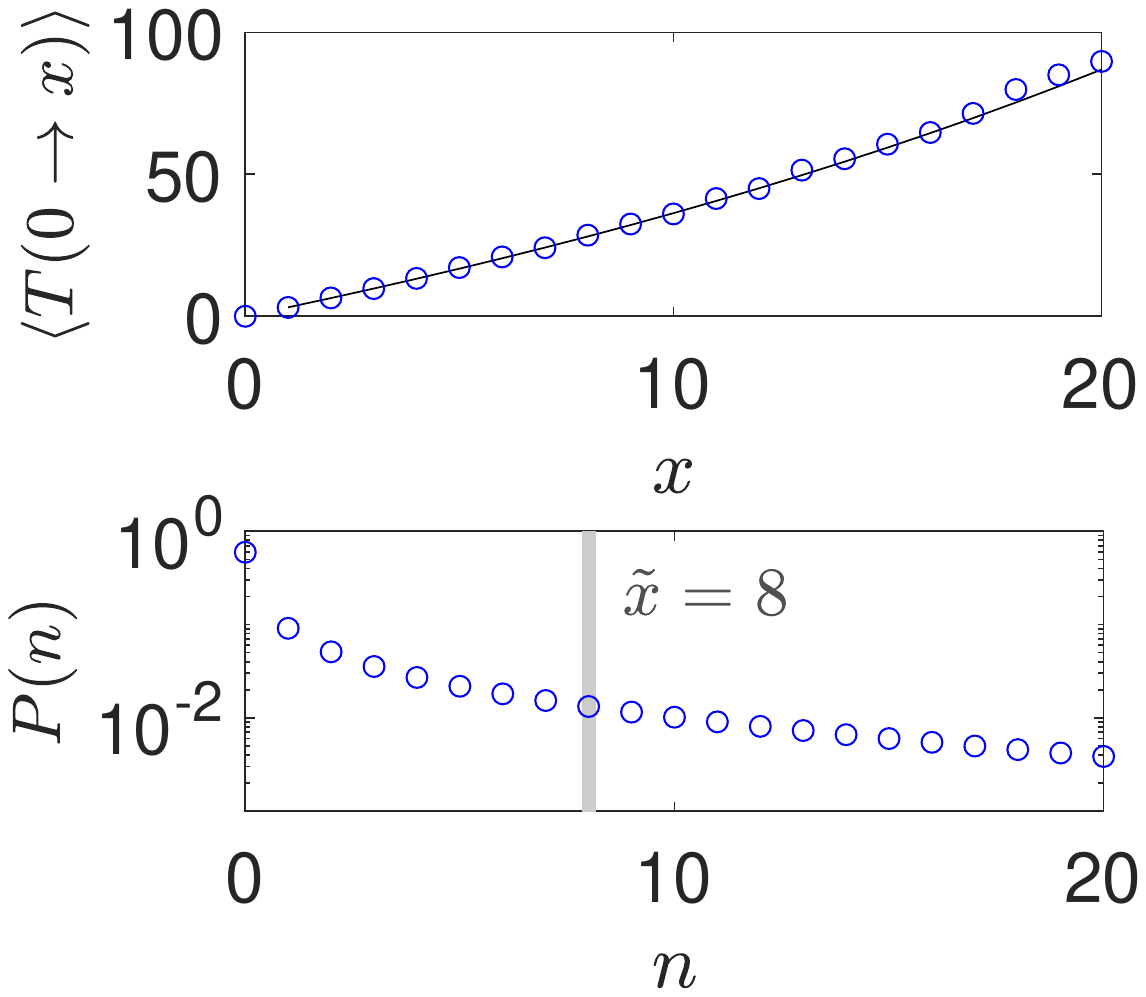}  
\includegraphics[trim=140 230 140 230, width=0.24\columnwidth]{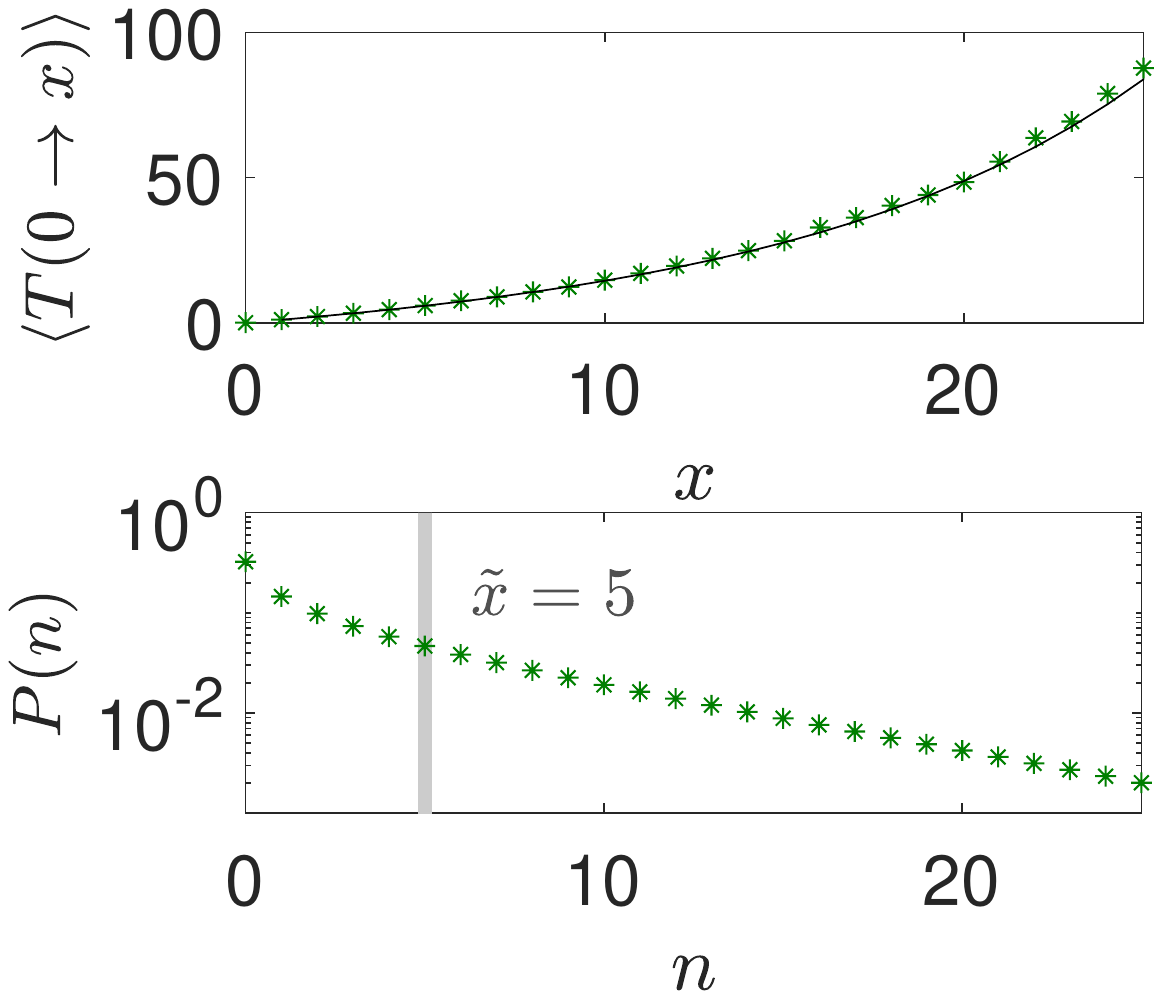}
\includegraphics[trim=140 230 140 230, width=0.24\columnwidth]{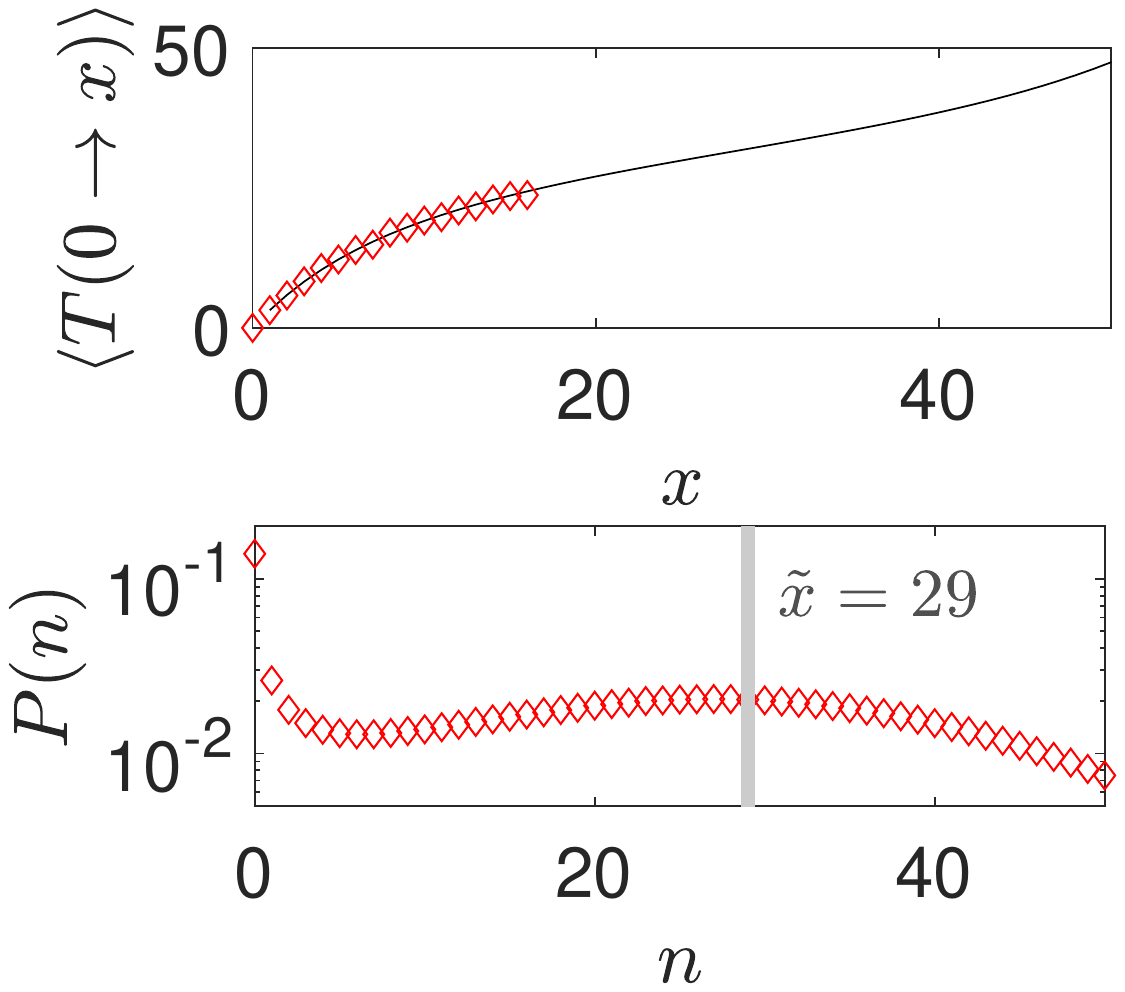}
\includegraphics[trim=140 230 140 230, width=0.24\columnwidth]{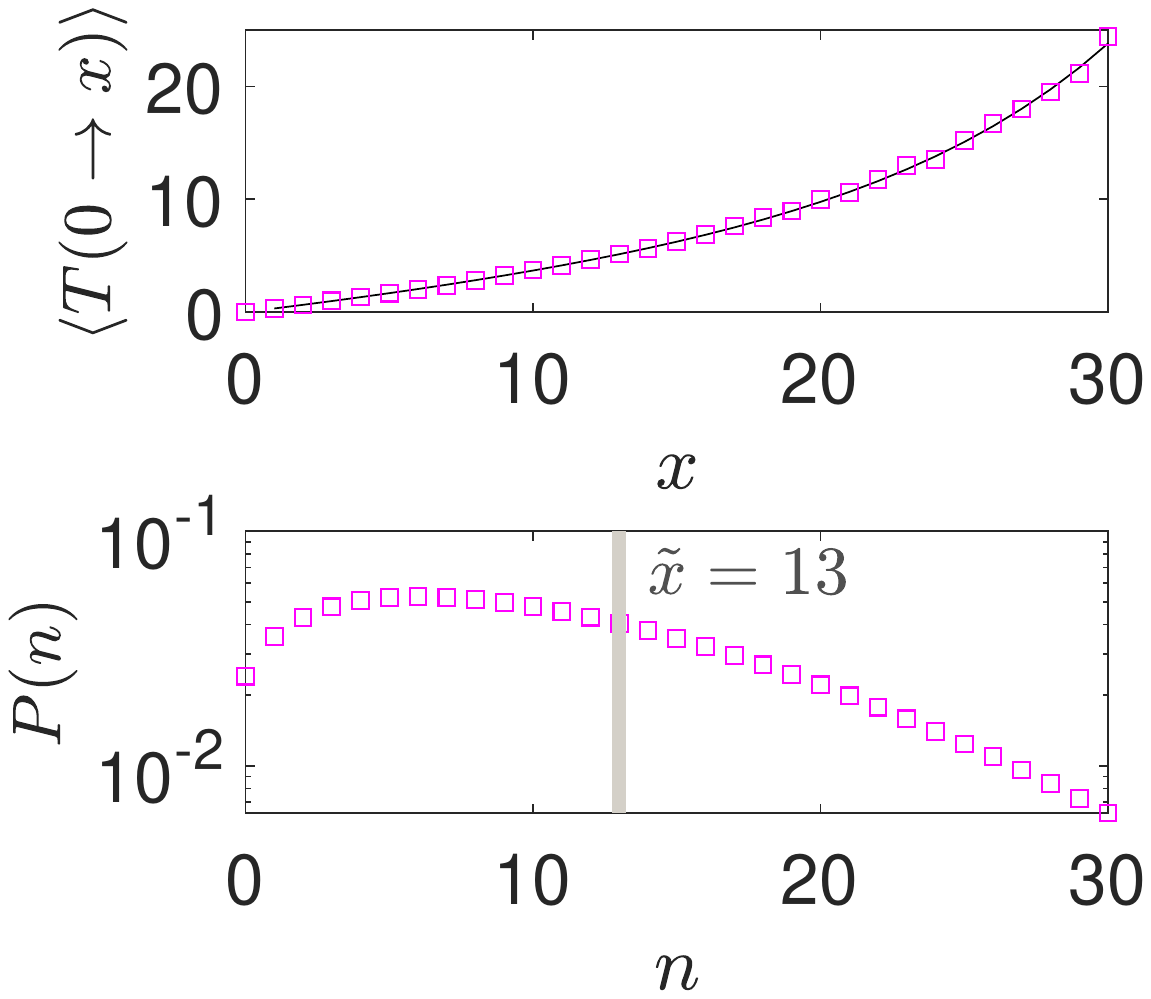}
\caption{Turnover time statistics. { Upper 4 panels: The raw data of the four MFPTs mentioned in the main text and used for the ratios figures(see Fig. 4 main text). Here $\tilde{x}$ refers to the deterministic LV fixed point. Importantly, these heatmaps were generated from Eq.~\ref{eq:MFPT_all} for a unidimensional process using the marginal distribution found from simulation. Lower panels: color markers represent  direct observations of the MFPTs of the full multi-dimensional system. We track a single species mean-first-passage-time from initial to final state. Black solid curves are the predicted MFPT from the unidimensional approximation. Note that the MFPTs $ T(x\rightarrow 0) $, $T(x\rightarrow x) $, and $ T(0\rightarrow x) $ are given as a function of $x$. 
The bottom row shows the corresponding simulated SAD.}}
\label{fig:MFPT}
\end{figure}

As mentioned in the main text, we concentrate on time scales associated with mean-first-passage-time (MFPT) from some initial abundance $i$, to a final one $f$. This MFPT, denoted as $T(i\rightarrow f)$, is inversely proportional to the transition rate from $i$ to $f$, where the first-passage times are exponentially distributed (see Fig.~\ref{fig:MFPT}).
Similarly, $T(i\rightarrow i)$ refers to the mean time of return to an abundance $i$ having left that same abundance.
In a unidimensional process over the interval $[0,\infty)$, the MFPTs we consider are  
\begin{eqnarray}
    \label{eq:MFPT_all}
     T (\tilde{x} \rightarrow 0) =   \sum_{y=0}^{\tilde{x}-1}\frac{1}{q^+(y)P(y)}\sum_{z=y+1}^{\infty}{P(z)}, & &
  T (0 \rightarrow 0) =  \frac{1}{q^+(0)P(0)}=  \frac{1}{\mu P(0)},\\ \nonumber 
  T (0 \rightarrow \tilde{x}) =  \sum_{y=0}^{\tilde{x}-1}\frac{1}{q^+(y)P(y)}\sum_{z=0}^{y}{P(z)}, & &
  T (\tilde{x} \rightarrow \tilde{x}) =    \frac{1}{P(\tilde{x})[q^+(\tilde{x})+q^-(\tilde{x})]} .
\end{eqnarray}
These expressions are exact for processes in one dimension \cite{gardiner1985handbook}, i.e. for a single species {whose dynamics are unaffected by other species. Using simulations, }we have found that these MFPTs agree with the multi-dimensional case, where many species evolve (see Fig.~\ref{fig:MFPT}). Hence, we substitute the SAD obtained from simulation, $P(n)$, in the expressions above. 

{In the main text, we discuss different MFPT ratios that reveal the different modality and richness regimes.
Note that these MFPT ratios can alternatively be understood as the reciprocal of the rates ratios which describe how much more frequently an event occurs than the other.}

{Overall, the `niche-like' Regimes I, II and IV  are characterized by a relatively stable behavior; generally species stay longer about the dominant species abundances, punctuated by the occasional crossings between dominance to nearly-extinct states and the reverse invasions from extinction into the dominance.
With regards to transition in richness shown in main text Fig.~2A, the transition between partial coexistence (b) to regime of competitive exclusion (c) is captured by the non-monotonic behaviour of the ratio as shown in Fig. 4C in the main text.
Surprisingly, unlike the other regimes, increasing the competition strength in the competitive exclusion regime (c) increases the stability of the dominant species abundance: return times to the dominant abundance are much shorter than the time to extinction for the single species in the frozen `quasi-stable' state.
Conversely, the `rare-biosphere' regime (III) features rapid dynamics where species cycle rapidly between extinction and a broad range of abundances without establishing `quasi-stable' states with slow turnover.
These different dynamic types are demonstrated by illustrative trajectory plots in Fig. 4A in the main text. 
}

\section{Species Rank Abundances vs Species Abundance Distribution }
\begin{figure}[ht!]
    \centering
    \includegraphics[trim=140 220 100 230, width=0.49\columnwidth]{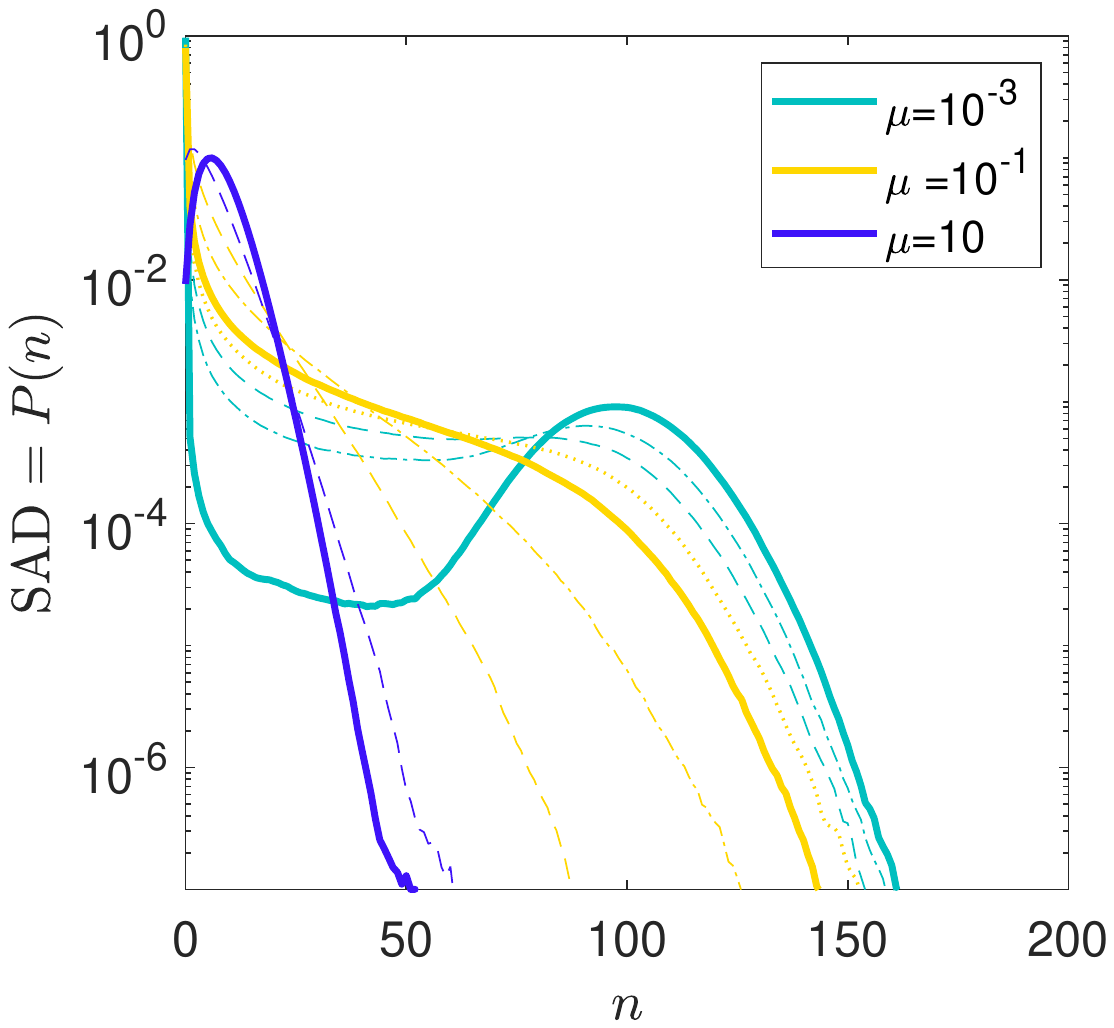}
    \includegraphics[trim=140 220 100 230, width=0.49\columnwidth]{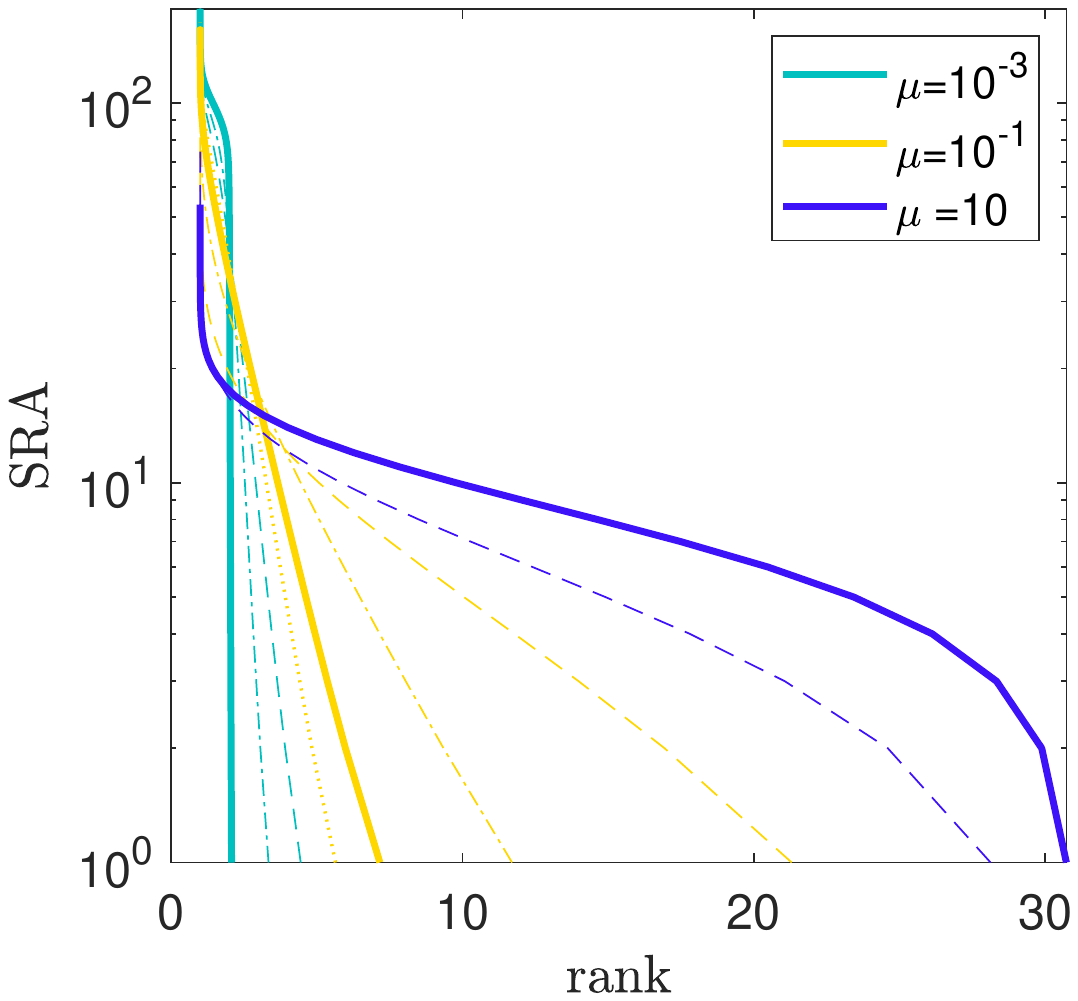}
    \caption{A comparison between the species abundance distribution (left panel) and its corresponding species rank abundances (right panel). Here $\rho=1$ is fixed, and the immigration rate $\mu$ varies. Solid lines represent $\mu\in [10^{-3}, 10^{-1}, 10]$ following the legend. The dashed, dotted and dashed-dotted lines are in-between $\mu$-s. The color scheme corresponds the modality classification of SAD; teal (bimodal, very low $\mu$), yellow (rare-biosphere, intermediate $\mu$) and blue (unimodal, high $\mu$).       }
    \label{fig:SRAvsSAD_rho1}
\end{figure}

In the majority of the manuscript we use species abundance distributions (SADs) along with dynamical properties to examine and classify processes into different regimes. 
However, in many experimental studies the species rank abundances (SRAs) are frequently reported instead, e.g see \cite{ser2018ubiquitous,mora2018quantifying,hubbell1979tree,descheemaeker2020stochastic,jeraldo2012quantification}. 
{We shall briefly show that t}he SRA is closely related to the cumulative distribution corresponding to SAD.
First, the cumulative abundance distribution is computed with ${\rm CAD}(n)\equiv\sum_0^n P(n')$. 
The most abundant species, namely species with rank 1, has abundance between ${\rm CAD}^{-1}(1-1/S)$ to  ${\rm CAD}^{-1}(1)$. The second most abundant species, i.e. rank 2,  has abundance between ${\rm CAD}^{-1}(1-2/S)$ to  ${\rm CAD}^{-1}(1-1/S)$, and so on. Therefore, the x-axis in Fig.~\ref{fig:SRAvsSAD_rho1} is computed with $1+S(1-{\rm CAD}(n))$ and the y-axis are the abundances $n$. Using this approach, we generated the SRAs corresponding to various SAD, see   Fig.~\ref{fig:SRAvsSAD_rho1} for for $\rho=1$. 
However, as is shown in Fig.~\ref{fig:SRAvsSAD_rho1}, classification through the SRAs is less discernible { than with the SAD}.

\clearpage

\newpage
\bibliography{bibliography}